To cite this paper

*Article*

# Design and Construction of a Brain-Like Computer: A New Class of Frequency-Fractal Computing Using Wireless Communication in a Supramolecular Organic, Inorganic System

**Subrata Ghosh, Krishna Aswani, Surabhi Singh, Satyajit Sahu, Daisuke Fujita and Anirban Bandyopadhyay ***

Advanced Nano Characterization Center, National Institute for Materials Science, 1-2-1 Sengen, Tsukuba, Ibaraki 305-0047, Japan; E-Mails: ocsgin@gmail.com (S.G.); krishna.aswani29@gmail.com (K.A.); surabhi.singh.mec11@itbhu.ac.in (S.Si.); satyaphy@gmail.com (S.Sa.); fujita.daisuke@nims.go.jp (D.F.)

* Author to whom correspondence should be addressed; E-Mails: anirban.bandyo@gmail.com or anirban.bandyopadhyay@nims.go.jp.

**Abstract:** Here, we introduce a new class of computer which does not use any circuit or logic gate. In fact, no program needs to be written: it learns by itself and writes its own program to solve a problem. Gödel's incompleteness argument is explored here to devise an engine where an astronomically large number of "if-then" arguments are allowed to grow by self-assembly, based on the basic set of arguments written in the system, thus, we explore the beyond Turing path of computing but following a fundamentally different route adopted in the last half-a-century old non-Turing adventures. Our hardware is a multilayered seed structure. If we open the largest seed, which is the final hardware, we find several computing seed structures inside, if we take any of them and open, there are several computing seeds inside. We design and synthesize the smallest seed, the entire multilayered architecture grows by itself. The electromagnetic resonance band of each seed looks similar, but the seeds of any layer shares a common region in its resonance band with inner and upper layer, hence a chain of resonance bands is formed (frequency fractal) connecting the smallest to the largest seed (hence the name invincible rhythm or Ajeya Chhandam in Sanskrit). The computer solves intractable pattern search (Clique) problem without searching, since the right pattern written in it spontaneously replies back to the questioner. To learn, the hardware filters any kind of sensory input image into several layers of images, each containing basic geometric polygons (fractal decomposition), and builds a network among all layers, multi-sensory images are connected in all possible ways to generate "if" and "then" argument. Several such arguments and decisions (phase transition from "if" to "then") self-assemble and form the two giant columns of arguments and rules of phase transition. Any input question is converted into a pattern as noted above, and these two astronomically large columns project a solution. The driving principle of computing is synchronization and de-synchronization of network paths, the system drives towards highest density of coupled arguments for maximum matching. Memory is located at all layers of the hardware. Learning, computing occurs everywhere simultaneously. Since resonance chain connects all computing seeds, wireless processing is feasible without a screening effect. The computing power is increased by maximizing the density of resonance states and bandwidth of the resonance chain together. We discovered this remarkable computing while studying the human brain, so we present a new model of the human brain in terms of an experimentally determined resonance chain with bandwidth $10^{-15}$ Hz (complete brain with all sensors) to $10^{+15}$ Hz (DNA) along with its implementation using a pure organic synthesis of entire computer (brain jelly) in our lab, software prototype as proof of concept and finally a new fourth circuit element (Hinductor) based beyond Complementary metal-oxide semiconductor (CMOS) hardware is also presented.

## 1. Introduction

The turing tape concept was introduced in the 1940s [1–3]. It suggests that all events around us could be written in the form of a tape or series of logically defined steps. All major brain-building projects [4–9] and the unconventional computing [10–16] follow the Turing path of computing either by suggesting that eventually their novel device is a logic gate or could be reduced eventually as an output of a series of sequential events. Starting from Quantum Computer (QC) [17], Cellular Automaton (CA) [18,19] to the Echo State Model (ESM), [20] always, it has been a trend to construct an equivalent of a logic gate, which is a reduction protocol for multiple choices. Sometimes, randomness is restricted to generate decisions, though apparently robust restrictions are imposed on the randomness in reservoir computing [21,22]. The problem with the Turing tape based computing is that all possible arguments should be known beforehand and entire processing scheme should be defined strictly, as an output of step-by-step logical reduction process. In the 21st century, the first problem we face in computing is that the amount of data sets we wish to process is incredibly large. Therefore, if we search entire database one by one to find a suitable data, instantaneous decision-making would be impracticable, so we have to complete the search without searching. Hypercomputation or computing beyond Turing came in 1999 [23].

**CTC and use of multiple clocks in a single hardware:** Looking beyond quantum computing, it is shown that the use of Close Time-like Curve (CTC) could enable solving NP complete problems much efficiently [24]. Recently, it has been argued that the CTC does not require time travel in the past, the only thing we have to do is to have clocks running at different speeds on simultaneously existing physical worlds where the same events are taking place. To solve the problem, the system point of a clock that has "one second" resolution moves to the world with a faster clock say "one microsecond" resolution, gets the information and returns to the "one second" resolution world where to an external observer computing is being performed with no detectable time lapse in the "one second" resolution clock as shown in Figure 1a [24,25]. This is similar to harnessing "negative time" of quantum mechanics but instead of one here we have multiple imaginary worlds, each with a different clock-speed. Instead of CTC we can use the fractal made of resonance frequency bands and investigate the possibility of realizing a similar advantage [26]. We consider multiple concentric spheres generating a multi-layered architecture, each layer with one type of clock, for each layer, the layers above and below are imaginary spaces. One important question is why do we consider the layers beyond in the imaginary spaces, even though they exist in reality? The reason is that we also consider that the materials of a layer is used as a seed that construct the next layer as shown in Figure 1b, hence, the dynamics of each layer are totally different. Say, atom makes molecule and molecule makes crystal, all three, atoms, molecules and crystals are very distinct materials in terms of their dynamics and resonance properties. When we probe molecule in a crystal, the dynamics of crystal and atoms do not appear, they become non-existent; thus, came the concept of imaginary spaces.

**Figure 1.** (**a**) Resonance frequency limits construct clocks at different layers, layer 2 = A (highest frequency region, clock is very fast), layer 3 = B; layer 4 = C; layer 5 = D; layer 6 = E (lowest frequency region, clock is very slow); (**b**) A is basic seed, it assembles into B, several B makes C, several C makes D, several D makes E; (**c**) Development of computing speed, current situation and the ultimate speed.

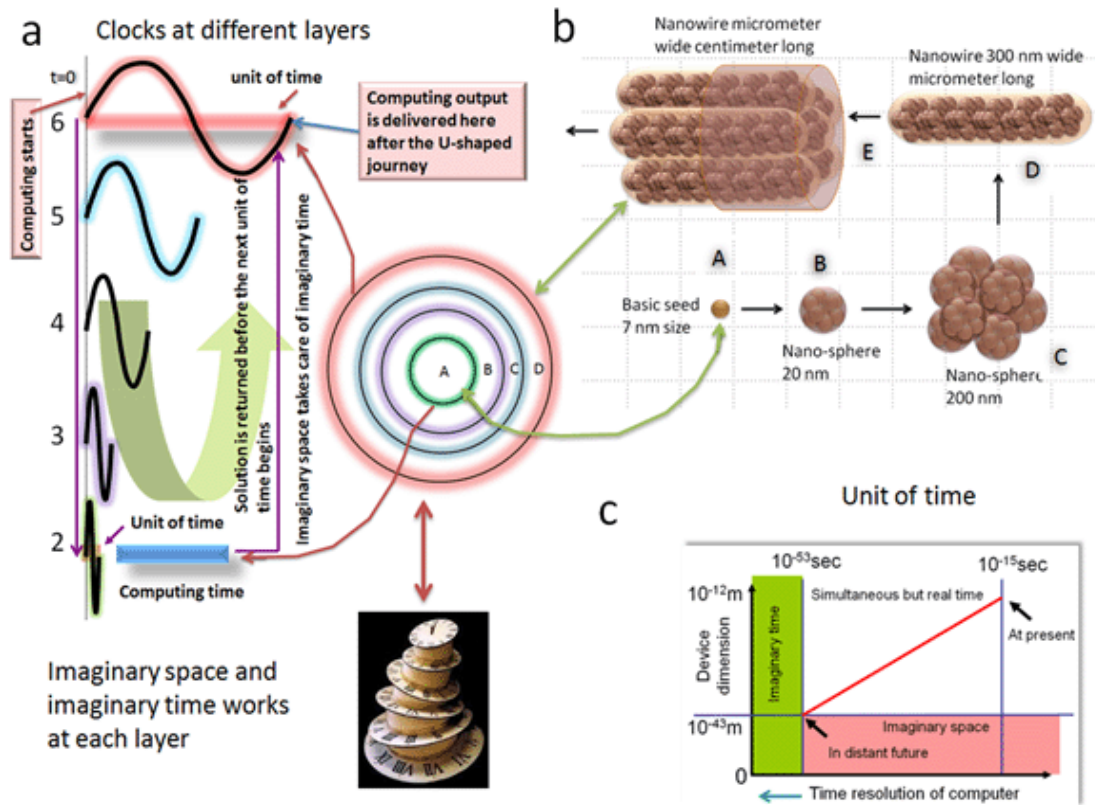

**Devising a Gödel's incompleteness harvesting machine:** The second problem is that the incredibly large data sets change continuously following a protocol that also evolves with time [27]. These data sets are correlated with unknown relationships and deliver unpredictable outputs at any random time. This situation is like Gödel's incompleteness argument and in this manuscript we have addressed the incompleteness concept from an engineering perspective. Since we need to write a software program for any kind of Turing based computing before the computation begins, instantaneous decision-making is not viable. When reading the entire database takes so much time, it is trivial to first identify the rules of large dataset evolution and then code them instantly. This problem is very different from the "simultaneity" issue. In the far distant future when if we have a technology that can differentiate between two events with a gap of Plank time, then all simultaneous and massively parallel events of todays will be converted into a series of sequential events in the future with the advent of new technologies as shown in Figure 1c [28].

**A new class of second order and first order fusion logic: Infinite set is not the key, classification into multiple space-time worlds is the key:** However, a class of problems would remain unresolved wherein the hardware enables a continuous coupling and decoupling of arguments at higher levels. In this situation, several "if-then" arguments made of a set of resonance peaks as shown in Figure 2a get coupled and form a cluster of arguments, then those clusters couples further in a complicated way as shown in Figure 2b, several coupling routes between clusters of arguments exist at a time and none of these situations are written in the hardware. The situation looks like a column as shown in Figure 2c where the base part is visible since those are written in the hardware as basic "if-then" statements and an astronomically long column of coupled "if-then" statements remains invisible. The column is an invisible reservoir, and is never a part of the classical Turing tape, every cell of the new tape that could incorporate this column has several imaginary cell space above and below and is related to them at

various imaginary times. Thus, in the manuscript, we name it "Frequency Fractal tape". When an input pattern of "if-then" arguments is applied to this column, the closest similarity region in the column is projected back as output. Since column part has never been programmed into the hardware, the solution comes from a region that is not logically embedded in the system. Only real visible part of the system is the column base. This is a perfect example of Gödels incompleteness argument where the decision is made in a hardware from an analysis that is not defined in the system in its real space or real time (in number system: consistent effective theory T containing Peano Arithmetic, the formula CT expressing the consistency of T cannot be proven within T). Search and find an astronomical amount of data in seconds does not mean that we will be able to configure the rules to simulate the future course events [29]. Thus, we begin our adventure of computing within the domain of Gödels incompleteness argument, which remained neglected as mere philosophy for nearly a century [30]. For us, we do not say complexity makes a system incomplete, rather, it's something beyond real space time (imaginary space time values ~ Platonic values). Thus, it is not classical second order logic that has an infinite set of arguments, it is about the classification of arguments in different space-time worlds. Irrespective of the order of logic assigned to our computing protocol, it follows fundamental features (i) just like conventional second order logic, it cannot be deduced to a first order logic and at the same time, infinite set is not an essential requirement to be in the second order logic class; (ii) since entire resonance chain computes together, it is possible to project a very logically defined protocol from the astronomically long 3D column of arguments unlike second order logic (second order logic cannot satisfy three attributes simultaneously, soundness, completeness and effectiveness), though major part of the column remains undefined. Thus, its a unique generic system of first and second order logic combined, which part of the 3D column would follow first order logic and which part would follow second order logic can not be determined in a finite algorithmic way. Thus, it is a purely new class of computation.

**Figure 2. (a)** Resonance band of several materials from several layers couple and form a set of peaks, this is called "if" and after some time if a new set of peaks get activated this is "then". This event is called phase transition; **(b)** How coupling = self assembly and phase transition occurs **(c)**. Two columns, left, column of if-then arguments and right, column of rule of phase transition connected to each other.

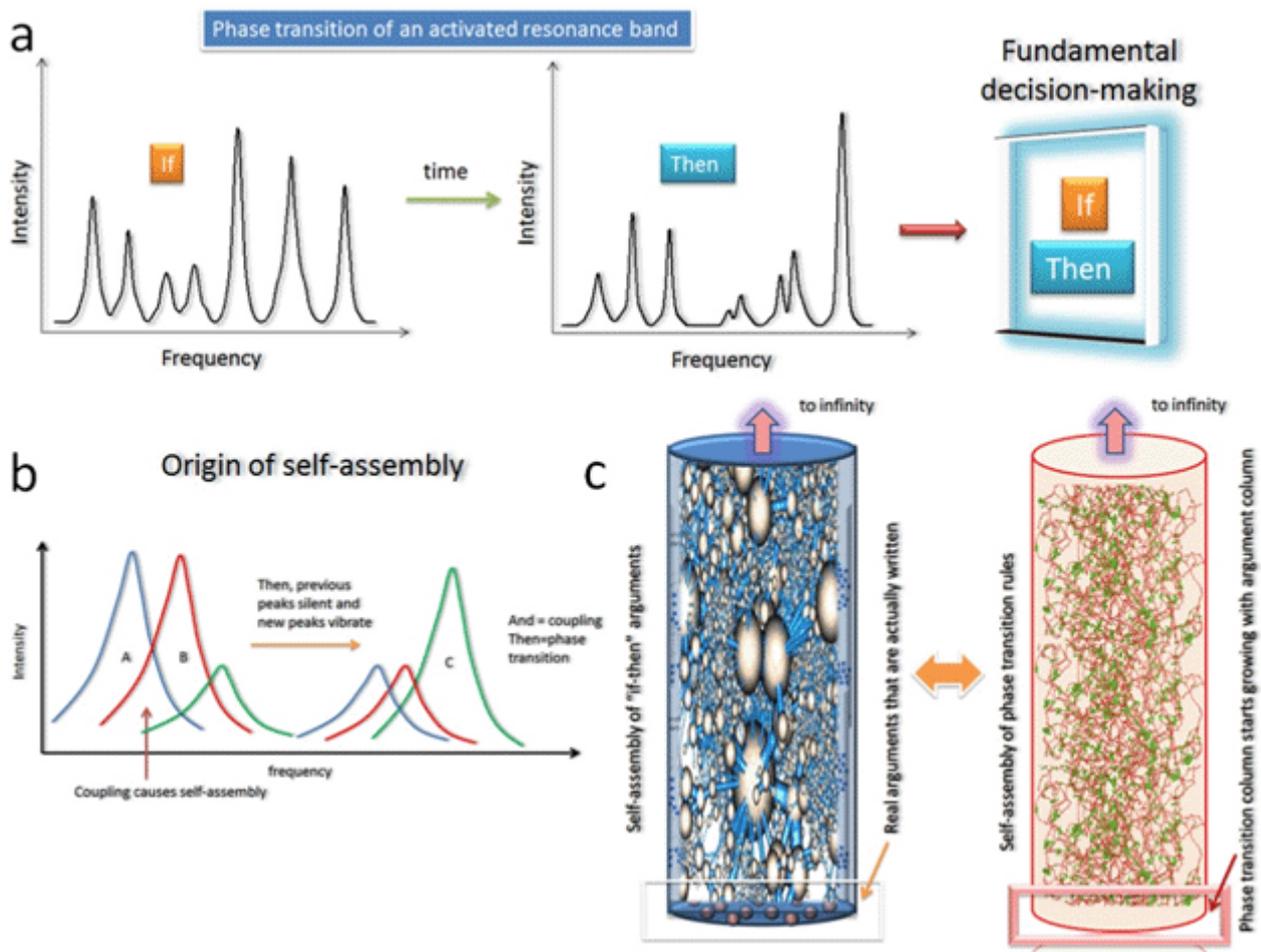

## 2. The Outline of the New Class of Computer: An Analogy with Existing Von-Neumann Computers

The computer has no circuits. It has no switches or logic gates. It's a multi-layered hardware, means, if A is a basic seed material, several A would self-assemble to form seed B, several B would self-assemble to form seed C, and this would continue to say up to eight layers, A to H, now seed H is the computer, which has G, F, E, D, C, B and A layers inside. Each layer has at least three electromagnetic resonance sub-bands, one sub-band is common to the upper layer and another is common to the lower layer. Thus a chain of resonance band is formed extending from A to H, a complex function AJ = F(x, y) + iG(x, y) represents the chain comprehensively but not completely. A resonance peak means a sharp current is allowed to pass through a material at a very small frequency window that keeps the material oscillating electromagnetically and/or electromechanically. Since, self-similarity of resonance bands (triplet) from A to H is derived from the expression of this function AJ, we call it frequency fractal (resonance chain = frequency fractal). Before learning begins, the entire resonance chain looks identical at any level A to H, self-similarity is absolute. As the computer learns, "if-then" arguments are stored by shifting resonance peaks via a conformational change in the seed and the absoluteness of self-similarity disappears. We get an inhomogeneous resonance chain. The difference between the ideal pristine resonance chain (serves as background) and the modified resonance chain is the real information or argument content of the computer. Resonance chain plays a primary role for searching a pattern without performing a real "search".

The computer solves only an intractable "Clique" or "pattern search" problem, so converts all other problems in to this class. It is a pattern based computer, where no software is required to be written. It converts any input information in terms of a time series of 2D images, each pixel in an image is a spike pulse, thus, the time series is a 3D wave train. Irrespective of the nature of original information, sound, visual, taste, touch or smell, the geometrical associations of the frequency pattern are constructed in the form of polygons ranging from triangle to sphere, thus, angle, 3D perceptions are automatically taken care of as linked sets and subsets as repetition of basic geometric shapes. This conversion is called fractal decomposition. We call these basic geometric shapes as fractal seeds, several fractal seeds couple via resonant coupling to define an "if-then" argument. Several such arguments self-assemble and form a complex column of arguments, direct measurable real arguments lie at the base of the column, and spontaneously self-assembled astronomically large number of arguments reside above this layer. If an "If" set of arguments are resonantly triggered more than a certain time period, "then" set of resonance is activated, such phase transition rules are encoded automatically during self-assembly of "if-then" arguments, thus, another complex column is created. Since each resonant vibration has several harmonic and inharmonic overtones, heights of both columns are astronomical. Both the columns interact continuously with a unique sequential dynamics. Thus, at the end of the analysis, a 3D dynamic map is created (3D network changes with time, hence dynamic) and saved. In this map, independent basic fractal seeds construct the base, and their subgroups/supergroups generate a network around this base. The entire 3D map could represent a sound, picture or even a taste, or even combination of them generating a higher-level perception. Now this 3D map converted in terms of fractal seeds without compromising the fundamental features of spatial or temporal information encoded in it, is the equivalent of an algorithm used in a von-Neumann computer to solve a problem.

When a problem is asked, a fractal seed based 3D network (pattern) is created for that problem as described above and that is matched via synchrony with the column of arguments noted above. Both synchronization and de-synchronization of fractal seed map continue during the matching process, the matching of a dynamic map of fractal seeds actually means matching of the resonance peaks only. The matching process continues along with the phase transitions (from its own column of rules) between clusters of different sizes of fractal seed sets with a motivation to track and activate the maximum density of coupled resonant oscillations in the argument column, and deactivate the routes with very low density coupling. This is the fundamental driving mechanism for computation. Once an equilibrium is reached, a 3D map is derived and it is sent as a solution. This 3D map is a projection from the dynamically changing large columns of arguments and rules of phase transitions. Thus, we could alternately view the process with a von Neumann eye: the internal 3D map or the software program writing engine writes a code for solving a problem when it matches with the external information generated 3D map, it executes the program when its unique phase transition protocol is derived. To sum up the analogy with conventional computer, we state: inside the column of arguments, astronomically large numbers of software code writing protocols are embedded with a unique technology, which emerges depending on the problem asked.

## 3. Spontaneous Reply-Back: Performing a Search without Searching

Majority of computational algorithms developed in the last three decades have considered that the devices that holds the optional solutions of a query could listen to the question but could not reply-back to the questioner simultaneously and spontaneously. Thus, to learn the location for addressing each option specifically, the wiring of computing elements became necessary in the computer chips [31]. The circuit is a liability even in a quantum computer. Additionally, we need a program that coordinates the process for a system point to reach to the individuals and retrieve the replies one by one, several protocols are adopted to decrease the computing time, a summary of the physical principles are noted in the Figure 3a. The quantum protocol only decreases the number of queries, with $Log_2 n$ advantage over classical, n is the size of the search space, but "reply back" requires an antenna and receiver attached to the memory elements and new kind of the identification code. Grover's algorithm suggests that due to entanglement any number of people in a group could be considered as one object/choice, if any classical route allows such group-test such that classical computing would match the quantum computing [32]. This is shown in Figure 3b. Additionally, except for a few problems (factorization), the quantum protocol does not provide sufficient speed up. The reason for exponential speedup is the sharing associative matrix D, which requires a peculiar requirement in the nature of the problem. Moreover, for pattern search, the matrix D needs to be redefined for each network mode, hence entanglement needs to be broken, which would collapse the speed up. There will be no difference between a classical and a quantum search. Exponential speedup is not the prerogative of quantum entanglement; it could be realized in a pure classical systems too [32,33]. Here, we look beyond exponential speed up and suggest "Spontaneous reply back" that supersedes the exponential speedup promised by a quantum computer, we explain the reason below.

Instead of $Log_2 n$ attempts, we want to make only one query, and the solution would reach the questioner, then the size of a search space becomes irrelevant. If we use a multinary switch (with more than two decisions) [34,35] that has an antenna for each state in addition to the sensor, it can radiate out the solution in all directions, so irrespective of n, the questioner gets the answer in one attempt (Figure 3b). We have already demonstrated this technology [36]. Fundamentally, our basic information processing device will be an oscillator attached to an antenna and a receiver, the oscillator is so designed that we can write/erase multiple resonance states. When an electromagnetic signal is applied to the device, multiple resonant oscillator circuits absorb energies specific to the resonance frequencies and start oscillating. This oscillation turns the system unstable; as a result, the energy is radiated outside via non-radiative coupling as shown in Figure 3c, noise does not mix with signals as the absorption occurs only for signals with the perfect matching frequencies, at the same time, the emission is always quantized.

**Figure 3.** (**a**) Time taken for computing per step for increasing population or search space N is plotted for computing when "coherence" is used by various means; (**b**) A comparison between classical, quantum and "reply back" routes; (**c**) Energy transfer for two inducting coils, far distance communication are presented in different orientations; (**d**) A wave train in our computer, four frequency domains.

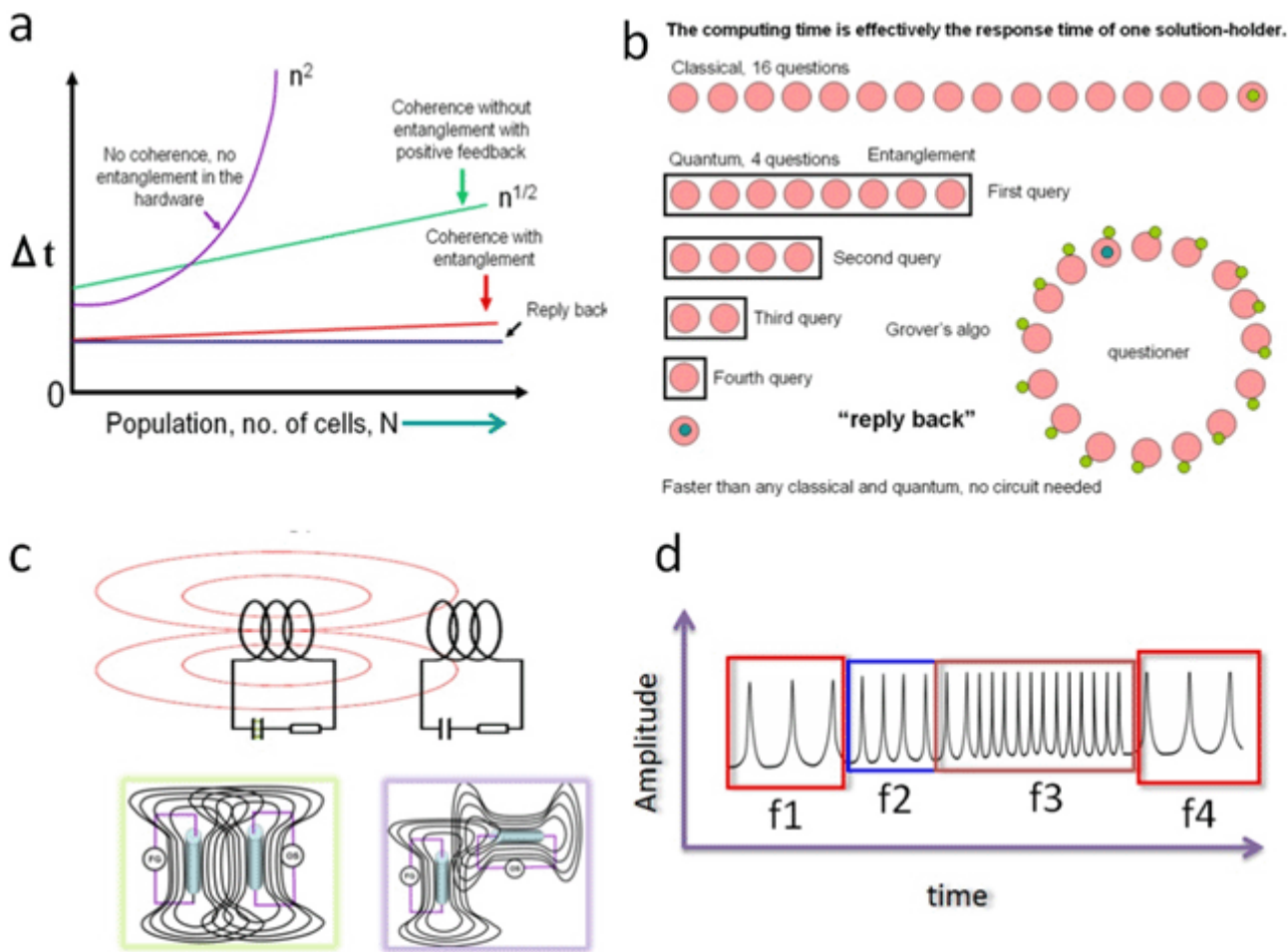

Since, irrespective of the nature of information, processing and communication should occur via coupling the resonance peaks, thus, operational and computational language of our artificial brain is written using only one parameter, "frequency". This principle is shown in Figure 3d. A wireless resonant non-radiative power transfer ensures reliable communication even when the noise amplitude is more than the signal [20,37]. An accurate frequency match is the key, and natural vibrational frequency of a material cannot change by an external noise, thus, solution holder spontaneously sends the reply even under noise, and penetrates any material that does not match its frequency [38]. Even under massive noise, these resonance frequencies act as an attractor during synchronization, when the system is dragged away, this basic natural property pulls back the system to the particular frequency state. Therefore, modulating the natural frequency and harnessing its advantages are the keys to our new class of computing, that explore materials "electromagnetic transparency".

**Why a chain of resonance for scale free "reply back"?**

We have explained above, reply back via non-radiative energy transfer. This technology relies on the "electromagnetic transparency" of the material, however, due to large reflection coefficient the transparency develops opacity. Thus, screening effect restricts wireless communication beyond certain limits of the immediate neighborhood. For this reason, we cannot rely on an antenna-receiver concept to scale up the "reply back" technology, where radiation energy passes through the air between two materials, the philosophy requires a fundamental change. Alternatively, we have introduced a multi-layered structure where output structural product of one layer is used as seed for the next layer.

**Figure 4.** (**a**) Two layers A and B, common resonance band exchanges energy; (**b**) A random form of energy given to layer C is eventually distributed among all layers and resonance bands of those layers get activated; (**c**) The resonance chain, energy transmission both ways; (**d**) A blue ball is kept at the topmost layer, if each layer has two seed structures then how energy will transmit, the path is shown here.

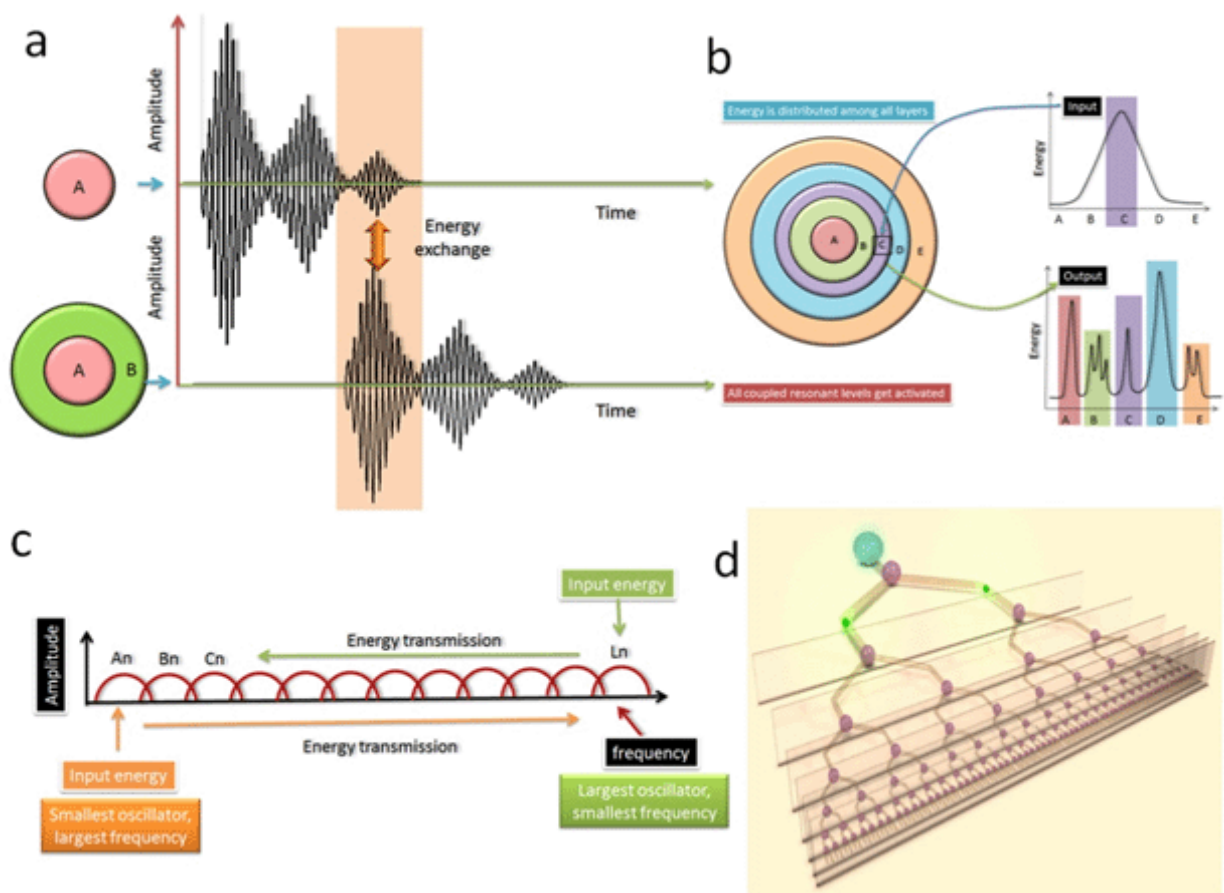

For each layer, the resonance band has three distinct domains, one domain is used to communicate with the inner layer, and one with the external layer, one layer is kept for its own information processing, as shown in Figure 4a. Thus, all layers are energetically connected by a single chain of resonance band, irrespective of the size of the device architecture now a wireless communication can transmit without getting screened anywhere. Energy given as an input at any layer transmits to the entire chain of resonance bands, both ways, towards the lower and towards the higher frequency regions of the chain as shown in the Figure 4b. Any form of energy is suitably absorbed and then transmitted across the resonance chain as shown in Figure 4c. A resonance chain connects every single computing seed in the system, thus, zillions of seeds are wired into a massively complex yet a single network as shown in the Figure 4d. It means energy given or radiated out from any part of the chain of materials does not need to leave the material, pass through the air and then return to the material, instead, resonance chain provides a safe passage, reflection/transmission becomes irrelevant. Thus, we realize a new class of wireless communication. It is remarkable that due to the basic material design for maintaining the single chain, screening effect does not affect the communication even though billions or zillions of devices are assembled together.

**No need to know location, hence, connecting circuit is not required:**

Our artificial brain does not use any meaningless "bits", and then converts the suitable information with it, like in a conventional computer, it directly captures and uses "if-then" arguments, writes every image and its groups, compressed into a fractal form of those arguments, and those fractal seeds and their association fractals are stored as a complete network of arguments. Even triggering a single "if-then" argument expands enormously in scale, since far distant sensory parts are also coupled (like the image of a lotus, associated sound and its smell) by a new fractal function, then entire function activates and responds synchronously to the smallest argument, and all associated dynamic routes are triggered. This is the most remarkable advantage of using a complex frequency space based computing, information located in any layer in any seed would eventually "reply back" simultaneously and spontaneously. Any local frequency map say located in a seed triggers one layer above and above finally covers the entire device via resonance chain, and if necessary, then synchronization could activate other functions located entirely in a new region, down to the entire local nets associated with it [39]. This is called an umbrella path, which we will describe in details later. Therefore, we do not need one to one wiring for the spontaneous reply neither inside a sensory region, nor outside. We can also suggest that whole computing hardware acts like a single molecule oscillator, hence the location of the seeds in the womb of another seed is taken care of in an unprecedented manner.

**The NP complete intractable "Clique" problem and a universal "reply back" protocol:**

The Clique problem is a very well defined intractable problem [40], we have chosen this problem for designing our computer because we want to construct a pattern based computer just like a human brain. The clique problem needs to be solved in a finite time for any advanced cognitive or creative intelligence observed even in primitive neural networks, this is our perception about brain engineering.

In a Clique problem a 2D or 3D random network made by connecting large number of points as shown in Figure 5a. Obviously, the number of possible patterns that could be generated from this composition is astronomically large. Now, if we want to search a given unknown pattern in that resource pattern, it is not possible to find that pattern with any computer within a finite time. However, if those points have the properties to reply back together spontaneously then we can get the search result without searching as noted in Figure 5a. As noted above that to avoid the screening effect we need a new kind of material that would follow the resonance chain throughout the architecture. There exist several classes of the clique problem originally proposed in 1949, as frequently observed in the classic intractable problems, certain constraint are imposed to simplify the complex network and then an algorithmic route is found to solve that problem. However, in this particular case, we take any sensory data, visual, sound, touch, taste or smell in the form of a 2D pattern and from that image we transform the single image into several layers of images, each containing several different classes of "frequency fractal" seeds as shown in Figure 5b. During transformation each layer distinctly represents a particular type of fractal seeds, based on size of the basic geometric shapes used that incorporates global relationship of elements in a pattern. The network between various different kinds of fractals are generated due to the typical conical architecture of the entire computing network as shown in Figure 5c. Finally, as noted above, all layers co-exist with a complex connection between the fractal seeds inside a layer and between multiple layers as shown in Figure 5d. The magnitude of many to one and one to many connections and directivity of connections ensure that even a single image to be treated as a 3D tensor not a matrix, thus, even the simplest pattern with 20 pixels belongs to the intractable class.

**Figure 5.** (**a**) A 3D clique network, the red noted part replies back to the query; (**b**) Five sensory columns hold a large number of fractal seeds, they get connected to form a complex argument; (**c**) Number of oscillator increases incredibly as we move higher frequency scale, conical column = AjoChhand computer; (**d**) A single image is divided into multiple layers, each path is a network of fractal seeds.

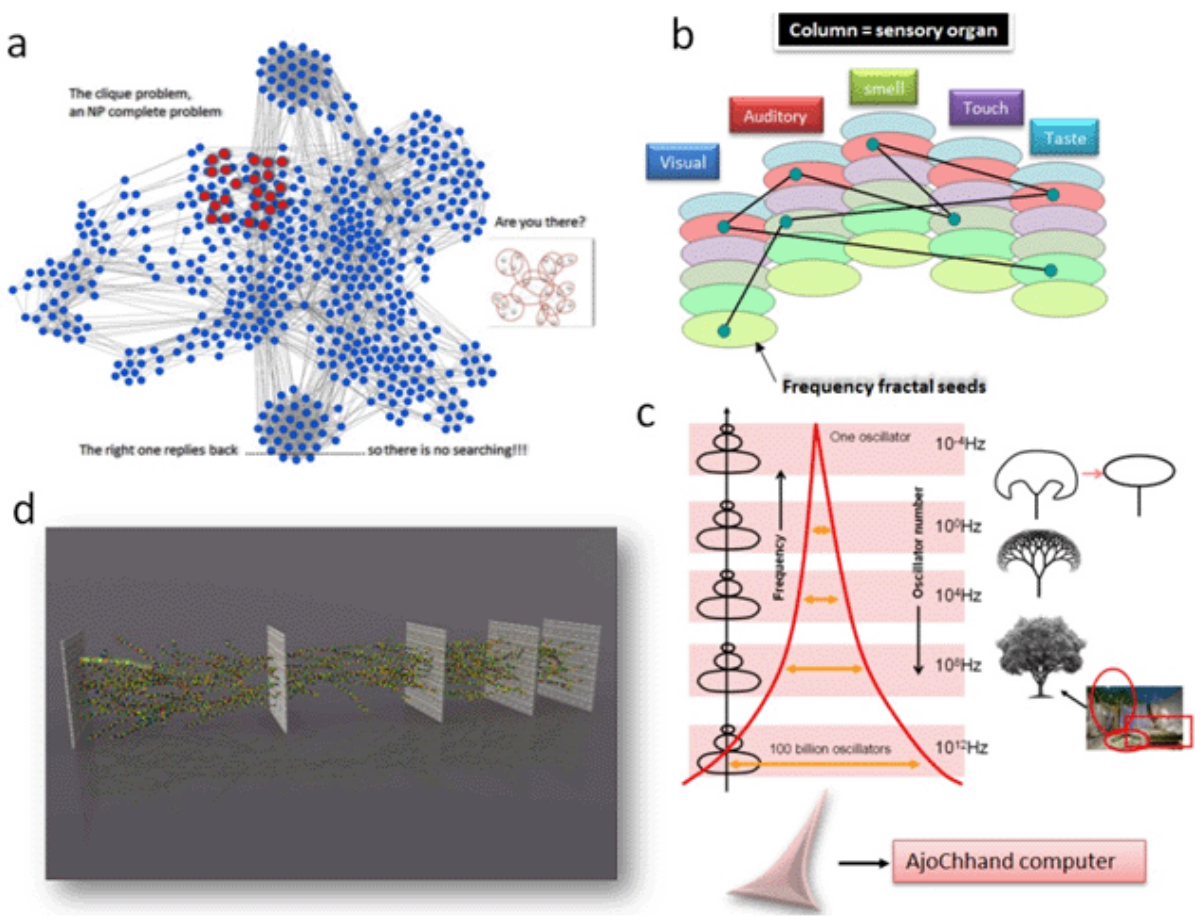

**Simplest form of computing in our computer: The engineering principles of a seed structure that can spontaneously "reply back":**

Creating liquid state models or echo state models [20] using the seed device is not practically feasible, since "reply" would automatically get masked within a limited distance. Therefore, we devised the resonance chain to take reply from the atomic scale to the meter scale and it is partially implemented using an innovative composition of wired and wireless technology [38]. One important aspect of computing with "spontaneous reply" systems is that an exposure to a specific external frequency signal triggers a natural vibration of the system, the nature of vibration is determined by the frequency of the signal. We cannot completely re-write the resonance states of our choice in the material, because it is the fundamental property of any material, we can only make a provision to shift the peak by changing its conformation or re-arrange electron density distribution in the seed geometry by modulating it symmetry [41]. As noted above that in our computer we save the "solutions" or "decisions" as a connected pattern where each point is a decision, so that we can solve the Clique problem in the hardware always and automatically any kind of given problem will be solved. Now "if-then" argument is the basic decision, if we save "if-then" as a single point in the pattern, then externally when "if" will be applied, "then" will be sent back. Using the ac signal with near the resonance frequency values, we shift the one or more resonance peaks such that if one set of peaks is triggered, for a certain time, the other peaks corresponding to "then" are triggered automatically. This is the simplest form of computing in our proposed computer. Therefore, we need to have a protocol inside the seed structure to introduce a coupling between particular peaks. If all resonant circuits packed inside a single seed are coupled together, then there is an automated coupling among all fundamental resonance peaks (FRP). Any injected energy is divided among all circuits according to their coupling factor. Since the coupling is not homogeneous, the input energy is distributed non-linearly among all resonance energy levels. We need a tool to change this coupling such that a new energy distribution law or coupling is imposed. This could be done by a conformational change in the structure or re-defining the electronic density distribution. Now, to trigger the spontaneous self-assembly of an existing seed, the existence of three distinct bands of FRP is a sufficient criteria. We have argued above the necessity of three bands to create a chain, now if the fundamental seed device is kept in a proper environment (solution, temperature, pH, electric or magnetic field), and all three bands are triggered by some means, then the uppermost band with lower frequencies would also get activated and couple neighbors roaming around in the solution, if provision of flexible bonding is kept, then, the fundamental seeds would self-assemble such that the upper resonance band remains strongly activated. Now, by trial and error in the design, we make sure that the seeds assemble and forms a new structure that has the triplet resonance band in the next frequency range. Recently we have synthesized such a material [42], at higher assembly levels (shape is large) the shape may appear like fractal antenna [43].

### How accurate is to suggest that chain of resonance is a Frequency Fractal?: Fundamental necessity to form a fractal

The generic form of a reply should contain a positive and a negative note, which is the basic criterion of a binary argument. Here we set the language of our hardware as "frequency", "amplitude" provision is kept only to modulate the coupling strength, hence we cannot set "zero" signal as the ground state. Therefore, the direction of signal propagation becomes an important criterion for defining an argument. If multiple resonating circuits (here circuit means simply an organic structure that resonates) are assembled to form a basic seed and an ac power is pumped, at particular ac frequencies, one particular direction is favored and for particular frequencies, the reverse direction is favored. All circuits can transmit distinct signals simultaneously and cause multi-modal mechanical vibrations of the system at the same time, which is often referred here as harmonic and anharmonic overtones.

Therefore, we need to classify the fundamental resonance band in two different ways, one set is positive, p-FRP (say x), and the other set is negative n-FRP (say y) as shown in Figure 6a. The same hardware will act as simultaneously operating multiple parallel worlds. Secondly, if we consider harmonic or inharmonic overtones naturally generated in the seeds due to the triggering of the resonance frequency, then the positive resonance frequencies and the negative resonance frequencies (x and y) try to generate two different sets of vibrations in the same hardware. As a result, we get only a few allowed frequencies for the x and y to survive in the "reply back hardware". However, some of these survived frequencies will "reply back" in the positive direction and some of these will "reply back" in the negative direction. Therefore, we need two functions to represent the generic "reply back features". In other words, we state that the set of frequencies {p-FRP} and {n-FRP} their overtones interact and generate F(x, y) and G(x, y) and since we cannot represent both of them in one axis, we create an imaginary axis for one function, this is explained in Figure 6b. Most interestingly, if we start from x+iy then we can generate the F(x, y) + iG(x, y) form partially, note that all values of the overtones cannot be generated from the complex number, since it is an infinite series. This function is a conservative map of a generic "reply back" pattern, which does not include all terms of the series, which physically exist in the hardware.

**Figure 6.** (**a**) Signals propagating in an opposite direction in a seed oscillator; (**b**) Escape point hardware means a given layer is a single cell of another layer B and at the same time has enormous seeds A inside; (**c**) Escape time hardware, produces basic geometric fractal seeds; (**d**) Harmonic and anharmonic overtones, one is integral multiple and another is non-integral, there are infinite levels.

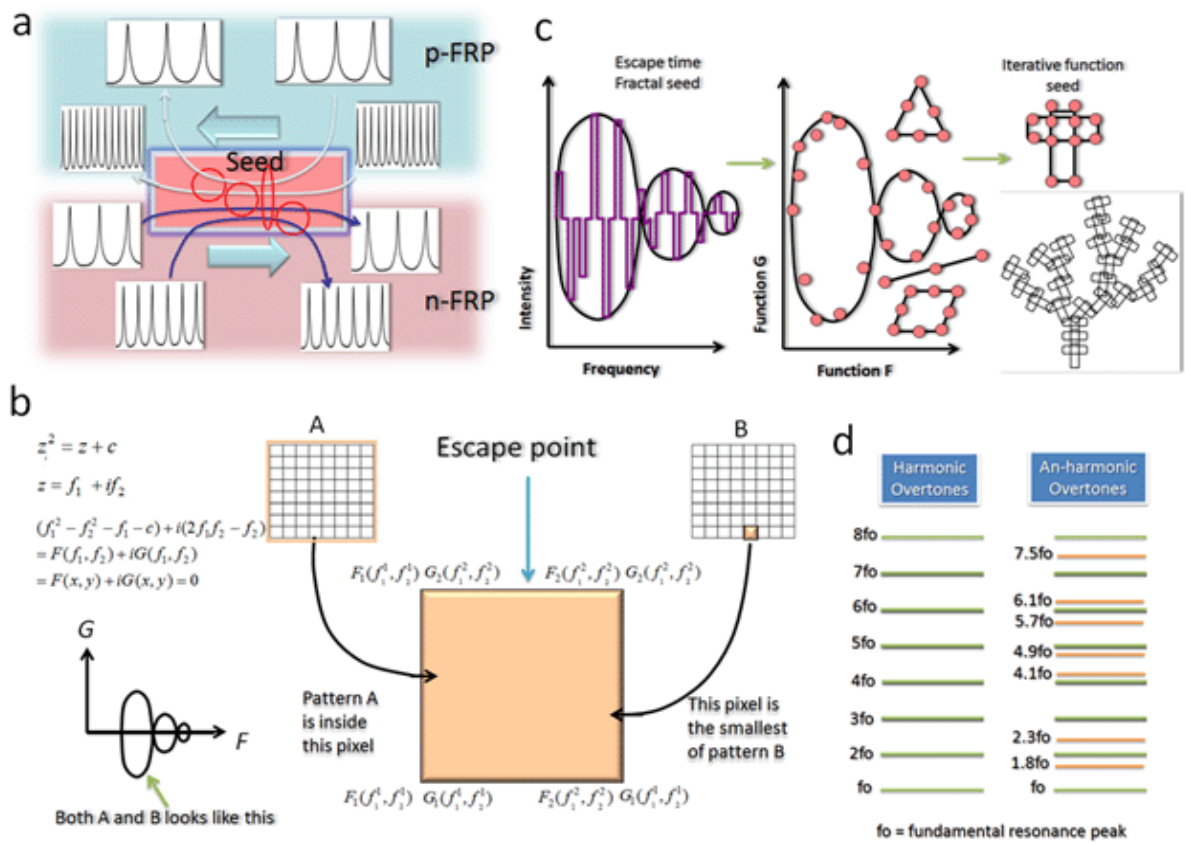

**The actual physical picture of an "if" statement in the hardware:** If there is a change in seed conformation (when an "if" statement is written), which redefines the electron density distribution, we get a modified reply pattern, which is again cut short into the reply back expression F1(x, y) + iG1(x, y). An essential query is that "if there is a peak shift, then how could a modified function represent the modified fractal, after all fractal means change in the every single layer?". The answer is "yes", first, as our experiment in the brain and in the artificial material shows that there are a large number of resonance peaks in the band, this is the reason we coined the term Fundamental resonance peak or FRP. In practice, some non-FRPs which have very low intensity rise up, silencing one of the FRPs, this event we term as shifting, this is shown in Figure 2b. Now, we would clarify a very important fact about this computer here to avoid any confusion, "When an "if" statement is written, it does not mean shifting of one peak at one seed, every single "if" statement forms a network of resonance peaks distributed over multiple layers to the top layer". Thus, our mathematical presentation of a fractal expression for a single "if" statement is logically justified. Mathematically formation of "if" is explained in Figure 6c, its physical equivalence is shown in Figure 2a,b.

If we self-assemble two such seeds wherein two different "if" statements are written, one with F1(x, y) + iG1(x, y) and another with F2(x, y) + iG2(x, y) then it could be possible that some frequencies of these two terms are common. Then, simply by triggering one "frequency" point in the 2D plot for device 1, one can trigger the other device 2. Now, if device 2 conformation change is such that as soon as the common frequency points are triggered, the device 2 triggers with a unique F2(x, y) + iG2(x, y), the operation is the simplest form of computing, by triggering the resonance chain. However, even in the simplest case the decision is an infinite series with several harmonic and inharmonic overtones similar to the ones shown in Figure 6d. Thus, Frequency Fractal is not complete, but a fairly conservative presentation of actual self-similarity clearly observed in the resonance chain. It should be noted that in an infinite series, $10^{81}$ th term can represent a solution and/or affect the solution whether we can write it or not is irrelevant.

## 4. Two Classes of Fractals: Escape-Time Frequency-Fractal and Iterative Function Systems

Spatial fractal describes self-similar geometric shapes at different levels of magnification [43]. Our computer uses mainly two classes of fractals, escape time fractal and iterative function systems. Self-similarity in a spatial fractal is visible, it generates beautiful patterns found in nature. First we explain the escape time fractal, because the hardware we build for our brain-like computer is an escape time fractal. A function y = f(x) means, expressing the plotting function as y in terms of x, one example is y = Logx. In a 2D spatial coordinate system, (x, y), we plot x and y values by varying x and determining the corresponding values of y. For a fixed value of x, we always get a well-defined value of y. As a result, the pixel size is a constant number thus fixed and irrespective of the boundary values of x and y, the nature of the curve remains the same. However, for fractal, we always get F(x, y) + iG(x, y) = 0, hence, each pixel or the smallest box of the graph is defined by two functions not by constant numbers. For example, in this case, the x axis of the pixel is from F(x1, y1) to F(x2, y2) and the imaginary i axis of the pixel is from G(x1, y1) to G(x2, y2). Therefore, if we change the boundary values x and y, within that pixel, the same pattern would appear as shown in Figure 6b, which is not possible for the normal functions. The reason for self-similarity or the beauty of the patterns lies in the fact that unlike the real world where the smallest pixels of a graph are simple constants, here, even the characteristic and nature of the smallest pixel is determined by the function itself. One can zoom into a pixel forever and zoom out forever. This is the basic concept of an escape time fractal. Now unlike conventional escape time fractal, we do not want spatial self-similarity, we want self-similarity in the information content so that we can construct a resonance chain.

**Our hardware is an escape time fractal:** In our computer, the hardware is an escape time fractal, we need to open up the womb of a seed to find billions of other seeds inside which construct the lower layer, then we can take any one of those seeds, open up its womb and find billions of other seeds inside as shown in Figure 7a. This is not the only reason why we call it an escape time fractal. While spatial fractal is apparently self-similar, the frequency fractal's self-similarity is very different. We consider that the real term of a complex number representing a fractal is "frequency" and the imaginary term is also a "frequency". Therefore, x and y distances are replaced by two variable frequency terms in the functions F and G. The graph paper is a 2D frequency space. Even if one looks at the real pulses generated by the hardware, no apparent similarity would be visible [44], when those pulses are converted in terms of "frequency" and fitted with an expression of a complex equation then the data would start making sense. For example $z^2 = z + c$, z is a complex number (z = x + iy) and c is a constant, this is an equation for mandelbrot fractal and we plot it's real and imaginary terms to visualize the fractal. If we plot that in the real and the imaginary space parts for the complex number of the equation that defines the resonance chain of our computer, then we will see the frequency fractal or resonance chain, and when we say escape time fractal, we mean this information perspective of the escape time fractal not the spatial perspective noted above. Therefore, for every seed we get a 2D frequency space and by entering any of the cells inside we can enter into another layer as shown in Figure 7b. Until now, fractal structures were studied for their ability to sustain coherence for quantum communication, *etc.* [45], but here, we are not interested in the structure, rather, self-similarity in the measurable property that is not visible in the object shape.

**The escape-time fractal hardware produces the software, which is a hierarchical network of iterative function systems of "Fractal seeds":** Interestingly, always, to explain the occurrence of self-similarity in nature, iterative function systems are used. It means a set of simple geometric shapes like triangle, square, and different polygons, straight lines are combined in a special orientation and that is repeated several times to generate the architectures observed in nature as shown in Figure 6c. Our hardware is not essentially an iterative function system; however, it does a remarkable job to decompose any pattern in terms of fundamental "fractal seed" patterns and network of those patterns are also connected by elementary geometric "fractal seeds", this process goes on and on. Suppose we are looking at a beautiful picture of Rama and Sita wedding ceremony, every single equal intensity contour's geometric shape to the nearest polygon is identified and then those are created by the coupling of resonance peaks. This happens at the ground level. At the next layer, centers of larger, brighter, geometric shapes are connected by polygons, straight lines, *etc.* The pattern of these fractal seeds evolve in such a way that we could represent entire phenomena as if several concentric spheres are there and on the surface of each sphere the patterns made of fractal seeds are evolving with time. Figure 7c.

**Figure 7.** (**a**) More is the "if-then" arguments, bigger is the sphere. Arrows denote coupling (top). Seed A makes B and so on (bottom). Resonance band and 2D graph of arguments are same; (**b**) Self-assembly of arguments form a cluster, these clusters self-assemble again; (**c**) Spheres of different radius denotes nature of "fractal seed" assembly. All patterns plot together do not overlap. This is the principle of superposition of fractals.

We have carried out an extensive research on image cognition and an advanced state of the art software was already developed for the proof of concept, which is described below in the AjoChhand-Soft section. In this way, a single picture is dissolved into a multi-layered compositions of "fractal seeds". This decomposition looks like our resonance chain hardware, we open up a seed and a large number of seeds come up. This automatic conversion can be imagined but cannot be fully decomposed using a conventional computer programming. A simple calculation would show that even within 3 layers of a picture decomposition we reach intractable domain with 10 fractal seeds inside a fractal seed and each seed is connected to others by 9 wiring ($10^{10^9}$ connecting point, $10^9=10^9$). At least we conclude from the above demand that our hardware resembles the demand of the hierarchical network crudely, 10 seed structure inside a seed structure and each seed with 9 resonance peaks generate a network of $10^{10^9}$ connecting points.

**An unprecedented mathematical relationship among allowed frequencies of resonance peaks:** The resonance peaks in the bands of a resonance chain hardware as we have derived in the human brain (the experimentally derived resonance chain values are given below in our model of a human brain below) follow a unique relationship. We simply plot the resonance peaks of the brain along the frequency scale. We find that even if we take log of frequency in the primary axis, the plot looks like as if the resonance frequencies are separated by a log scale, normally peaks should appear equidistant after taking the log scale, as shown in Figure 8a. The distribution of resonance frequencies is a log function inside a log function. To eliminate the log distribution completely, since log values are separated by a linear distance, we take the linear values and then plot the derived resonance frequency once again, we find, it is a log distribution once again, the log feature or the non-linearity cannot be diminished. This means the frequencies are separated by a log function inside a log function inside a log function. Possibly this would continue. Recently, we have succeeded in synthesizing an organic supramolecular architecture wherein a multi-layered self-assembly was triggered naturally, we observed a similar log function behavior for the resonance frequency plot. Since a natural product, human brain and the artificially built organic supramolecular architecture both exhibit a unique resonance frequency correlation among them we investigated the origin and found that it originates from the power scaling law. If there is a homogeneous distribution of power among all resonance frequency values when the architecture of the multi-layered seed structure (or escape time hardware) is being formed, then the architecture should adopt a symmetry that allows it to maintain equal power loss throughout. If equal power loss is maintained, the lower frequencies would be spaced much nearer, now the power law is a conservative claim, the exponent of the power relationship holds an infinite series, thus generating a log inside a log inside a log function (this is not log(log(log(frequency....))), it is undefined).

**Figure 8.** (**a**) Triplet band made resonance chain plotted in a Log scale; (**b**) Entanglement causes immediate collapse, however, we need spontaneous synchronization and de-synchronization hence we need time Δt, however, it includes programming; (**c**) The mechanism of fusion and fission of fractal seed networks during computation.

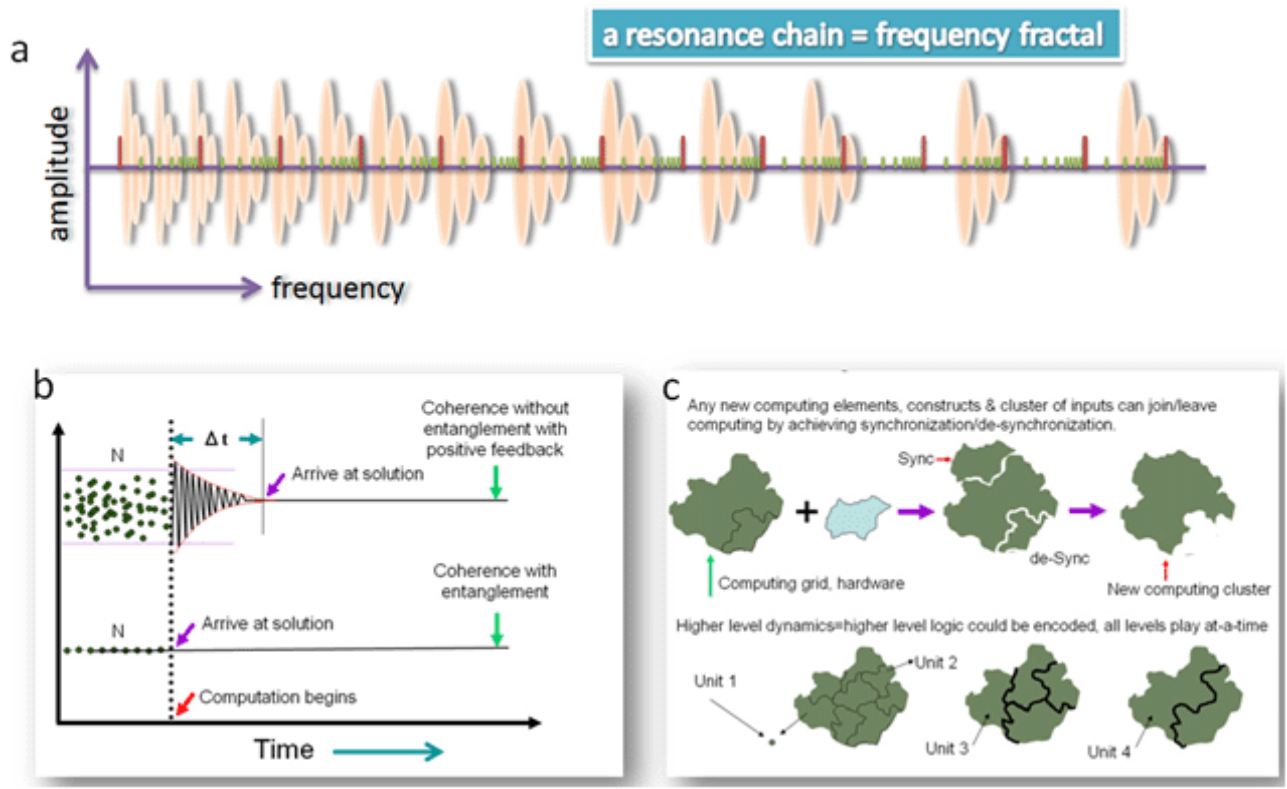

**Start from a Lie algebra and end in an unknown world of mathematics: Why do we need an imaginary space-time world (F+iG) that defines the imaginary term for another such imaginary space-time world (F+if(x+iy..))?** The physical significance of log is that a physical parameter varies depending on how much a parameter weighs at that point. Now, this is a very interesting situation even if we have a single log relation. Lie algebra [46] developed in the 1930s nicely address this issue. Kac-Moody Alzebra (1960) an extension of Lie Alzebra developed over a complex space addresses infinite dimension [47]. The linear space transformation that governs the "rate of change", could be a complex number if there is self-similarity, in our case we have self-similarity. Now, the interesting part is that the linear space transformation cannot define a rate of change which contains a parameter that is defined by another complex number's space transformation. In such situations, without a debate in mathematics we consider that a function becomes an undefined mathematical entity. However, we leave with a major conclusion that since in our case the complex number's real and imaginary parts are frequency, therefore, we have an imaginary space-time world inside another imaginary space-time world and so on. Already a part of this theory was formulated in Riemannian manifold [48].

Iterative function systems are resolved in the images given to the new class of computer in the way described above, however, during computation, at different layers of the hardware, synchronization and de-synchronization of the resonant oscillation continues [49]. Computing time is the synchronization time of the fractal seeds as shown in Figure 8b. Synchronization leads to coherence which means Fractal seeds oscillate in the same phase and frequency. Spontaneous switching between synchrony and de-synchrony is essential, thus, entanglement is not a pre-requisite, the overall situation is explained in Figure 8b. The information perspective of that physical process of computing is that several fractal seeds of iterative function systems form the network, coupling and de-coupling of large networks is a generic event that happens during computation as shown in Figure 8c. We have noted above that there exists two columns, one for the "if-then" arguments, and the other for the phase transition rules. In fact the "if-then" arguments are formed by the construction of networks of fractal seeds of iterative function systems from the resonance peaks as shown in the Figure 9a. Below, we describe how these iterative fractal seeds get connected to form the network.

(i) **Fusion and fission of frequency fractals:** Mathematically synchronization means, several iterative fractal seeds get fussed and de-synchronization means several such seeds get disconnected as shown in the Figure 9c. In the fractal theory three types of fusion and fission of fractal seeds have been developed.

**Type I:** Suppose we are looking at a tree, then, the entire tree could be made of a square and a rectangle put together as a seed of the fractal and then by copying this geometric shape several times and then by rotating and connecting with it in very different ways, we can reproduce the entire tree. By combining and rotating basic structures, open and close versions of triangle, square to all polygons including a circle or curve, or straight line, every single structure found in nature could be created [50].

**Type II:** From a basic straight line all primary structures like triangle, square, any type of polygons could be created using a simple fractal relationship. Therefore, the elementary filters for a complex pattern need not to be created specially and stored separately in the hardware, a generic frequency fractal generates all possible polygons from a straight line to circle, and all patterns co-exist. This is our background fractal F(x, y) + iG(x, y). Any modification to this fractal stores the arguments in the form of new fractals and again all possible patterns of those newly stored fractals co-exist. Co-existence physically means change in the F and G co-ordinates of the 2D frequency pattern to create a superposition of all images, just like several traveling paths of electrons around nucleus generates a diffused orbital perception [51].

**Type III:** Several type I fractals when evolve with time in the 2D frequency space, at the high frequency layer where we can see the evolution of the frequency fractal A in a large number of pixels, the entire pattern might appear as if it is a simple straight line or curve. Now, at this situation, if other fractal B evolves similarly in the same frequency space with typical common points so that A and B together appears as if a single circle or rectangle, then type II fractal may be born. If AB fractal is born which starts evolving together at all frequency space with AB as their seed we might get birth of a new thought that never existed. Similarly, several groups in multiple different regions of the same hardware might spontaneously get coupled just like AB, due to similarities in the dynamic evolution then a higher level perception fractal is born. These two types of fractals are called type III fractal [52].

**Figure 9.** (**a**) Resonance bands for three iterative function systems, common peaks couple and form an argument, set of resonance peaks denoted as yellow balls; (**b**) Self-assembly of arguments form a cluster, common values form "loop", "expansion" "fractal", *etc.* local

networks; (**c**) Square 2D images make complex 3D net; (**d**) How resonance band change shape, when memory is stored.

**Different ways of modifying the frequency fractals inside our artificial brain:** The first route is to modify the antenna geometry, so that two independent frequency-controlling parameters (x and y), inhibitory and excitatory resonance frequencies change their interactions. As a result, F and G would be re-defined and scales would change. The second route is at the sensors, where a fractal frequency is created and in the cerebellum (computational output center of our artificial brain) where information-processing fractals regenerate the sensory signals. A fusion of fractals means addition of real and imaginary functions, F1 + F2 and G1 + G2 to generate a new fractal (type III), while fission is just the opposite. A fractal could be expressed as the sum of two different Fractals, due to a common physical phenomenon bifurcation in synchronization this kind of situation might arise AB fractal breaks into two parts A and B. In the hardware, when the basic resonance frequency of the hardware is fixed, fusion/fission of fractals would mean a change in the very basic combination of excitatory/inhibitory resonance frequencies (x and y). Does it mean a change in the basic hardware parameters? The answer is both yes and no, when we change the antenna inside the oscillator network of a particular level, actually we modify the seed partially for the next stage. Entire hardware thus undergoes a change collectively.

**(ii) Co-operative superposition of frequency fractals:** A frequency fractal F(x, y) + iG(x, y) for the hardware means pulses of very specific frequencies as described by this equation would only resonate with the existing hardware. As soon as an input pattern Q(f1, f2) enters into the particular region of the artificial brain, say visual control region, due to the existence of multilayer hardwares, a replica of the image is created in all the layers [53]. As a result, fractal compression and expansion occurs at various levels. For example, if we see a tree, the basic pattern that evolved to generate the entire tree is automatically resolved in the lower frequency domain (larger sized oscillators). Thus, one could identify a tree even if the input could be given in an astronomically large number of distinct ways, it is already explained above that even a simple input image is transformed into stacks of images wherein each image has a large number of iterative fractal seeds, there are connections among seeds inside the decomposed images and between the images.

Thus, a massively interconnected network of fractal seeds is generated, one such example is shown in Figure 9c, where we demonstrate that 2D fractal images are connected by wiring. This is not the complete picture, every single geometric shape that includes all polygons could be generated from a straight line with a kink (a straight line is not a fractal), this is already shown in this tutorial for students [54]. Now, we have incredibly large number of oscillating lines in the hardware. If we consider all sensory signal processing domains of the entire brain, everywhere, we would find only oscillating straight lines. Thus, in our multilayered hardware, we multiple clock-control paves the way to superimpose astronomical choices, one top of another. Quantum and classical computing algorithms do not explore the higher and the lower levels of any pattern, which enables spontaneous creation of several different kinds of groups and simultaneous recovery, hence our non-argumentative approach does everything that "decisive computing" paradigms promise, without an algorithm and a circuit. Moreover, we can do something that existing computers cannot do, we can save the image of an "icecream" and retrieve its unique 3D perception,—a look-alike tree without writing a single line code.

When all fractal shapes created by a tree are being generated in the hardware, there is always a little difference with the basic background fractal of the hardware. Therefore, the junctions between the seeds at all levels undergo a little change so that natural vibration matches with the external input. How the resonance band changes with the input pattern is shown in the Figure 9d. Not only for a tree, even if there are a large number of objects in an image, the artificial brain identifies abstract geometrical relationships among different objects and creates an equivalent fractal for that abstract relationship also [55]. In case of several different kinds of sensory input data, due to the natural property of frequency fractal, the patterns in different parts of the artificial brain hardware (entire brain is a single fractal object) get correlated and a new fractal is formed. The brain circuit undergoes subtle changes to incorporate these features. In this way, visual, sound, taste, touch and smell data get correlated in the hardware. It should be noted that for the highest level of the brain fractal hardware, the basic seed pattern of the new input fractal is the highest-level perception data, this is saved in a very particular region, we call it frontal cortex region. All these basic fractal patterns for a single object, assembly of objects, to complex events are made of square, triangle or basic shapes, but with a unique feature, this is what we call co-operative superposition of various different fractals [56]. Note that new fractals are stored in the brain only when it does not match the existing patterns; if it matches there is no question for the hardware to store anything new.

## 5. The Collapse of Turing Machine: Advancing Gödel's Incompleteness Argument to a New Class of Computing Engine

The Turing machine concept [1] is based on a clear description of information, which is the logarithm of the total number of distinct symmetries possible in a system, Turing machine needs very well defined sets of arguments. Turing patterns are widely seen in nature, the reason is that the pixel size is fixed by the organic molecular structure, hence computation is finished in a finite space and the product is delivered. However, if a mother machine decides to create the smallest pixel of the daughter fractal using another fractal, then the problem is un-decidable, as if an infinite growth process [57]. For three reasons we can scientifically argue what is the information content, however, cannot estimate it.

First, here we create a column of arguments where the base part is real, could be measured in the system using machines, but the upper layers, which are formed by the coupling of the harmonic and anharmonic overtones of the resonance vibrations, those levels reach to an astronomical heights. Those coupled arguments do exist; however, since we cannot estimate fully, we simply say that the column of arguments extends to infinity. Now, this is not all.

Second, there is another column of phase transition rules for the column of arguments, there exists rules, which cluster would undergo before, which one would undergo later. Due to natural self-assembly of arguments the rules of phase transitions are also written in the system and obviously the number of rules are astronomical in nature and cannot be written down.

Finally and most importantly, resonant vibrations have temporal dynamics associated with it, therefore, the column of rules for phase transitions do change continuously at all levels. The entire hardware has multilayered concentric sphere like architecture (say A, B, C), and each layer has a clock, while the clock moves extremely fast in the innermost layer A, and slowest at the topmost layer C. Now, a chain of phase transition could propagate from the top layer C to the innermost layer A and adopt a new path and then propagate to the top layer C before the top layer detects any significant change in time [24,25]. Thus, for three reasons, information content in any layer at any given time can be defined qualititatively but not estimated by the system properties of that layer. If information cannot have quantitative expression, then no Turing machine could be constructed [23].

Now we explain the above reasoning in terms of frequency fractal. The resonance chain that defines our computer is a fractal that is realized by its typical pixel, which is itself a function not a constant number. At every upper and lower limits of fractal hardware, there are two singularities where any Turing machine would become undecidable. It is not a series or parallel connection of two Turing machines, each processing a particular fractal, we need several simultaneously operating invisible Turing machines in the upper layer and in the lower layer to explain the output of the existing layer. The limiting pixel of the fractal is not a simple junction of two different hardwares representing two different operational domain of a single fractal, singularity points are junctions of simultaneously vibrating multiple resonance bands. The situation is like co-existence of multiple distinct identity of a single information, which has been proved mathematically undecidable by a Turing machine [57]. The information content of the lower layer A and upper layer C are completely inaccessible by middle layer B, though B defines A, and C defines B. This situation does not fit the definition of Turing decidability. At every layer, we observe only the effect. In summary, controlling logic for the world B is located in C and the output of B is delivered to A, while A and C are inaccessible by B, this is against the principle of Turing machine.

**An attempt to explain all features using Turing machine network:** Since conceptually we can put a Turing machine at every single layer and at every single places in the hardware, now we will try to do just that in a thought experiment and try to explain the entire event in terms of a Turing machine as shown in Figure 10a. During computation, synchronization and de-synchronization of resonance frequencies continue among all layers, between clusters of different sizes with a motivation to track and activate the maximum density of coupled vibrations, and deactivate the routes with very low density coupling. This is the fundamental driving mechanism for computation and decision-making. If we look at a single oscillator at any given time, it can physically couple with large number of other oscillators, clusters of different sizes, and trigger multiple pathways of phase transitions across the hardware. Therefore, one can define its information content in various different ways at any given time, it physically carries out many-body interactions and that interactions play vital role in governing the decisions of the system, hence, even for a single oscillator or a single cluster, information or logical set of arguments cannot define its status. We explain the same situation in another language. For every higher level of coupling, a cluster of oscillator's unique compositional symmetry would define a distinct information content. In this way, logarithm of symmetry for all invisible layers has to be taken into account, which would simultaneously coexist. This situation is almost like a quantum system where 0 and 1 states coexist.

**Figure 10.** (**a**) Resonance chain represented by a network of Turing machines working independently for the imaginary and real numbers; (**b**) Godel incompleteness based engine for our computer that should replace the Turing machine.

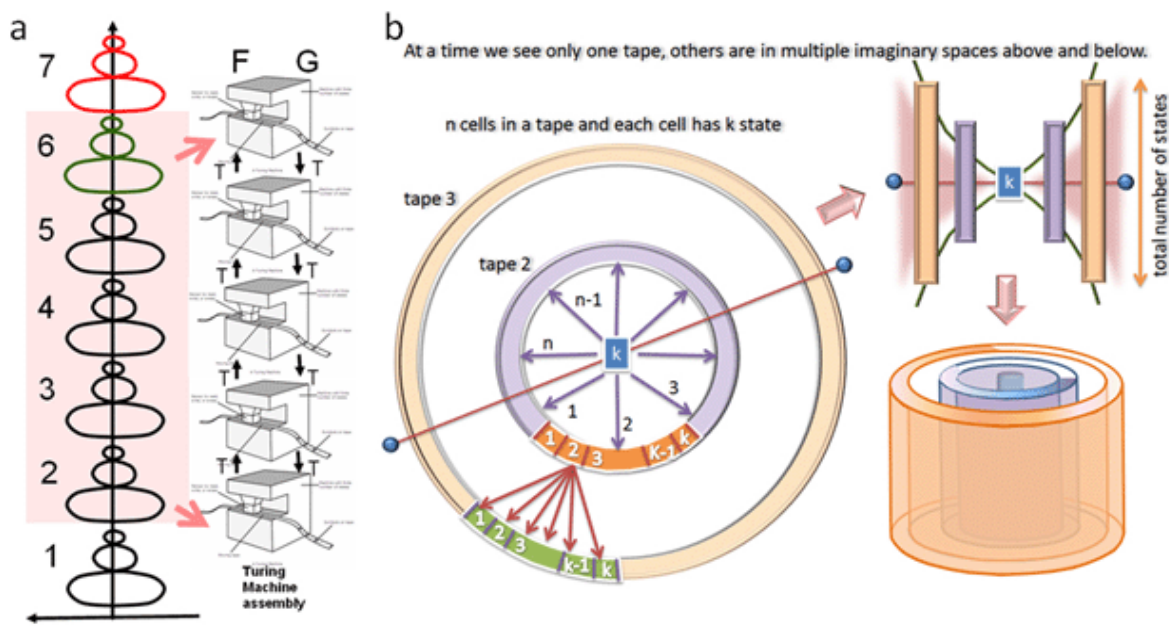

**Looking beyond the quantum world: the necessity of a Frequency Fractal like Turing tape:**
Since imaginary time and space is an essential element of this hardware, the query often comes to our mind is it related to a quantum world? We may define logarithm of the product of possible symmetries in all layers as the information content of a single oscillator or a cluster of oscillator located in a particular layer and try to construct a Turing machine just like we do, in a quantum Turing machine [58]. Two Turing machines one calculating the real term another imaginary. However, as described above, we have a different situation here because we are unable to estimate the information content and set logical argument here. Apparently, we need several tapes coexisting in the modified Turing machine, if it happens only once this is fine, just in the quantum system. However, when basic information has several faces, and all distinct faces simultaneously define particular kinds of information then, every single cell of the modified Turing tape should contain a Turing tape inside as shown in Figure 10b. Both sequentialization and parallelization are destruction of that information itself, to define it, we need to go inside any state at any given time, and a new world of information processing would open up, which we would represent with a new Turing tape and this journey would continue forever. We will try to explain below more elaborately that even in quantum computing with one imaginary space and one imaginary time, we were very fine with the Turing tape. What goes wrong now when we consider multi-layered architecture, where one set of oscillators is the womb of another set of oscillators and each layer contains a new clock [24,25]?

This situation is not like quantum where several simultaneous facets collapse into one real value, even at the smallest unit scale we have imaginary terms which does not disappear as it happens in the quantum world. For quantum, we have one real world that is the classical one which we use for measuring and then there is an imaginary space and time world which gives a projection on the real space. When we study quantum mechanics we know very well that in the real world, whatever is coming, is arriving from the single space-time world which cannot be seen or measured using clock hence thats imaginary, but defined. The situation changes dramatically, when we add another imaginary space-time world in the bottom and that also makes its projection in the real space. This simple addition of another imaginary space-time world has enormous implications. First, two imaginary worlds can interact with each other, ignoring our real world, which is sufficient to argue that the information content in the real world cannot be defined anymore and no argument could be constructed. The situation turns very interesting because now, in our real world the output of two imaginary space-time worlds cannot be said as "collapse", as it is defined in a quantum mechanics textbook [17]. Basically, our real world then disappears and it becomes just another imaginary space-time world, hence, as soon as we add one imaginary space-time world below and one above, every single observation would remain unpredictable with the information of the three imaginary space-time worlds. We would clearly have some other interactions never taken into account. This is how a radical and a dramatic fundamental shift occurs when we consider a multilayered imaginary space-time world network, not parallely, not sequentially, but one inside another. The quantum-like world is a local observation in this generic world view. In our nature, quantum supersedes light speed, however, here clock speed is provided by maximum resonance frequency of a particular band.

## 6. Self-Assembly and Phase Transition of Columns of Arguments

Our artificial brain is a pattern based computer, every single information enters into the computer is converted into a 2D matrix or picture and then from this image a larger number stacks of images are constructed as a preparation to further processing. The first basic information that we need to learn about our artificial computer is that at the starting point the hardware is simply a jelly of organic molecular neurons. With learning, the jelly forms a semi-solid structure, a circuit where the neuronal molecules could still change their configuration and re-define the fundamentals of the circuit. The most important aspect of this learning is automated self-assembly of arguments. This is a very interesting aspect of our computer, since this enables us not to write any software program for the computer. In our computer, a "if-then" argument means, first a set of resonance peaks are coupled to construct "if" and another set of resonance peaks get coupled to construct "then", as shown in Figures 2a and 9a. When our artificial brain learns, it stores "if" set and the "then" set such that, if from outside "if" set is triggered, then it causes the phase transition of the hardware and the "then" set gets activated. By storing this entire automated phase transition event, our artificial neural network spontaneously stores the "if-then" argument. Note that, "if" and/or "then" can have multiple peaks of different frequencies connected by a fractal relationship. This is the basic computational principle of our artificial brain-like computer.

**Creation of the column of arguments by self-assembly follows a power law:** The brain starts learning from the first arguments and then more and more arguments are added. This addition process is very special. When two arguments are added to the hardware, new higher-level coupling rules among them are created and similarly more and more high-level coupling rules are born automatically since the addition of a new "if-then" hardware follows a particular protocol as shown in Figure 11a. If we make a simple calculation that only one resonance peak represents a "if" statement and only one peak represents a "then" statement, then, for the nth addition of "if-then" statement we will find ~$n^2 + 1$ number of higher level coupling rules are automatically added, this is shown in Figures 10b and 11b,c. These automatic additions of higher-level couplings follow fractal relationship and develop their own phase transition rules during learning. Therefore, during self-assembly of arguments by the brain jelly (our organic computing architecture, see Figure 1b, an astronomically large number of phase transition rules are also created and stored as shown in Figure 2c. The most interesting aspect of this feature is that a simple trigger at the ground level (or simple input data) could activate the phase transition at the topmost level or any other levels depending on the condition-match. This can lead us to a situation where we can plan for a very complex form of computing that Turing class computers cannot deliver within a finite time [28]. For example, if we simply consider that "if" and "then" states are made of a pair of exhibitory and inhibitory resonance peaks then $n^2 + 1$ additional coupling increases to $n^4 + 1$ and so on as shown in Figure 11c. When 2 clusters of "then" arguments generate 1 new cluster of argument, then we get a converging triangle, however, when 3 or more clusters couple, immediately from the first layer the arguments expand massively and then converge like a triangle as shown in Figure 11c. Eventually we reach a situation where the system could face an astronomically large number of new "if" situations and due to coupling by fractal thread, "spontaneous reply back" would identify "then" path instantly, entire path will respond to the query. The synchronization/response time will depend on the number of different frequency fractals used, not on the complexity of the path.

**Figure 11.** (**a**) The basic principles of "self-assembly of arguments". Each circle is a cluster of "if-then" argument created as shown in Figure 9a; (**b**) The complete triangular network of arguments, the bottom layer is the actual arguments written in the hardware, the right end arrow of the triangle is where the growth occurs; (**c**) Three cases show how arguments grow in reality; (**d**) The basic relationship between "if-then" argument and phase transition.

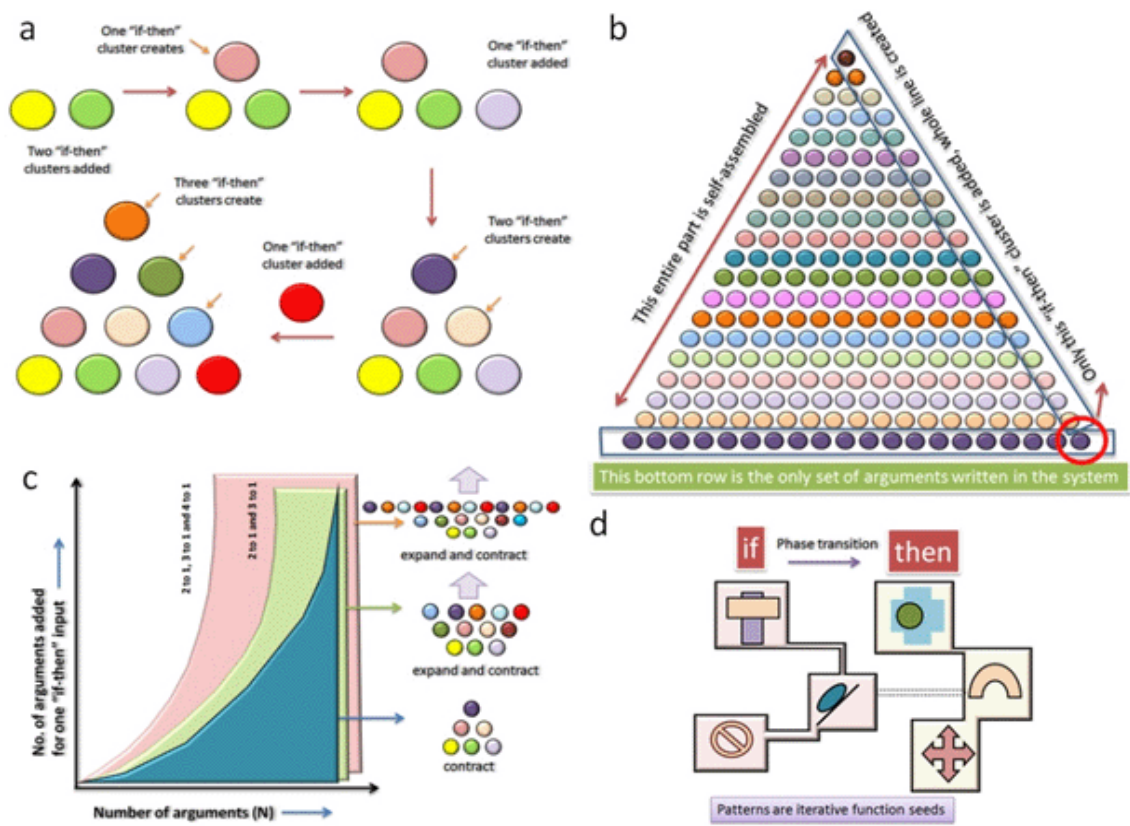

**Three fundamental rules for the creation of the column of rules of phase transition:** Self-assembly and phase transition of arguments are two interplaying fundamental processes of computing in our artificial brain, both the columns are constructed simultaneously, however, apparently two columns would appear very different. While the column of argument (Figure 2c) provides the network of "if-then" argument clusters, the phase transition column (Figure 2c) which stores the data for "if" cluster transforms to "then" cluster relies on three essential considerations as explained by Figure 11d. First, the occurrence of phase transition is determined by the duration of the resonant vibration of a column of arguments. Second, the larger is the size of the cluster of "if" arguments, the smaller is the threshold time for the onset of a phase transition to the "then" cluster, the situation is described in the Figure 12a. Third, phase transition is a continuous process, the system moves towards a direction of those local resonant vibrations wherein maximum coupled resonant vibrations are located. In other words, the system point moves towards the higher density of states and expands that domain globally to dilute the density. This is the fundamental phenomenon that drives the computation in our artificial brain.

**Figure 12.** (**a**) The self-assembly of phase transition rules, fractal seeds of iterative function systems form clusters of clusters, it starts from sing seed to the largest brain system. (**b**) Phase transition or symmetry breaking via conformational changes is the mechanism of storing a complex "if-then" argument cluster. (**c**) The common part between the writing hardware (here hippocampus) and the argument/computing space is rejected, rest part triggers "beating".

**A remarkable resource management: store astronomically large data without using space:** Our self-assembly concept is also fundamentally different from the conventional self - assembly of materials, here, randomly distributed basic geometrical elements made of frequency points in the $F(x, y) + iG(x, y)$ map are coupled physically and a new fractal is born. This is an integration of information in a separate identity keeping the original pattern intact and without using any new hardware resources. Therefore, in materials self-assembly resource is essential for growth, however, when arguments self-assemble by creating new fractals then for its complete formation (fully expanded and compressed form) only the new seed needs to be stored. Since, the seed already exists in the patterns, only minor conformational changes in the hardware would couple different geometric elements of the fractal seed. Only changes that the hardware accounts for, are the conformational change to store superposition of all possible evolved forms of a fractal seed, this is explained in the Figure 12b. The next argument that arises, whether this particular self-assembly process would run out after a certain time, because, in a fixed hardware, the total number of conformational changes allowed, whatever big it is, after all would be finite. It could be shown using mathematical calculation that the possibilities are astronomically large and even if billions of conformational changes occur at every second, the 100 billion neuron devices or computing seeds can run for trillions of years without being exhausted. Thus, an alternative, better resource management than that demanded for quantum computing is provided here [33].

**How the artificial brain decides where in the system actual changes need to be made for self-assembling the arguments: The necessity of beating:** One of the fundamental aspects of resource management during self-assembly of arguments is that the associated conformational changes to couple basic geometries of the frequency map are made at one-step higher frequency range where the basic pattern is detected. The conformational changes are permanent in the sense that even in the absence of the basic criterion for the formation of the argument the conformational change is retained. However, if necessary, say one circle turns to two basic circles or triangles in the F(x, y) + iG(x, y) map (fission of fractals), then reverse conformational changes might occur in the system. Finally, conformational change can happen several parts of the hardware simultaneously, however, always minimum changes are made, if that does not meet the criteria then the hardware decides not to self-assemble that argument in a particular domain. This is how it is decided where to self-assemble a particular argument. In practice, if two frequency points are nearby, then a new phenomenon is generated in the hardware, this is called "beating", this causes a unique response in the localized neighborhood. This "beating" effect identifies the exact location of the hardware needs to be changed and how much changes are required to nullify the "beating" effect [59,60]. We have explained with a schematic in Figure 12c how in a real human brain this "beating" effect would identify which neuron needs to be changed. This is the hardware mechanism of self-assembly of arguments.

## 7. Complete Operational Model of the AjoChhand Brain-Like Computer

Since our computer does not require programming, we need to have a hardware arrangement so that all input sensory data are converted in a very particular format that is processable by our brain-like computer. Our large-scale model of the functioning brain is very different from the existing models that capture the neuro-anatomy features [61,62]. First, we capture the brain functions in much more details and secondly, both in our hardware and software versions, the system creates its own arguments, we never design or program anything that is function specific for a particular sensor or behavior. The system learns those features by themselves. The use of frequency fractal instead of normal fractals is beneficial for several reasons. In the real human brain, (a) the distribution of functional connections; and (b) the probability of finding a link *vs.* distance are both scale-free [63], this could be achieved using a frequency fractal naturally.

**Mathematical representation of the computer: universal law for constructing multilayered seeds:** The entire computing hardware is the physical representation of a complex resonance frequency map $F1(x, y) + iG1(x, y)$ as shown in Figure 13a. Thus, any form of noise matching the frequency at any part would trigger the vibrations associated with the hardware. Even if a small part of it is triggered, at any frequency domain, the energy would be canalized to other frequency channels. Thus, entire hardware is designed to remain always in an activated mode under noise. Noise turns it to a start up mode and no external energy is required. We have already mentioned that from the innermost layers smallest seed to the largest seed (a tree of complex frequency patterns), the defects in the F(x, y) + iG(x, y) map (resonance chain) is the memory. We have specifically noted that the role for the fractal antenna, which is connected to every single seed of every frequency range, is not to send signal all over the brain, just in the neighborhood, only up to the limiting dimension seed for the next range of the frequency map. First, the orientation of antenna constructs the F (x, y) + iG(x, y) map and then the antenna continuously evolve structurally to generate the modified form of F(x, y) + iG(x, y) into F1(x, y) + iG1(x, y). Therefore, at this point, we try to underpin specifically, what are the exact changes that an antenna needs to reflect when it evolves its configuration [64,65]. At every level of the hardware, the angle between the oscillators and their length determines the resonance frequency and the coupling behavior. We have explained it in the plot Figure 13b, depending on the hardware the allowed values change, it should be noted that irrespective of the layer, in a particular computer, all seeds starting from the atomic scale to the largest scale would follow the particular nature of the plot in Figure 13b. This is the most important code or universal engineering principle that is followed during construction of a hardware.

**The information content of the hardware: Role of synchrony and de-synchrony:** (i) Synchrony and de-synchrony [39] shown in Figure 13c, among large clusters of pulses automatically convert and save data into a pattern of frequency fractal, which is the fractal computer analogue of a software program. It constructs the circuit for a new problem as shown in Figure 9b, which is a change in the local fractal antenna to incorporate the modified F(x, y) + iG(x, y), learn relationships among geometrical shaped groups of frequency patterns and save the operational rules by the system itself, which are simply fractal seeds; (ii) The system computes based on its temporal dynamics, which is recovering all associated fractal seeds via synchronization and de-synchronization of pulses belonging to the fractal pattern executes the generation of entire "then" pattern for the input "if", which we term as "phase transformation of the "If" pattern". As a result of this transformation, the system decides automatically, simultaneous, parallel and sequential parts of a problem,—modified Turing machine is trivial to construct [66]. In the Figure 8b, c we have explained that "synchronization time" is the computing time. However, synchronization time depends on two primary factors, Quanlity factors of the resonance peaks and the coupling area, the coupling area relates to the frequency and intensity of the resonance peak controls the quality factor. These four parameters are key controlling factors, which arguments would survive, and which arguments would disappear inside the computer, hence we explain the interrelationship explicitly in the Figure 13d; (iii) Informative content [2] on this computer is the time-evolution of a 3D pattern of pluses, which is fully expanded and compressed form of a frequency fractal at all possible frequency ranges. Every single fractal seed is correlated particularly in some groups via another fractal seed. The addition of a new fractal seed or an argument or memory first finds a suitable seed-encoding region, all associated seed based correlations are found and then the seed is included in the hardware along with all of its correlated seeds. Therefore, no equation can alone describe this situation. The co-existence of positive and negative resonance frequencies, drive synchronization in two parallel directions at the same time and the conformational changes occurring in the hardware at every possible frequency range of the frequency fractal modulates the process so much that "symmetry breaking" would not be a sufficient explanation for the entire situation, rather we would say, "phase transition" of the system (Figure 12a, b). Therefore, an imaginary term that is invisible in the hardware plays a fundamental role in computing, and this situation cannot be programmed. Note that phase transition is fractal equivalent of random network where we see "explosive percolation" [67].

**Figure 13.** (**a**) The resonance chain showing the frequency limits, the chain could be a moderate output of a single fractal equation; (**b**) In a seed, the angle between helical

oscillators and the length of the oscillator, defines an allowed area, that remains constant for all layers; (**c**) The definition of synchrony and de-synchrony; (**d**) Two plots (above) shows how four inter-related parameters coupling factor, quality factor, intensity and frequency determine the synchronization or cumputing time. The plot below explains the basic definition of coupling factor and Quality factor.

**Fractal antenna enables all frequencies to have equal intensity:Power scaling law:** In the start up hardware which has no information or argument, represented as a pure function $F(x, y) + iG(x, y)$, all the fractal antenna are ideally connected, no deformation is observed. Therein antenna reconfiguration is the key to information storage and processing. Due to fractal shape, entire artificial brain follows a power scaling law, which ensures a homogeneously distributed power consumption throughout the system, and it ensures frequency invariance [56]. However, frequency invariance undergoes a little change continuously due to the evolution of the circuit.

(**i**) **The principle of coherent superposition of sensory fractals:** All information entering into our artificial brain-like computer via our specially designed sensor attachments first converts entire sensory information into a stream of pulses in a well defined architecture as shown in Figure 14a [62]. The time gap between two pulses is the frequency, however, there are several kinds of time gaps between groups of pulses. Every group is characterized with a particular frequency range and distribution. The "collection of frequency" is shaped in a 3D form as a function of time, (2D surface changes with time so we get a 3D pattern), which is transported via nerve bundle to the original brain. Each class of sensor operates in a different frequency domain and cooks its own fractals in a very particular way. However, there is a strict principle of defining sensory fractals in a living body, we call this "cooperative superposition principle". We have already noted before that if we plot F and G terms of all fractals of all sensors in a living body or attached to the brain, then in the combined graph paper no point will be common and no point will cross between any two fractals. This is very essential to design the fractal-based computer like ours to operate, because eventually, when all sensory fractals will fuse in the artificial hippocampus, then, there should be no conflict. The protocol that is followed for the superposition is demonstrated in the Figure 14a right, using a line, several fractals in the five sensory domains located in different parts of their distinct resonance chains are connected to form a network.

(**ii**) **Automated sensory data editing in a Fractal way: The necessity of linear circuiting:** Linear wiring is used in the brain to execute sensor-specific manipulation of the entire information content, in our computer, we do not rule out the necessity of linear processing. Only point we keep in mind that all individual components which are circuited, form a single molecule like oscillator even after the connection is made. Each component is like atoms of a molecule and the wiring acts like orbital bonds. On the way to the sensory signal-processing region of the brain, the different kinds of signals are synchronized; filtered, manipulated and thus, initial processing is performed. For example, distance between two objects are processed in the synaptic cord (touch sensor synchronizer), some parts of the visual data from left and right eyes are exchanged, *etc.* Since all sensory data is converted in terms of fractal seeds right after they are captured, coordinates of each neural cable does not require to be fixed in the bundle, it is transport of a single fractal that evolves with time. Depending on the sensory signals, particular manipulations are carried out: (i) Determination of particular kinds of correlations among certain groups and fuse them into a single fractal or divide them into multiple fractals; (ii) Synchronize particular kinds of groups if they occur repeatedly within a particular time interval [44]; (iii) Isolate some parts of the unified fractals physically, or physically couple with other sensory fractal-seeds by wiring manipulation. The advantage of fractal based data manipulation is that fusion, fission or other mathematical operations are carried out even in the presence of massive noise and since we use frequency fractals, we do not need to superimpose two sets of signals at a perfect time. Even if there are delays, particular patterns overlap depending on their mathematical geometry.

> **Figure 14.** (**a**) The plot on the left shows how input image distributes as fractal seeds, in five layers, before sending the output. The plot right shows how multiple sensory domains with distinct resonance chains hold a complex "if-then" argument; (**b**) The complete operational protocol for the decision making in our computer; (**c**) Tree is resolved into a geometric shape at the top which triggers several similar objects (umbrella route); (**d**) Each layer has its own dynamic stack of layers, and those dynamics layers couple during computing, the network between dynamic layers are not shown for clarity. The temporal stacks couple in astronomical ways.

**(iii) Expansion of "If" clusters and phase transition to "Then":** All 3D sensory fractal data triggers the matching frequencies in the artificial cortex region of the brain, all associated fractals triggers and the expansion is massive, global, several fractal seeds get activated. Equilibrium is reached. During expansion of the synchronized fractals, several indirectly coupled seeds, when they form new fractal seeds, an attractor loop [68,69] works naturally as a direct consequence of multi-directional synchronization process. This attractor is also an essential feature of the decision making, because even when we look at a single fractal seed (of a tree), depending on the conformational changes involved in the process, the probabilities of all the fractal forms arising out of that seed are not equal, some forms are favored, there is a probability distribution. Similarly, when several fractal seeds are coupled to create another seed, probability distribution of the associated dynamics triggers an attractor and the system when activated spontaneously tries to take the system in a particular direction. This is what we call the execution of an "if-then" statement or symmetry breaking. Local scale transformations of "if" into "then" is a continuous process, it starts at different places locally and then spread globally. It was earlier mentioned that the rules generated in the topmost layer or lowest frequency regime of the frequency fractal hardware is stored in a special place, namely artificial forebrain. The local scale symmetry breaking spreads all over the fractal hardware and it shifts from local to global scale phenomenon.

**The fundamental steps of computing: The cyclic loop:** The complete computing path of our artificial brain-like computer is shown in Figure 14b. Beyond a certain threshold, artificial forebrain rules or fractal seeds matches the fractal-seed compositions of global "if" pattern and then a large global scale symmetry breaking takes place and from "if" the "then" pattern is generated. We define it as a phase transition. Entire "if-then" system, including those fractal-seeds, which are not found in the brain, fuse to form a single fractal and then the unit is sent to the artificial hippocampus. The device that has a spiral path, which automatically integrates all fractal-seeds arriving from different parts of the brain and it sends the fused fractal back and forth to different regions of the brain in a cyclic loop. Neighboring artificial amygdala, thalamus, hypothalamus, limbic system, pre-frontal cortex all eight vital mid brain components act a "fractal eraser" and "fractal transformer" which causes another phase transition of the fused fractal in the hippocampus. The transformed fused fractal is then sent back to the cortex region and another expansion of "If" cluster and phase transition occurs. After a few cyclic processes, there are not many changes in the eventual fused fractal in the artificial hippocampus is the final solution of the "if" fractal entered into the brain through various sensory organs, which is then sent to the artificial cerebellum for generating an output.

(iv) **The "umbrella rule of perception":** The strength of our computing lies in the encoding protocol, since situations are encoded as pattern of fractal seeds, therefore, highly linked interconnects are saved as is in the 3D network as a new fractal which naturally includes its inherent dynamics. Later, if we cook any new "if" argument say a circle made of frequency co-ordinates located inside the 3D map of an extremely complex fractal seed, the corresponding "then" is automatically generated due to the physical process triggered by "if", even though that "if" is not willingly programmed in the system. The "then" is also a geometric shape say a square made of frequency points and a particular "then" geometry is selected based on the coupling between different geometric shapes in the fractal. One can create an astronomically large number of "if" in such a space, and answers to those "if" are always ready. Due to spontaneous generation of coupling continuously inside the fixed fractal hardware, more and more fractal seeds are generated at every moment. For one "if-then" argument or a fractal seed, an astronomically large number of correlated seeds are triggered and one can create new compositions made of basic elementary geometries as an input, the device would be ready to find a "then" composition for that unknown problem, instantaneously. Unknown means "not programmed or circuited intentionally". For the entire processing, we start from the input matrix of resonance peaks in the form of a fractal seed, which are the real values of a problem. However, our brain never search the information one by one from the real space, just like the umbrella fractal seed initially expands towards a higher perception level or global transformation rules, spread all over the fractal's frequency space through the uppermost layers. This is shown in Figure 14c. Eventually, once a convergence is reached then only the system point comes down from top to the bottom layer where the real information or arguments reside. As a result, we could drastically reduce the complexity of the search process.

(v) **No static information, the basic unit is temporal dynamics; spontaneous sequential programming:** When a sensory data enter into the brain as a 3D picture, due to the multi-layered seed structure resonantly oscillating like a single molecule, the entire hardware acts like a filter that carries out multilayered fractal decomposition unit as described above. As a result, all fractal seed groups in the picture are automatically separated in a multilayered architecture (see superposition of fractal section for details), thus, a single image is itself the origin of an intractable "Clique" problem [40].

When several kinds of information are linked as argument and forms a 3D column of arguments and a 3D column of rule of phase transition, even if we do not consider harmonic and anharmonic overtones. The situation is described in the Figure 14d. When the astronomical large number of temporal stacks shown in Figure 14d couple, even a simple static 3D column architecture (left column of Figure 1d) could generate a multistep sequential processing. Some elementary parts of an object or some groups move faster and some slower towards synchrony with other similarly shaped groups. Physically, it also means that the associated resonant oscillations undergo a synchronization process [49]. The hardware, during synchronization would bifurcate or trifurcate into multiple paths while recovering a new complex pattern say Fu(x, y) + iGu(x, y) and a variation in the synchronization speed would automatically generate an extensive sequential processing. Therefore, systematic argumentation is a natural byproduct of massive parallelism via synchrony. Additionally, if we consider that 3D column of phase transition rules, basically suggests which cluster of arguments would undergo synchronization faster, and that decision is also driven by the fundamental computation driving principle, "preferring the maximum density of coupled if-then argument domain" when there are several choices of synchronization paths. Thus, the sequential programming originates not only at the very elementary scale where a few fractal seeds are coupled, at the same time, it is also driven by a cooperative management of two columns (argument and phase transition rules) under the guidance of fundamental computation driving principle. This leads to a very interesting situation. During computation, even at a very local level, at any layer, any subtle finding of a unique high density path would change entire computation pathway and re-write the whole sequential programming.

**(vi) A comparison with existing artificial brain building adventures:** Izhicovich model [8,70–72] saves images as a limit cycle attractor [68,69], thus, image or pattern is a database of pulses, which requires synchronization for matching, and consequent memory retrieval. Izhicovich model is far better than reading the image pixel by pixel as followed in any existing brain building models of todays [4–16,61], however, synchronization dynamics are used as a tool to fit data, which is far inferior to the "reply back" concept noted here. In a sharp contrast to any existing brain building model, including fractal holography, for us, the objects get automatically isolated or filtered in the flow of time due to the differences in the synchronization speeds, thus, incredibly large number of images are packed in a single set of geometric shapes, which we name as fractal-seed. Such an automated classification and creation of a massively complex network is repeated during recovery of the entire information from the fractal seeds; therefore, the dynamics associated with multiple groups are stored and retrieved automatically. During retrieval, our computer matches this 3D pattern, if there is a match during a sync process, which could run for a long time, a new sync process begins. Due to the fractal-seed based storage, the hardware spontaneously encodes astronomically large number of 3D space-time units, small defects in the input image, which, hangs conventional programs, are not an issue here. Finally, at every stage of analysis, one could point out that existing computer hardware would face intractable problems, right from accessing the information in any given set of inputs to the delivery of the decision, every single technology implemented here is brand new, never proposed or implemented earlier.

   **(vii) Protocols to increase its creativity, intelligence and machine consciousness:** The strength of our computer depends on the length of the resonance chain starting from the smallest seed to the largest seed, for example, we have determined experimentally that starting from DNA to the largest oscillator that is the entire human brain has a bandwidth of $10^{30}$ Hz, DNA resonance occurs in the range $10^{15}$ Hz and entire brain oscillates in the range $10^{-15}$ Hz (a few hundreds of years ~$10^{-12}$ Hz, our concern is not biological death of a human, we take the slope of EEG). The resonance bandwidth of the universe is $10^{80}$ Hz, which starts from Higgs boson ~$10^{40}$ Hz, and to the gravity wave of the entire universe ~$10^{-40}$ Hz together $10^{80}$ Hz. Now, a human brain has ~$10^{30}$ Hz, means it covers a significant proportion of the resonance bandwidth of the universe. Thus, it has enormous bandwidth in a small space. Secondly, the density of resonance peaks is also enormous. The architecture of the brain is so beautifully designed that it has a large number of peaks in every region of the $10^{30}$ Hz bandwidth. Therefore, the brain hardware has the fundamental capacity to simulate a significant amount of events of the universe, thus, it is intelligent. However, maximizing the bandwidth and density of resonance states are not enough, for creativity, the hardware must be able to self-assemble astronomical columns of arguments and rules of phase transitions. Finally, since no one has defined consciousness yet, we do it here, based on antenna theory, which suggests that a suitable antenna-receiver system should be one sixth (14.3%) of the operating wavelength, since the universe is one wavelength of a gravitational wave, a conscious machine should have the resonance chain bandwidth (or length of resonance chain) of $10^{12}$ Hz. Now, for operation, this bandwidth must be packed in a small space, we inspire from neural networks in nature, hence it is 8 fundamental resonance peaks per two order difference of 10. It means between 1 MHz and 100 MHz, there should be at least 8 FRPs (discussed above). As noted above, these two are only crude parameters, ability to generate columns of arguments and phase transitions is the third and the final requirement.

## 8. Our Unique Model of a Human Brain (AjoChhand-Bio Mod)

   We have constructed a new computing model for the human brain by considering each composite material of the biological brain as electromagnetic resonance oscillators, based on our published [73,74] and recent experimental results. Our model considers that every single biological product e.g. any biomolecule say protein, DNA, enzymes and their bi-products used by nature is designed to exhibit unique electromagnetic resonance and associated mechanical vibrations. In simple terms we can say, all biological systems have a light (electromagnetic) and sound (mechanical) effect, and both electromagnetic and mechanical effects play an equal role in governing the neuron firing while neuroscience ignores the electromagnetic part and considers mechanical ionic movement controls everything in any living cell or neurons. The squarely parallel chemical and physical protocol is then implemented in creating an entire model of the brain, we believe that this step would fundamentally change the way biology is being studied today. We have noted above that the genes are linked to proteins electromagnetically and chemically as a route to synthesize and evolve entire architecture in a programmed manner. Though existing neuroscience [72] has concretely established the chemical link, it does not explain how self-assembly begins at the atomic scale and ends up in the meter scale, therefore, our electromagnetic interaction model is also an additional framework to provide the missing link.

**Fractal processing in the neural network:** During computing, the human brain receives pulses of different frequencies wherein various different kinds of sensory data (visual fV, auditory fA, *etc.*) are encoded as a 2D pattern of different frequencies. Right from entering into our brain, all kinds of sensory data are converted into fractal seeds of distinct nature so that when we plot those patterns from multiple sensors (fV, fA, *etc.*) in a single graph they do not conflict or overlap. The collections of different input frequencies at different sensors first transform into a pattern based on the sensor's typical spatio-temporal and intensity-phase variations. This image is automatically resolved into different groups by frequency-space hardware and associations of different classes of groups are automatically resolved in different layers of the hardware naturally. To store information in our hardware we utilize the natural complex frequency map of the hardware with subtle modifications as per the critical limits set by the complementarity criterion of the sensory data. After filtering of sensory input we use linear circuiting to send the signal to the region of the multi-layered hardware where particular kinds of sensory arguments are stored. However, afterwards we do not require wiring every element of protein or any of the higher level clusters. Similarly, transmission from sensory processing regions to the main computation-controlling center (artificial hippocampus region, which is also frequency-fractal hardware) should be done following linear circuiting data. Therefore, "reply back" in the operational human brain occurs via linear and massively parallel wireless computation circuits. The brain processes and stores fractal-seeds in the form of arguments, for example, if the A seed is triggered then an associated B seed will be triggered irrespective of its location inside and outside the layer. The input argument parts or many "If"'s when reaches a particular sensory region, we should take into account the fact that the entire processing region is part of a single integrated function (mathematically as explained earlier $F1(x, y) + iG1(x, y)$, or f1). The multi-layered network of devices a complex frequency map is stored (f1, f2, .... fn), therefore, we do not need to search each of them within the sensory region, all parts of a global function f that is the solution of the problem are triggered simultaneously [55].

If the observed similarity in the neural pattern spread all over the brain hardware is truly related to memory, the frequency fractal or the chain of resonance would signify that the memory is everywhere [53]. Thus, we see a human brain as frequency fractal hardware. This is based on our microtubule research. Though we mentioned earlier that "frequency" is our only control parameter, but intensity variation within a material during conformational change can redefine the gaps between neighboring pulses, thus, changing the "frequency". Consequently, we cannot explain this behavior as real space frequency modulation; rather, an imaginary or non-real parameter needs to be incorporated. When axon-cooked fractal frequency packet passes through the fractal antenna of the neuron, (microtubule bundle [73,74]), it evolves again due to the typical architecture of the antenna arrangement. But, we do not need to map with accurate details, the frequency fractal modulation at the local scale, that would generate enormous complexity in the information processing (there is no need to speed up the communication channels like a Turing machine, [66]. Note that nano cylinder bundle forming artificial axon is also construction of another fractal antenna and as described before in the context of "spontaneous reply back" that local changes follow a global function map $F(x, y) + iG(x, y)$, in the brain we always need to step up one level in the frequency range of resonance chain and then analyze the computing top-down.

**The decision-making protocol of the human brain:** In our model, the human brain is neither a classical nor a quantum computer, it operates among multiple imaginary space-time worlds at the same time, generating a column of self-assembled arguments as described above. Figure 15 describes operational model of a human brain based on our resonance chain concept. In that column "all possible networks of arguments exists, but different relative weights are assigned only when a query is made" [75], simply put, depending on the query which is a 3D pattern of "if-then" arguments a solution is projected by the column, which is also a 3D pattern of "if-then" argument. This is how a human brain solves a problem continuously, in fact, the column of arguments and the brain-jelly made of neurons have one to one correspondence and these two protocols perform only one computing in a human life, it begins with birth and ends with death (one life is ~$10^{11}$ s, it is a single pulse of a $10^{-11}$ Hz oscillator). There is no reduction protocol like classical or quantum computing, no deterministic decision for any query, resonance based projected set of vibration by the column resembles more with a holographic engineering. The human brain therefore is an automatic fractal correlation analysis, storage and retrieval machine, since relative weights of different arguments are assigned in the column of arguments only after the query is made, therefore, we cannot explain the human brain using a Turing machine. Since any input signal, visual, voice, taste, smell or touch senses are taken directly as 3D pattern of pulses, eventually converted in terms of basic geometric patterns like circle, triangles, several kinds of polygons, close and open U-shapes etc by "fractal decomposition", therefore, we can find some past correlations for any input fractal, processing does not require framed logic. The decision-making in the brain means transformation of one argument pattern into another. Thus, computing acts smartly when due to unknown parameters, the problem is even difficult to define, the faintly correlated pattern from the 3D column of arguments is projected as output, thus, we get instant decisions continuously for a series of inputs where even the problem is not defined.

**Figure 15.** The complete operational model of a human brain based on the resonance chain concept. Existing models consider "Fire and evolve neural circuit" part.

**For human brain, the most important part of computing is done at the sensors itself:** In the biological human brain, as soon as an image enters into our sensors, the neuron bundles are so designed that at the first step, very peculiar kinds of pulses are generated with typical slopes, both positive and negative intensity variations *etc.* In the second step, in the sensor itself, the typical features are removed; instead, pulses of equal lengths are generated; only information that is carried forward to the brain is frequency. If we look at the cross section of an entire sensory bundle containing a million nerves, it will appear as a 2D pattern of frequency changing as a function of time say it is Q(f1, f2). However, it should strictly be noted that the very particular geometric relationship of the input image is not destroyed. Our fractal frequency approach can generate the quasi-periodic oscillations in the brain described by Izhicovich as a direct output of a generic frequency fractal map [62].

**Learning and spontaneous brain circuit evolution:** As explained in Figure 14a, that the brain uses maximum energy in the central layers. The situation is more explicitly explained in Figure 16a. A gaussian plot suggests that conformational changes are maximum allowed in the central layers, where the neuron resides. The eventual fused fractal in the hippocampus, which is the solution of the input problem (layer 6 of Figure 16a), does not disappear instantly from the hippocampus after sending its mirror copy to the cerebellum. It sends the final solution fractal back to the cortex regions and the process to write the evolved fused fractal in the brain jelly starts. The majority of the fractal frequencies is found common, hence they neutralize vibration, however, some neurons vibrate continuously to meet the requirement of new frequencies. The minimum frequency difference causes "beating" and the antenna [59,60] of the neuron generates "electronic signal burst" which is then detected by all other neurons in the neighborhood. The stream of bursts continues until suitable neurons create wiring with it properly and/or microtubules inside the axon re-wires to accommodate the new fractal part [70]. Figure 16b shows the "beating" based modulation principle, in this diagram we also argue when brain decides to fire and why firing occurs at a very selective region of the brain. The mathematical foundation of the beating process is explained in the Figure 12c. Bursts in neurons having subthreshold oscillations are already studied extensively [71,72]. Automatically, higher-level phase transition rules are also generated in the forebrain. Once the hardware meets the requirement, computation stops and the artificial brain is ready for the next computation. Entire process completes around 200–300 ms. Artificial cerebellum sends a solution to the sensory organs for response, after every 200 ms. Within that time, the computation output of the brain is static, in principle, if one looks at our human brain at a gap of say 1ms then he will find that since computation is going on, the response is null.

**Figure 16.** (**a**) Resonance chains are located different parts of the human brain, inside those individual regions, several resonance chains exist for the smaller seed components; (**b**) The both routes of energy transmission paths in the human brain; (**c**) The mechanism of "beating", how brain decides writing and when actually brain decides to fire.

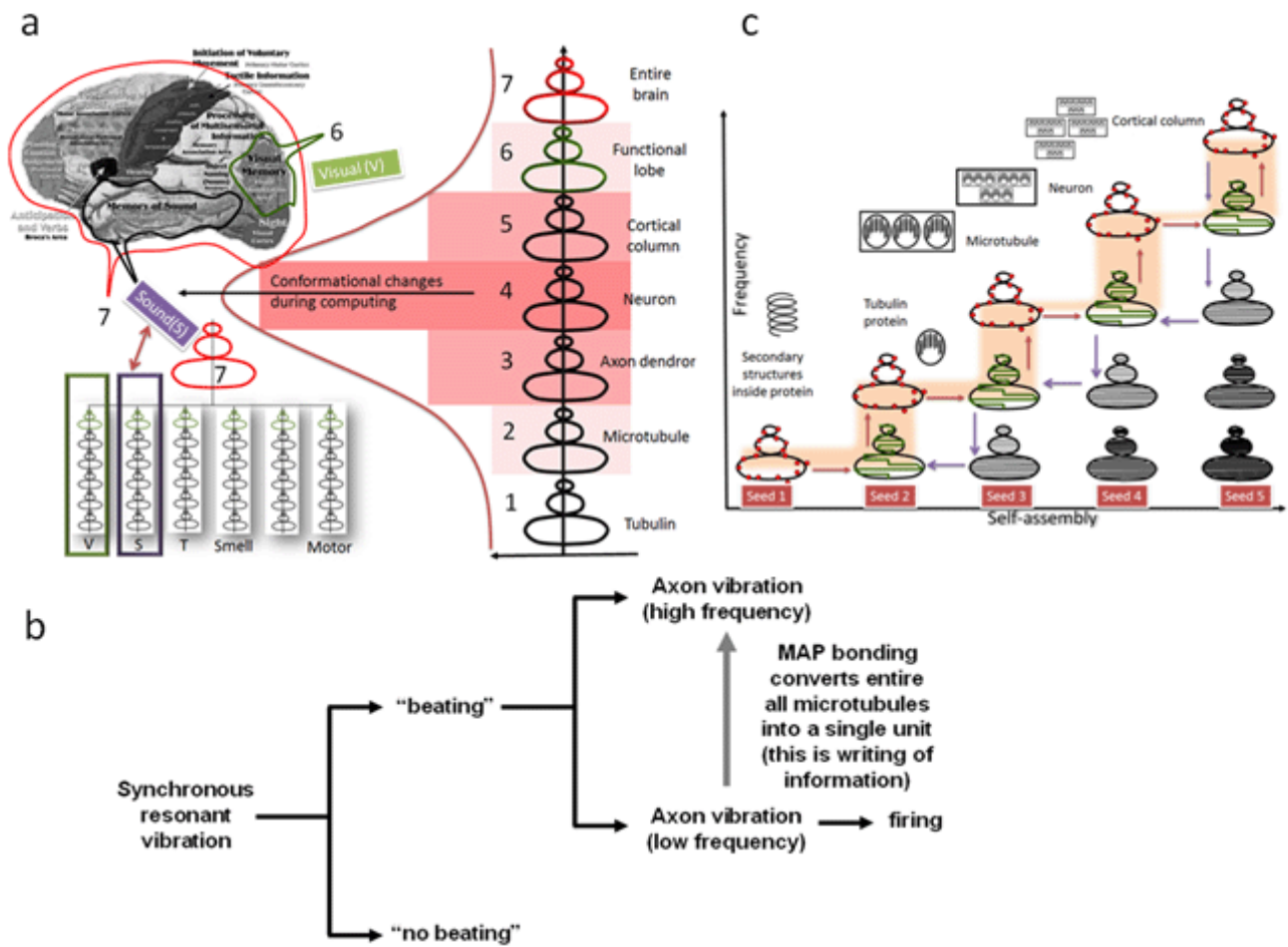

**Does it require any energy to compute?** Brain operates only with 15–20 watts and current artificial brain builders are demanding for 1000 Megawatts of power supply. We have been discussing on creating a complete organic version of a brain-like computer with 100 billion neurons for nearly a decade spending a very low power [76]. We have already developed a functional neural network [77], and a full-fledged massively parallel organic computer at the nano-scale that uses extremely low power [78,79]. Here, we finalize the human brain model by considering the human brain as the jelly made of organic molecular neurons kept inside an electromagnetic wave reflecting cavity, that is the skull in a human brain. The required power for computing in a human brain is drastically low due to wireless non-radiative power transfer that increases negligibly with the complexity of the network. the mechanism of this energy transport is explained in the Figure 16c. Right from DNA to the largest brain oscillator, the energy transport occurs both ways. We trigger natural vibration of the oscillators for computing; therefore, we do not require power supply even if we want to resonantly vibrate entire frequency fractal hardware made of 100 billion neurons. The only reason for energy expenditure is the circuit evolution, changing the neuron orientation, constructing new synaptic junctions, rejecting old connections, which are essential to write a new fractal seed in the column of "if-then" arguments. Neurons need to be fed, since cells are alive, we repeat, following our model, the energy required for computing is negligible.

**What is the information content in our human brain model?** In general, in our model of the human brain, basic sensors like eye, ear *etc.* are basically 3D "if-then" pattern construction engine and an automated "resonance state writing or erasing" machine are attached to every single component of the brain just like the way resonance states are encoded. Thus, all resonant oscillators, starting from DNA, protein enzymes microtubules in neurons and their clusters, every single component learn, memorize, and react by itself chemically (ionic and or genetic processing) as an observable part of global electromagnetic resonance based information processing ongoing continuously across the brain. Resonance state writing is a unique process in the brain, at different levels, it occurs in a unique way; at protein level, it is a change in the relative orientation of secondary structures; at the neuron level, it is selection of microtubules of different lengths at the neuron level; at cortical column or rhythm control level, it is the organization symmetry of the neurons. Thus, before any information is written the chain of resonance bands starting from DNA, protein, enzyme, microtubule, neuron to the neuron clusters and ends with the largest oscillator that is the entire brain.

**Complete resonance chain of a human brain:** We carried out direct experimental electronic resonance band measurement for DNA, proteins as shown in Figure 17a, microtubules and organic structures, neurons and their clusters, took EEG and other data for the global scale measurements. Microhertz resolution could be measured without noise trouble. Below microhertz, large time domain data was collected and based on the slopes nano hertz to femto hertz data are produced. Triplet of triplet resonance bands are observed every single layer, in each of the three sub-bands in a single triplet there are eight "fundamental resonance peaks" and numerous other resonance peaks as shown in the Figure 17b. Finally we have generated the resonance band of the entire brain as shown in the Figure 17c. Here are 12 bands of a brain, first six bands are experimental data, rest six bands are derived from other researcher's brain data.

(1) **EKAM:** A DNA molecule acts just like a single molecule oscillator, has three resonance bands ($10^{10} \sim 10^{16}$ Hz, gap in order 6) Triplet 1 (1–15 GHz, 16–40 GHz, 50–75 GHz), Triplet 2 (10–19 THz, 50–80 THz, 100–228 THz), Triplet 3 (1–5 PHz, 7–10 PHz, 12–18 PHz). 400–800 THz are visible light region, Peta hartz is in the extreme blue domain.

(2) **DITIYA:** A single Tubulin acts just like a single molecule oscillator, has three resonance bands ($10^7 \sim 10^{13}$ Hz, gap in order ~6) Triplet 1 (50–140 MHz, 180–250 MHz, 300–400 MHz); Triplet 2 (12–18 GHz, 25–50 GHz, 100–300 GHz), Triplet 3 (8–20 THz, 22–30 THz, 35–60 THz). 300 GHz to 1 THz is the inaccessible THz band, wherein for a long time we had a technological gap. Terahertz radiation is emitted as part of the black-body radiation from anything with temperatures greater than about 10 kelvin, so does DNA and Tubulin, both DNA and Tubulin resonates with IR and UV.

(3) **TRITIYA:** A single microtubule, acts just like a single molecule oscillator, it has resonance bands ($10^4 \sim 10^{10}$ Hz, gap in order ~6) Triplet 1 (15–20 kHz, 25–80 kHz, 100–300 kHz), Triplet 2 (10–19 MHz, 20–40 MHz, 100–228 MHz), Triplet 3 (1–5 GHz, 7–10 GHz, 15–30 GHz).

(4) **CHATURTHI:** Microtubule bundle inside a neuron say axon, synapse, the local core skeletons, which is made by coupling multiple microtubules by MAPs, acts just like a single molecule oscillator, has the following triplets ($10^2 \sim 10^7$ Hz, gap in order ~5) Triplet 1 (100–200 Hz, 250–400 Hz, 500–800 Hz), Triplet 2 (15–20 kHz, 25–80 kHz, 100–300 kHz), Triplet 3 (500–800 kHz, 1–5 MHz, 10–19 MHz).

(5) **PANCHAMI:** A single neuron is made by coupling several axon bundles acts just like a single molecule oscillator, it has the following Triplets ($10^{-1} \sim 10^4$ Hz, gap in order 5) Triplet 1 (0.1–1.2 Hz, 1.3–2.5 Hz, 3–7 Hz), Triplet 2 (8–13 Hz, 14–80 Hz, 90–300 Hz), Triplet 3 (800 Hz–3 kHz, 4–10 kHz, 12–30 kHz).

(6) **SASTHI:** Neuron bundle like cortical column is made by coupling several axon bundles acts just like a single molecule oscillator, has the following Triplets ($10^{-4} \sim 10^1$ Hz, gap in order 5); Triplet 1 ($1 \times 10^{-4}$–$18 \times 10^{-4}$ Hz, $25 \times 10^{-4}$–$80 \times 10^{-4}$ Hz, $120 \times 10^{-4}$–$260 \times 10^{-4}$ Hz), Triplet 2 ($1 \times 10^{-1}$–$8 \times 10^{-1}$ Hz, $10 \times 10^{-1}$–$25 \times 10^{-1}$ Hz, $30 \times 10^{-1}$–$50 \times 10^{-1}$ Hz), Triplet 3 (1–10 Hz, 10–15 Hz, 18–30 Hz).

(7) **SAPTAMI:** Cortical column bundle like fractal unit is made by coupling several cortical columns or rhythm clusters acts just like a single molecule oscillator, has the following Triplets ($10^{-6} \sim 10^{-1}$ Hz, gap in order 5); Triplet 1 ($6 \times 10^{-6}$–$25 \times 10^{-6}$ Hz, $30 \times 10^{-6}$–$80 \times 10^{-6}$ Hz, $105 \times 10^{-6}$–$260 \times 10^{-6}$ Hz), Triplet 2 ($0.5 \times 10^{-3}$–$1 \times 10^{-3}$ Hz, $2 \times 10^{-3}$–$12 \times 10^{-3}$ Hz, $15 \times 10^{-3}$–$40 \times 10^{-3}$ Hz), Triplet 3 ($0.8 \times 10^{-1}$–$1.2 \times 10^{-1}$ Hz, $2 \times 10^{-1}$–$4 \times 10^{-1}$ Hz, $5 \times 10^{-1}$–$12 \times 10^{-1}$ Hz).

(8) **ASTAMI:** Functional module made of several fractal-like-cortical column assemblies acts just like a single molecule oscillator, ($10^{-8} \sim 10^{-4}$ Hz, gap in order 4) ; Triplet 1 ($9 \times 10^{-8}$–$16 \times 10^{-8}$ Hz, $19 \times 10^{-8}$–$28 \times 10^{-8}$ Hz, $30 \times 10^{-8}$–$55 \times 10^{-8}$ Hz), Triplet 2 ($3 \times 10^{-6}$–$15 \times 10^{-6}$ Hz, $16 \times 10^{-6}$–$26 \times 10^{-6}$ Hz, $35 \times 10^{-6}$–$65 \times 10^{-6}$ Hz), Triplet 3 ($7 \times 10^{-4}$–$16 \times 10^{-4}$ Hz, $18 \times 10^{-4}$–$25 \times 10^{-4}$ Hz, $30 \times 10^{-4}$–$55 \times 10^{-4}$ Hz).

(9) **NAVAMI:** Sensory and and sub-functional-modules (sensory organs, nucleus, mid brain sub organs) and organizational components (hippocampus, cerebellum) are formed by circuiting several functional modules by massively complex linear wiring of neurons acts just like a single molecule oscillator, ($10^{-10} \sim 10^{-6}$ Hz, gap in order 4); Triplet 1 ($5 \times 10^{-10}$–$12 \times 10^{-10}$ Hz, $14 \times 10^{-10}$–$27 \times 10^{-10}$ Hz, $32 \times 10^{-10}$–$57 \times 10^{-10}$ Hz), Triplet 2 ($9 \times 10^{-8}$–$17 \times 10^{-8}$ Hz, $18 \times 10^{-8}$–$31 \times 10^{-8}$ Hz, $35 \times 10^{-8}$–$63 \times 10^{-8}$ Hz), Triplet 3 ($8 \times 10^{-6}$–$16 \times 10^{-6}$ Hz, $17 \times 10^{-6}$–$28 \times 10^{-6}$ Hz, $30 \times 10^{-6}$–$53 \times 10^{-6}$ Hz).

(10) **DASAMI:** Brain functional modules connected by superhighway neuron bundles forms a single giant oscillator (e.g., spinal cord, forebrain, left and right brain, entire mid brain) ($10^{-12} \sim 10^{-8}$ Hz, gap in order 4); Triplet 1, ($7 \times 10^{-12}$–$13 \times 10^{-12}$ Hz, $15 \times 10^{-12}$–$29 \times 10^{-12}$ Hz, $33 \times 10^{-12}$–$56 \times 10^{-12}$ Hz), Triplet 2 ($5 \times 10^{-10}$–$18 \times 10^{-10}$ Hz, $22 \times 10^{-10}$–$62 \times 10^{-10}$ Hz, $64 \times 10^{-10}$–$69 \times 10^{-10}$ Hz), Triplet 3 ($0.8 \times 10^{-8}$–$2.5 \times 10^{-8}$ Hz, $4 \times 10^{-8}$–$11 \times 10^{-8}$ Hz, $12 \times 10^{-8}$–$20 \times 10^{-8}$ Hz). Here one period occurs at three years.

(11) **EKADASI:** All brain modules connected by superhighway neuron bundles forms a single giant oscillator ($10^{-13} \sim 10^{-9}$ Hz, gap in order 4); Triplet 1 ($8 \times 10^{-13}$–$15 \times 10^{-13}$ Hz, $17 \times 10^{-13}$–$22 \times 10^{-13}$ Hz, $29 \times 10^{-13}$–$46 \times 10^{-13}$ Hz), Triplet 2 ($3 \times 10^{-11}$–$9 \times 10^{-11}$ Hz, $12 \times 10^{-11}$–$22 \times 10^{-11}$ Hz, $25 \times 10^{-11}$–$40 \times 10^{-11}$ Hz), Triplet 3 ($0.7 \times 10^{-9}$–$1.1 \times 10^{-9}$ Hz, $1.8 \times 10^{-9}$–$3 \times 10^{-9}$ Hz, $3.1 \times 10^{-9}$–$5.5 \times 10^{-9}$ Hz). Here, one period is nearly 30 years.

(12) **DASOSHI:** Entire body sensory network interfacing with the brain as single oscillator, all distributed sensors all around the body integrates with the entire brain just like a single giant oscillator ($10^{-15} \sim 10^{-11}$ Hz, gap in order 4). Triplet 1 ($20 \times 10^{-15}$–$30 \times 10^{-15}$ Hz, $33 \times 10^{-15}$–$55 \times 10^{-15}$ Hz, $59 \times 10^{-15}$–$76 \times 10^{-15}$ Hz), Triplet 2 ($0.9 \times 10^{-13}$–$11 \times 10^{-13}$ Hz, $15 \times 10^{-13}$–$21 \times 10^{-13}$ Hz, $27 \times 10^{-13}$–$42 \times 10^{-13}$ Hz), Triplet 3 ($0.76 \times 10^{-11}$–$3 \times 10^{-11}$ Hz, $4 \times 10^{-11}$–$12 \times 10^{-11}$ Hz, $15 \times 10^{-11}$–$20 \times 10^{-11}$ Hz). Here one period occurs at three thousand years, it does not mean that it required 3000 years for the changes to be felt, the time gradient is 3000 years. In an atom, further we go outward from a nucleus, separation between energy level decreases, energy decreases, for resonance chain, it is just the opposite.

**Figure 17.** (**a**) Tubulin protein resonance measurement, intensity of electromagnetic resonance, positive and negative direction, as a function of frequency; (**b**) Main resonance frequency band for a particular oscillator, say, tubulin or microtubule. Background band is the natural band, after conformational/structural change, the band reformats and this information is stored; (**c**) The construction of frequency fractal, complete bands of microtubule is shown.

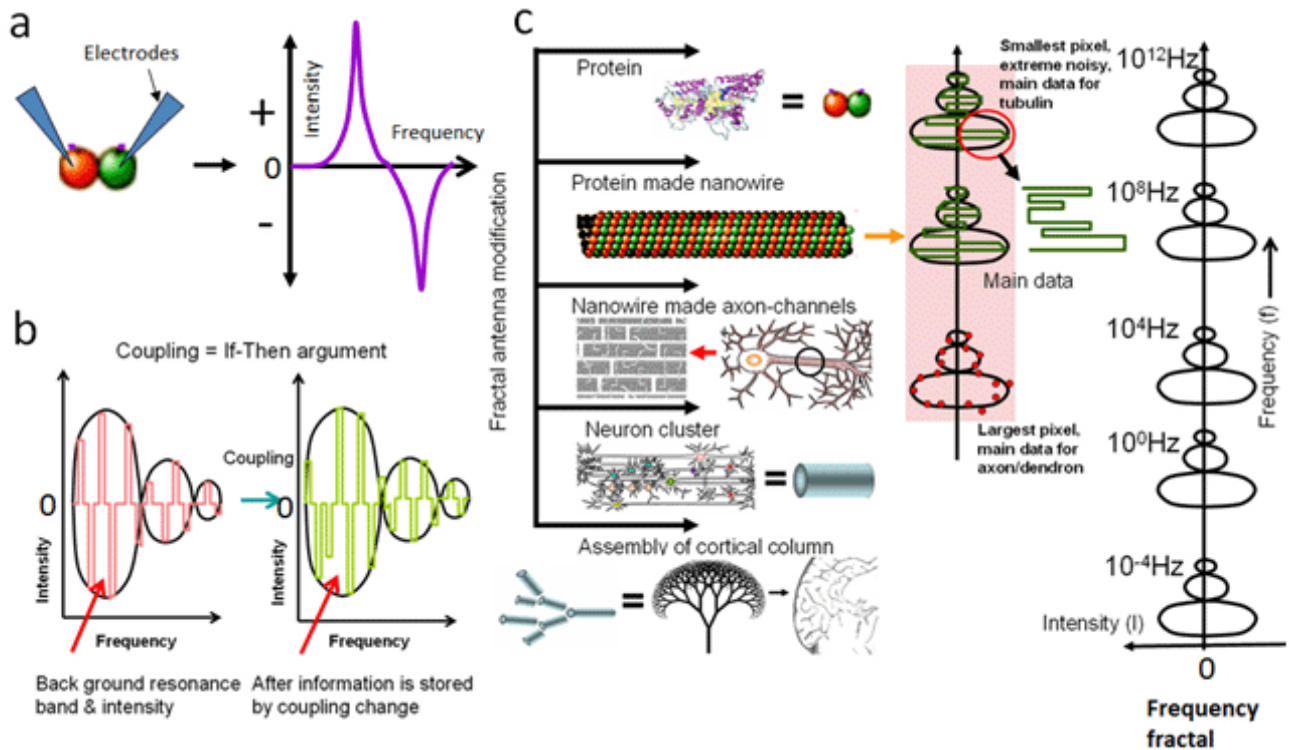

**Can we read absolute information in the Human brain EEG and f-MRI data using our model?: The answer is yes:** In general, the brain does not use an intensity variation for the decision-making to avoid noise. Brain uses only one parameter, frequency, therefore equal width and height pulses. This is the reason, that in the sensory, the cortex or any other region of the brain, the wave train appears as if only one kind of pulses are being transmitted, information could be retrieved only when one could read the original frequency fractal and estimate its deviation from the background frequency fractal. If one looks at the real pulses generated by the brain or any signal, no apparent similarity would be visible [44], when those pulses are converted in terms of "frequency" then the data would start making sense. Therefore, we have to read the pulses of all neurons in a region capture all the frequencies and identify the real and the imaginary parts of the fractal equation separately. We have generated the resonance chain for the generic human brain, that extends even to the f-MRI and EEG frequency scales, hence, those data could be read and absolute information content could be extracted. Note that the difference between the pristine resonance chain and the modified resonance chain is the information content of the brain.

**Where in the brain do we see "spontaneous reply back"?** As noted explicitly in the Figure 16b, in our model every biological organ system has a fractal like a signal receiver and a fractal antenna for the wireless communication and/or "spontaneous reply back". A DNA replies back to proteins, enzymes and all single molecules, these proteins and small molecules reply back to the supramolecular assemblies created by them like microtubules and the neurons to the final assembly, the brain. Therefore, spontaneous reply back is not restricted only to the organization one level higher and one level lower. An ordered microtubules forming inside the axons and dendrons work together as that receiver and antenna simultaneously for the neurons, while the shape and geometry of the neuron columns form antenna and receiver to contribute to the resonance chain to reply back. It should be noted that Microtubules "spontaneous reply back" does not mean that microtubules communicate with neurons at a far far distance apart, it is an entirely different kind of wireless communication as explained above in the "spontaneous reply back" section. We explain below the human brain perspective of resonance chain that enables the reply back. Any change at microhertz at the giant oscillator affects the DNA via resonance chain and then that very DNA would modify the giant brain oscillator. Thus, every single seed replies back to everybody at every moment.

**If it is wireless computing, then, why is there a visible circuit of neurons in the brain?** To create a resonance band in a protein say tubulin, we have experimentally demonstrated that by measuring tunneling current that its electron density distribution in the secondary structures (cluster of α helices in tubulin) of a protein molecule is closely related to the resonance bands. Similarly we have seen experimentally that the potential fluctuations on the microtubule surface is related to its resonance oscillations. The same pattern evolution is also observed in the microtubule bundle of the axon. We have concluded that the potential fluctuations among different seeds, (alfa helices in a single protein, tubulin in a single microtubule and microtubules in the axon bundle) are delocalized among many seeds to generate the collective resonance properties. This is just like an electrons charge distributed among various atoms in a single molecule. Now, we cannot expect similar orbital coupling at all scales, but, distinct engineering would result in the "distribution of a single resonance peak governing electromagnetic potential among multiple sub-seeds". Say, molecules self-assemble to make a larger seed, then, potential takes a different form, which controls the electromagnetic resonance, that very potential should remain distributed among all elemental seeds inside the bigger seeds. Hence, when multiple microtubules form a single neuron, they couple resonantly and MAP or golgi associated electromechanical plays the connecting role, between two microtubule seeds just like the orbital in a single molecule. Similarly, for the multiple neurons, the neurotransmitters play the role of sustaining the potential distribution among several neurons [80]. In the cluster of cortical columns the inter-column potential induction via diffused gradient of neurotransmitters, the composition of neurotransmitter river-flows combine multiple cortical columns as a single integrated unit, just like a single molecule. Several cortical columns or nucleus regions made of neuron clusters operate as a single molecule like system. When several such sensory control domains located at far distant places in the brain needs to be integrated and formed a single molecule like oscillator then bundle of neurons are need to function as wire, just what atomic bonds do in a single molecule. Wiring is a mode to connect large functional lobes (sub-seeds) so that the potential corresponding to a resonance peak of the ultimate giant oscillator that is the entire brain is distributed among multiple far distant lobes just like a single electron charge is distributed among different atoms in a molecule.

**Twelve hierarchical levels of integrated hardwares of Human brain:** Our model makes a significant shift in the concept of a neuron, because in the conventional model of a neuron, the Ca, Na and K ionic channels control the membrane potential and that explains neuron firing. We include an additional factor, the electromagnetic oscillations of microtubule bundle as secondary control parameter of firing. Microtubule of specific length has a specific resonance band (some peaks are prominent some gets dormant), thus, brain, during its circuit evolution choose compositions of different microtubule lengths to generate a complex signal transmission behavior that eventually governs the membrane potential distribution and sets the statistical distribution of the different ion channel percentage. The skeleton of the entire neuron body turns alive via microtubule and no matter how far the axons extend, entire pathway of a giant network of a neuron cell works as if a single molecule.

Spiral symmetry of DNA quantizes electromagnetic potential, however, three different unique dynamics, atomic, sub-molecular, and Pi orbital coupling delivers three resonance bands (**Level 1, EKAM**). Secondary structures of tubulin proteins make eight resonant circuits. The existing electric potential of the living cell is sufficient to create high frequency in the higher and in the lower GHz oscillations in the helical domains of these protein structures (**Level 2, DITIYA**). The generated oscillations do not die out when these proteins form microtubule nanowire, it only takes a new form in the MHz range. In a single microtubule eight independent simultaneous parallel routes of signal processing exist at any given point of time, classically (**Level 3, TRITIYA**). By changing the MAP bonding between two microtubules of different lengths, the complex modulation of resonant oscillations of microtubule are made in the neuron core such that from high frequency oscillations, a low frequency wave packet in the kHz to MHz range is constructed naturally. For permanent data storage and processing, the microtubule bundle has a provision for chemically bonding with each other using MAP (microtubule associated protein) at particular lattice locations. More is the bonding by MAP, the frequency of vibration of the microtubule bundle decreases, however, the time-response turns extremely sharp, Q factor of the resonator increases. More is the learning via MAP bonding in the neuron's axon bundle, the more is the creation of sharp peaks during antenna radiation (**Level 4, CHATURTHI**). The collection of entire microtubule bundle will work as an integrated fractal antenna to generate complex programming in the electromagnetic burst pattern [80].

Helicity in the molecular arrangement is a key factor in quantization of energy, both in DNA, protein, enzyme, microtubule and cortical columns, hippocampous, rhythm generating complexes. The lattice geometry of protein carpet on the microtubule surface oscillates wide ranges of frequencies, exchange energy with all other microtubules in the same neuron to convert entire microtuble circuitry of the neuron behaves as a single molecule oscillator; note that microtubule residing inside a neuron does not radiate out power to a far distant microtubule located in another neuron. We do not require classical wireless communication in the brain. Entire microtubule bundle in the axons and in the dendrons oscillate as a single unit and this giant oscillator interacts electromagnetically with another such network located in another neuron. It also should be noted that neuron firing is a mean to orient and redefine the resonance energy band of the oscillator. This is how all isolated microtubule bundles generate a single neuron oscillator (**Level 5, PANCHAMI**). Firing pattern generated by a cluster of neurons forming cortical columns, rhythm generating complex and nucleus create 3D potential distribution again like a single molecule oscillator (**Level 6, SASTHI**). These clusters or cortical columns arrange in different local pattern and form functional domain containing visual and other function controlling activities. These regions even visually appear like a fractal or an ordered geometry, a functional domain's firing pattern, if converted to a frequency map we get an equivalence of the resonance frequency band (**Level 7, SAPTAMI**). Different functional regions of particular sensory or functional domains are connected by neuron bundles, the connections via circuit-like wiring makes sure that potential distribution over a large distance is like a single molecular oscillator ((**Level 8, ASTAMI**). For each function, speech, language, 3D shapr we get one oscillator for a typical functional module, different are connected by neural wiring to create one oscillator for each sensor, nucleus *etc.* the functional design delivers integrated potential distribution (**Level 9, NAVAMI**). At a higher level, in the cortex regions, sub organs of mid brains, hippocampous, cerebellum like organs are formed which are also made of neuron assemblies, at this level neurons are arranged in a peculiar symmetry to deliver particular kind of potential distribution and whole system acts like a single molecule oscillator (**Level 10, DASAMI**). those are combinations of local functional modules At the highest level all these organs are connected by bundles of neurons and those form a few single oscillators (forebrain, cortex, midbrain made of seven organs, etc), these oscillators resonantly communicate with the sensory organs, thus entire brain along with the spinal cord is a single oscillator (**Level 11, EKADASI**). Finally, the highest level is the integration between the sensors and the end oscillatory modules of the brain and to the sensory parts distributed all over the body (**Level 12, DADOSI**). This is the basic concept of the complete oscillatory model of the human brain.

At some resonance frequencies, the energy transmission is allowed from positive to negative direction and at particular frequencies, it will be just the opposite (x and y parameters). This particular criterion is very important for generating frequency fractal $F(x, y) + iG(x, y)$ and the trick to design it lies in the arrangement of a particular symmetry element in the structure. The second important feature that the artificial neuron should have is to change the coupling with neighboring neurons, for this purpose, microtubule bundles of two neighboring neurons are designed by proper choice of microtubule lengths and suitable orientation of the neuron-edge, chemical firing at the junctions of multiple neurons is a route to measure electric field distribution and decide how and where to orient the axon bundle, which would change the wiring with the neighboring neuron based on weak interaction or long range ionic interactions [81]. At every layer from the first to the seventh, potential distribution is the key that decides how to change the structure, which modifies the background resonance band and the difference between the modified resonance band and the background resonance band is the information content of the brain.

(i) **Neuroglia:** Biologically, a neuroglia performs several tasks (cleaning neuron cell deposits *etc.*), however, in our human brain model, wherein computing is the key, the neuroglia works as a synchronizer, we ignore other jobs as those are not directly related to the computing of the brain. In the human brain, neuroglia and neuron are in a ratio 2:1. To remarkably speed up the long distance chemical communication, brain needs signal amplifiers and in our human brain model the neuroglia is an essential buffer engine that delivers homogeneity in the potential gradient produced by individual neurons in the neural network. It is a multi-functional resonant oscillator is constructing the neuroglia. The fundamental electronic property of this material would be receiving any band of electromagnetic signal emitted from the neurons and then radiating it out all over the environment, thus working similar to an amplifier in the conventional power transmission lines. This is essential to maintain the homogeneity of the electrical potential distribution in the entire frequency fractal hardware. This is therefore computationally equally important to the neurons.

(ii) **Functional models of giant sub-Brain oscillators:** There are a large number of tiny brain components which plays a vital role in the human brain processing. In our artificial brain-building project, we are going to copy almost all of their physical computational impacts. Fractal like signaling and switching of chaotic attractor is very much present in the brain [82]. We have designed visual and non-visual sensors first, since cognition is one of the most critical challenges in the brain science.

Visual and other sensory control systems: Complete model of the human cognition and the sensory system from resonance chain-based engineering point of view:

A. Re-engineering the cognition via visual control system:

**[Vision Circuit → Optic Chiasm → Thalamus → Visual Cortex → V1 → V2(complex shape, color) → V3(angle, orientation) → V3a (motion + direction) → V5 (Perital lobe, fusion form gyrus)]**

We have explained the entire visual control system in the Figure 18a,b. As an essential part of image processing, we create a time gap between two streams of pictures for two independent visual contents, say, focusing an apple kept in the floor and a water bottle kept nearby. Brain cannot process all objects accurately in a particular visual image, and isolating them is equally important as conversion into arguments since one need to pick up what to process, at the same time, keeping an eye on the other parts of the image. There are several claims of artificial eye building by engineers; our approach is entirely different from any attempt made before, simply because we are not inspired by camera, *i.e.*, storing information in the pixels, we have discussed this issue in the beginning of this chapter. Rather, our objective is to preserve all information for the pulsed based processor brain, as arguments. For an eye, argument means a correlation statement, combining several visual parameters.

There are eight different kinds of pulses discovered so far by analyzing the human eye-cells with state-of-the-art instrumentation, which takes part actively in shaping the 3D waveform sent to the brain via eye-bundle of nerves. There are several different kinds of cells, which take part in the photon to particular electronic pulse conversion process. The 3D waveform is a 2D matrix with variable amplitude, the rate of change of pulse with time contains the phase information and the base of the 2D matrix simply provides the coordinates. 2D matrices are continuously sent as time t changes. We have argued above that all sensory organs keep particular places in the grid free for others to contribute; the same principle is applied for a large number of sensory cells in the eye too. Thus, if sensory cells reach threshold and fire, contribution automatically finds its place in the grid, thus, an automatic classification for the dimension of an object, color, shape, is made in the superimposed grid. Here we summarize the pulse management scheme. There are only three tools in a stream of pulse-based wave trains. First, gridding; it means 2D periodic pulses, each at the corners of the four-sided cells that groups a set of information. Second, all primary information is stored in the typical nature of the pulse (growth and decay rate, amplitude and coupling of more than one pulse). Third, several 2D pulse-streams are superimposed on the background waveform; a suitable filtering hardware could isolate each 2D pattern separately. Here we will use these three tools (gridding, pulse-shaping and fusion) to construct a model device that can encode everything that we see in terms of a stream of pulses.

**Figure 18.** (**a**) The operational mechanism of visual cognition; (**b**) The model of visual cognition, how a stream of pulses are produced when we look at a book (above), and when we resolve a picture, how the groups are made.

We have summarized the research of the last three decades into a protocol that is not exactly the way a human eye works but very similar. For example, any color could be expressed as a combination of three bits, red, green and blue, and one pixel of color could be expressed as a set of three ratio values of four consecutive positive and/or negative pulses A, B, C and D. The ratio of the amplitude between A and B would be, say, blue; the ratio between B and C would be green and the ratio between C and D would be red. Thus, a cluster of these pulses defines the color information in the brain, where the ratios give the percentage contribution of each primary color, say, $30\%\,R + 22\%\,G + 48\%\,B = Pink$, like this, we can create millions of colors. To rebuild it, we need coupled oscillators containing optical sensors and some analogue of horizontal cells found in the eye, which combines or couples the elementary pulses into a compound waveform.

The most important thing that an eye should encode is the boundary of an object. When we are looking at the book, the important corners or change in contrast should create a pulse in the 2D stream, since color confuses the intensity, brain cleverly isolates the processing of color part and the intensity part, using two different kinds of cells. We will follow the same, and have a grayscale domain for shape and 3D perception and a separate part for color processing. Therefore, a grayscale version of a colored image is constructed by creating pulse boundaries along the points of sharp contrast change, and if the average level of intensity in a region is particularly higher, then, signals from neighboring cells get added up and we get the background intensity layer along with the pulse boundaries. In the brain, this work is done by Ganglion cells, which send pulses to tell us about the boundaries of a certain object, for example, if we are looking at the book, the three sides will have straight lines in their boundaries and a background intensity region depicting the three sides. It should be noted that three classes of information, the pulse-group containing the color information, then finally, a grayscale image superimposed with each other where the object boundary and the intensity background is combined. This is how a static picture of an object is formed that tells us about its shape and color.

When we send several static pictures through our eye, we have to do something so that size, relative distance, movement, orientation, all sorts of information could be perceived. Our eye uses three simple tricks to use the conversion. First, there is a provision to have a relative electrical bias between the two eyes, when we move our pupil, what actually happens is that the baseline of the entire picture is either slightly tilted towards the right or towards the left. This generates integration between the two eyes, simply by moving the pupil knowingly or unknowingly, we combine two pictures as the part of a single potential baseline. Since, two static pictures taken by two eyes tell us about two different angular projections of a single object, this is just like creating a hologram, but only one integration via single potential baseline is not enough. We need to take the left part of the image seen by the right eye and right part of the image seen by the left eye, and combine them with the left eye and the right eye respectively. This causes complete integration of the images formed in the two eyes, and one could easily understand now the reason for creating a single potential baseline. If there had been no single baseline, the same cut-n-past by switching a part would have been complete fatal. Say, we want to create a species that has four eyes or eight, we have to create first a single potential baseline connecting all eight, and then, switching the common minimum region between two neighboring eyes.

The angular perception gives us a nice information as shown in Figure 18b, the relative size of an object and the distance between the objects. If an object is near to us, we see more of its angular parts, but if it is far, then observe less. Using this simple principle, if we add a detector that compares two fused images, one seen before and one later, the computing machine will be able to tell us which one is bigger and relatively by how much. The same principle also works for the measurement of relative distances too, more angular information, nearer it is. However, one can easily see that there could always be a conflict, is it smaller or far. This is the confusion we have always in our mind. At this point of time, we need higher level information processing where we learn to build groups or at least encode that information in the eye-network so that all possible grouping gets unraveled later on. This means, if we are looking at a mango in front of our eye, then, a boat coming to the river bank and the scenery on the other side of the river bank, then, everything nearby the boat, the other side and nearby the mango should form three groups. These grouping are done in the brain by two parallel technologies, one is to analyze the relative perspectives of a single object in a single static picture, which requires previous learning, and second, analyze the time-evolution of stream of pictures during live observation, which does not require previous learning.

This is done by comparing the static pictures, it should be noted that there are sensory cells in the eye itself that combines the static pictures produced in the eye at different points of time, since their response time and other properties are just defined like that. The important question that always comes to our mind, how an eye understands, that well one static picture is completed? Well, these same cells synchronize the pulse output all the pixels made of nearly a million in the nerve fiber that carries the image of a picture. Once synchronized, the entire picture is sent through. There is a physical limit of our brain, how fast it can send pictures, the minimum time gap between the two static pictures, tells us the speed that we can capture using our eye-hardware. Speed management throughout the brain is done by clustering neurons according to their preferred speed. Therefore, we have to create different classes of artificial neurons, classified based on their speeds.

Once eye gets the technology to create one complete picture, the next thing it does, is to use one dimensional time grid, it is like a ruler, strictly setting the minimum time resolution of synchronization as discussed above. In this 1D tracker, if we compare the scenery, mango, and the boat regions, we will see that (i) angular perception of particular regions does change with time keeping a fixed center. Therefore, the only job the eye has to do is to move its pupil, even if nothing changes in the world around us, our eye would capture more data confirming that what forms an object, where it sits, and about its relative dimension to an analyzer who is looking into a stream of pictures. The second point is that, (ii), change in focus, this particular action also changes the angular distribution of a particular object, so it becomes easier to identify an object by an external observer.

Non visual control systems enroute in a single category, we have explained all systems in Figure 19. Here we detail the engineering each sensors separately.

**Figure 19.** A diagram how actually the non-visual (taste, smell, touch, hear) sensors would operate under the new resonance chain protocol.

B. The fundamental engineering of the smell control systems:

**[Olfactory circuit → smell → Olfactory lobe of lymbic system (looped around primitive brain) → Amygdala → Cortex]**

Very interesting information about nose is that it was the first organ created during the evolution of the animals. Therefore, its nerves are very close to the brain-stem wherefrom the fundamental control of our body, heart-beating, lungs-breathing, and motor nerves are controlled. On another planet, if the environment is such that the touch is favored, one could find touch controls near the brain stem. Similar to eye, our artificial nose differs distinctly from any of the artificial nose built to date. Similar to human nose, we also use "lock and key" principle of supramolecular chemistry, where a single molecule could sense a particular smell-molecule. Each sensory cell has several fine hair-like cilia containing receptor proteins that are stimulated by odor molecules, however, there is no permanent chemical change. Hence, a resonant and wireless communication is triggered to sense odor molecules and then the locking is removed. However, it embeds a special pulse-cluster synthesizer attached to the molecular sensor, for a particular kind of smell a distinct spike-pattern is created. The number of activated receptors determines the intensity of the stimulus, the layer of neurons located beneath the sensory molecules in the olfactory bulb generates a synchronized wave. These waves have been recorded in the electroencephalogram (EEG) tracings, rise in the wave-amplitude indicates the intense sense, and fall is caused by inhibition. Note that EEG records information for thousands of cells at once.

Reinforcement technique is a beautiful feature of nose; some neurons remain as a buffer between the neurons in the bulb and the neurons of the olfactory cortex. These buffer neurons activate suitably during training, and strengthen the connection for generating a complex signal response, such that a species could have a peculiar sense of a complex combination of smells. Thus, if trained a complex smell signature could carry an extremely vital environmental change or threat or even a very particular event, this tool could alone serve as an extreme intelligent encoding machine.

C. The touch sensor controls

**[Touch detection circuit → Touch bud → spinal cord → thalamus → cortex]**

Touch sensors are distributed all over the body surface, and experiments have shown that the brain relies on the collective response of the touch sensors to create a packet of arguments regarding the body-shape that it needs to protect under external attack. Touch sense is used to determine size, shape and texture of an object, and according to the need, density, and pattern of distribution of sensors are varied to sense more or less pain, and touch discrimination. Slight movement of the body-hairs could sensitize the touch sensors distributed all over the body; the whole body is divided into particular regions in a very ordered fashion via neural network to transmit touch senses from individual parts of the cortex via the spinal cord. Temporal synchronization is the key to couple touch sensors across the body.

Touch sensors also require a permanent memory hardware mechanism, therefore, it has involved in the process dozens of chemical messenger and receptors to sustain pain if the condition continues, in our artificial brain, we could altogether replace it with physical permanent memory processing protocols as described in the chapter three. Information gathered from touch sensors requires three kinds of responses, first, immediate response by moving the muscle or the body-part, the decision is taken via spinal cord (Dorsal horn loop). Second, pain message is sent to the brain and instruction for inhibiting the excitation comes back, the process involves thalamus and cerebral cortex. Third, always, for all kinds of pain and touch senses a part of the signal is sent to the hippocampus and this particular brain stem region controls complete situational analysis by sending it out to the upper brain, cross-checking previous learning and then a comprehensive decision is taken, output is sent to the cerebellum to execute action.

D. Engineering of the taste sensor controls

**[Taste circuit → nucleus of solitory tract (brain stem) → thalamus → cortex]**

In the chapter nine we will describe training a nano-brain, arguing that an idiot baby brain gets trained by following three basic desires of a life form, first, acquiring food, second, defending the safety of the body and third, taking reproductive measures. When a body looks for basic foods, just to run the body, it does require the test buds to cover a range of tastes that provides signature that the food is suitable for the body or not, whether the food contains more calories or not, even, the concurrent necessity of the body could also be reflected in the tongue-sensory-response. The tongue is a chemosensory assay; different regions have distinct taste buds triggering both chemical and non-chemical responses. Adapting to the particular taste always decreases its sensitivity to a particular taste, and particular chemical concentration always helps in reaching maximum efficiency. The learning of the taste sensors has twofold importance, first, if the sensors are exposed to the polarizing pre-conditioner, it acts as an exhibitor, the sensitivity decreases, second, if the pre-conditioner is de-polarizer, then, it acts as an inhibitor and the sensitivity increases.

E. Engineer the auditory system

**[Auditory circuit → Organ of corti → Cochlear nucleus of spinal cord → Superior Olive (side of brain stem) → Inferior Colliculus → thalamus → cortex]**

Auditory signals are captured by giant tuning forks located in the ear that vibrate resonantly with the sound signal, the interesting part is that this is the only sensory organ where the relevance of our "oscillator based artificial brain development" project is distinctly visible. For all sensors, we are using resonant oscillators, since, wavelength is very less for light (400–600 nm), eye requires very small size molecular opto-electronic resonant oscillators, but for sound wavelength is large (300 m), therefore the drum required is quite large. A large-scale coherent neural assembly operates connected to the drum over a sub-second time frame (250–300 ms), voltage-sensitive dyes (VSDI) provide high-sensitive (both temporal and spatial) real-time imaging of the neural activity. While capturing visual signal human brain is most sensitive at the point of stimulation, but for auditory signal it is most active at the superficial layer, as the brain is not interested in the spatial distribution of auditory signal. In addition, neural assemblies are organized in such a way that, naturally, the visual neural assembly responds faster that the auditory counterpart in the first 120 ms response time. However, after 250 ms, the visual neural assembly response falls sharply, while a linear increase in the signal response of the auditory signal is observed. Signal integration time for the auditory response is much larger than the visual one.

Information is encoded in the neural assembly by spike intervals; there are three kinds of responses, mainly. First, "direct response to external stimuli", second, if the time gap is more between the spikes, then it slowly adopts and then starts responding. There are two types of adaptation, first, "build up to the necessity", wherein, signal response increases as the number of pulse increases; second, "bursting with a logical argument", wherein alternate periods of responsive and silent modes are observed. This is the pulse management grammar of a biological ear; in case of an artificial analogue, one could easily define its own grammar. While describing the grammar of eye-pulse, we described the delicate use of tri-phase signals, tri-phase means two peaks with a negative trough in between them, and the ratio between the crest and trough stores an argument. The similar signature is common to store and process arguments for all organs and neural assemblies.

**Non-sensory organ engineering:**

**Hippocampus:** The first and the foremost organ is hippocampus [83]. The basic architecture of hippocampus will be a helix shaped molecular assembly with large number of cables coming out of it radially at different parts of the helical length. The cables and all parts are made by neurons only in the real human brain, thus we in our artificial brain propose to build the same, the most critical part of AjoChhand-Org creation is the synthesis of a suitable equivalence of Tubulin that would automatically generate microtubule and neuron analogue following a similar resonance chain model, the rest of the synthesis of all mid brain components are controlled by particular frequency modulated environment. The basic principle of hippocampus design is "time delay is proportional to the length of the wire the signal will travel". Just based on this simple principle we can fuse all different kinds of frequency fractals in it and at the same time isolate them, categorize and send them back to the particular specific region of the brain. If we simply pump in all signals (as fractal seeds described above, the fractal seed is the unit of information in our brain) through the spring like structure, and we take out an output after every single loop, we can get signals with quantized phase differences. By self-assembling the neurons, in a network to generate a unique spring in the hippocampus region (all four layers), in addition to that it should be noted that pre-frontal cortex, thalamus, hypothalamus to all seven (or eight, we avoid any hardware debate at this moment) sub organs of the mid brain are connected by neuron fiber.

**Amygdala:** The second most important organ is amygdala [84]. Our functional model of Amygdala is a filter, which helps a human brain in selecting one among many possible choices of frequency fractal seeds. The fundamental drive for the brain for learning is automatically decided in the mid brain as several rhythm generators exist in this region, which play a fundamental role in putting fundamental life cycles in a sustained periodic loop. For a living life form the fundamental drive for learning are three folds, first, food for supplying energy to our body, second, sex for giving birth, third, safety of the body. This triangular filter is modified with the development of arguments in the artificial forebrain fractal-seeds. The drive for art, music and spiritualism are born which create complex drives for the brain and turn the triangular filter into a circular filter. In our human brain model nature has constructed a neuron based hardware that implements this complex filtering process. During a decision-making process, human brain receives the final fused frequency fractal in the hippocampus, which contains all possible decisions and solutions, the reduction of all possible choices to a set of preferred choices need to be done, and that very process is initiated by the seven sister organs of the mid brain, we consider one of the seven sister filter is Amygdala. The design of amygdala hardware is such that it generates a set of unique fractal seeds that reads particular signatures in any other fractals and according to its list of preferences, it deletes unacceptable choices or fractal seeds. In other words, amygdala sends a fractal to the hippocampus and that fractal checks coexistence of contradictory frequencies (multiple choices in the solution) and then it simply deletes the unfavored ones. The unfavored ones are created naturally in the neuron arrangement of Amygdala, based on previously adopted decisions. The hardware of Amygdala look like a seed with a large number of fractal antennas surrounding a complex assembly of neurons that changes its axon such that a very peculiar kind of fractal seed is constructed with remains saved in the Amygdala with birth and evolves with our life practices. Thus, the orange or mango is different for every single person on the planet.

**Thalamus:** Thalamus is a gateway to the sensors, a gateway to the brain and a gateway to the cerebellum [85]. It is a universal synchronizer. The machine looks like as if many springs are side by side, if one of them is triggered then a complex vibration is automatically triggered over the entire system. Each spring like structure accepts the particular kind of sensory signals. The nature has constructed this organ by self-assembling neurons and all input output connections are feed through vertically across this hardware.

**Pre-frontal cortex:** Pre-frontal cortex controls futuristic simulations [86]. This region is directly connected to the frontal lobe where the highest-level fractal seeds are stored. However, this region captures the final fusion fractal which is the solution of the computation carried out in the brain and concentrates only on the "then" related "action" parts, which the body will execute. Then it feeds higher-level arguments stored in the frontal cortex to generate "futuristic simulation" of the actions in the hippocampus. This hardware is made of two antenna and receivers coupled with each other. First with one antenna it reads the "action to be taken" parts of the frequency fractal, matches with the higher level rules "If" clusters, as soon as it matches, then those phase transition rules are sent to the hippocampus for transforming the final decision-making fractal. Thus, futuristic thoughts or imaginations are generated with this neuron made hardware. As this region expands the final frequency fractal in the hippocampus, the hippocampus sends it back to all over the brain and executes significant modifications in the final frequency fractals.

**Hypothalamus:** Hypothalamus in a normal human brain is used for vital motor controls and generate rhythms [83]. These programmed activities are not essential for the computer we want to build, however, it is the supreme authority, if we want highest-level transformation in the final decision-fractal in the hippocampus, then we can program those protocols here. The neurons are self-assembled similar to an LC coupled oscillator type periodic fractal seed generator and an antenna made of neuron that transform the final decision fractal within the certain limits. This is the interfacing point of the user modulating the computational process in the artificial brain.

**Cerebellum:** Cerebellum reads the final decision fractal from the Hippocampus and its hardware is a tree like fractal network of neurons [87]. It sends extremely synchronized signals to all sensors for better data capture for improving the decision-making process and actions for the organs. Since we use frequency fractal as a tool for information processing and in the beginning of our discussion, we noted that we construct frequency fractals such that if all kinds of fractals from all sensors are fused they should not conflict, therefore, we do not need additional hardware to identify which signal belongs to which organs. When the final decisions come from the hippocampus, frequency wise filtering is done automatically. Depending on the length of signal travel through the hippocampus if we connect the wiring with the particular wires of the cerebellum, right signal will reach the right machine output.

**Basal Ganglia:** Basal Ganglia perform procedural learning, or learning by practice [88,89]. This organ modifies the final decision-making frequency fractal of hippocampus so that a very systematic input is captured and the brain jelly could reconfigure the neural circuit concretely. Therefore Basal ganglia generates a new kind of frequency fractal that enables it to control multiple sets of solutions generated by the brain at an interval of 200 ms, in this sense this is one of the most important organ in the brain, it keeps continuation among discrete computational input-output process. The hardware necessary for this is the creation of a fractal and then run it in a loop longer than 200 ms so that it affects continuously the self-motivated data acquisition process. However, there is another important aspect of it, how would the fractal nature be determined? Unlike previously described organs, in this case the fractal nature is not pre-programmed. In fact, pre-frontal cortex simulates futuristic outcomes and if the final decision fractal before and after futuristic simulation suggests missing of significant "then" correlated data, then that event is captured by the basal ganglia and it generates a continuous relevant data capturing job.

## 9. Basic Engineering Principles of an Artificial Organic Brain: Model to Design and Organic Synthesis (AjoChhand-NanoOrg)

Helical symmetry is used as a basic device design as shown in Figure 20a, which transmits electromagnetic signals of a particular frequency (we call it resonance frequency), this passage causes electromagnetic oscillation in the structure, it might affect the associated mechanical vibration induced conformational changes, which could shift the resonance frequency. The conformational change in the helical structure causes memory state writing, which is essential for a distinct "if-then" argument storage. We design the smallest "reply back" seed device using multiple spiral oscillators, in which the length of a spiral is inversely proportional to its resonance frequency, thus, by suitably combining the spiral structures of different lengths one could generate a desired resonance band. Even in the larger seed structures, resonance peak amplitude varies with length, the majority of other electronic properties remain unchanged with length since the entire organic system is fundamentally an insulator.

Elementary helical symmetry structures self-assemble to form the seed for the next structure, the principle objective of the self-assembly is to create the next range of resonance frequency bands. There are three parts in the resonance band of any seed oscillator structure, one part is kept open for coupling with the next seed that will be formed after the self-assembly, one part is for its own processing, and the third part couples with the resonance bands of the seeds inside the structure. Thus, every seed has two hands in the resonance spectrum. Synchronization could trigger with one or more peaks singular or in a plural manner in the system. One single seed is synthesized, the rest forms automatically by self-assembly, as shown in Figure 1b. When the fractal hardware reaches to a classically perceivable time scale (and associated "frequency"), then we need a particular kind of circuit for a suitable "reply back" to user interface [90]. The complete design of an organic brain is shown in Figure 20b.

The system uses topological symmetry for the carrier transport, the helical symmetry could be visible or may not be visible in the apparent structure of the seed, however, carrier transport would ensure resonant transmission of electromagnetic signals across distinct paths for the core resonance frequency states, through the structure.

Electromagnetic and Electromechanical resonances operate in a cohesive manner in the structure. While electromagnetic resonance is responsible for the synchronization and de-synchronization of the oscillations, the electromechanical routes ensure memory creation, learning, generating evolving circuits.

Physically, the materials with higher resonance frequency resides inside the materials with lower resonance frequency. To make it a powerful computer, we need to maximize the density of resonance states and also the range of the resonance chain.

**Figure 20.** (**a**) Every single seed of the brain architecture has a helical symmetry just like the human brain; (**b**) The complete design of our organic supramolecular architecture based human brain. The egg-shaped complete brain has two parts (left), the upper part holds "brain jelly", the spinal cord and the mid brain is inside this egg, shown separately in the right. (**c**) The experimental set up where we study the circuit evolution of a brain jelly; (**d**) The microscope image capture of a video when 7nm seed creates visible brain circuits.

In every single seed, we want distribution of electromagnetic potential such that the basic element that configures an argument (instead of information we use argument), should not remain localized at one point of the seed, this electromagnetic potential should remain delocalized over the entire seed structure. Say, if an electron is the unit of an argument, then its cage should be distributed all over the immediate atomic arrangement just like that happens when an electron is added to a molecule. Some atoms take say 27%, some atoms take 8% some atoms take 19% and that very electron remains part of the entire molecule. This is the very critical aspect of quantum mechanics. Quantum architects try to create giant systems where single electron would remain fractionally distributed just like a single small molecule [34–36]. However, we do not want that. Say, molecules self-assemble to make a larger seed, then, potential takes a different form, which controls the electromagnetic resonance, that very potential should remain distributed among all elemental seeds inside the bigger seeds. This is a unique demand, and we want this because our density functional study suggests that only then we can create the kind of resonance band we have imagined. Thus a quantum like fractional potential distribution would integrate all elemental resonance states even in the largest seed that is the entire organic artificial brain hardware.

**(i) Construction of brain jelly:** The first step of the artificial organic human brain synthesis is to construct a protein like molecule which if triggered by an electromagnetic signal starts self-assembly. The experimental set up and how the "if-then" circuit evolves is shown in Figure 20c,d. Note that this brain jelly would be poured into the cortex domain of the completely organic brain architecture shown in Figure 20b. Recently, we have succeeded in self-assembling such an architecture, where a 7 nm organic molecular seed expands to another seed of 20 nm, which then is automatically coupled with several others to trigger another self-assembly of helical nanowire [42]. The schematic of the entire synthesis process is shown in Figure 1b. In this way, the self-assembly continues, this is one of the finest examples of multilayered hierarchical self-assembly. Once we synthesize the basic neuron structure, we need different versions of synthetic neurons. Modified versions of neuron are used only for self-assembling them in the form of a wire, spiral, glia like spherical, or semicircular assemblies [91]. These basic assemblies are essential for the mid brain structures. In our current phase, this is not a systematic rigorous synthesis of the entire hardware, but we change the electromagnetic field such that the neurons start sensing that they need to re-orient in the way want them to be to create the mid-brain organs, along with delicate wiring and other parts. This is an example of directed self-assembly, step-by-step complete organic synthesis of the entire brain is not possible practically. All neuron wires should grow one by one and extend to the final destination. From sensors to the editing regions of the brain and finally to the cortex domain there is linear wiring wherever we want to preserve the fractal-seed and do not want it to interact with anything else, we need to grow the circuits carefully there. Such a massive scale self-assembled supramolecular architecture was never done before. Now, other than the delicate places we use neuron jelly wherein neurons could re-orient and reconstruct circuits based on internal axon restructuring.

**(ii) Resonance modulation based organic sensors:** The sensors for the artificial brain are different from the existing world of sensors, in a complete brain design shown in Figure 20a, those are located at the bottom part of the egg-shaped architecture. We use a new kind of wireless sensor technology. In this technique, we use resonant oscillations of the sensing molecules following mechanical, chemical or electromagnetic energy in creating pulses of different kinds and then convert all complex-shaped pulses in to a wave train of rectangular pulses but with different time gaps. Thus, all kinds of sensory data are converted in terms of one and only one parameter that is frequency. There exists a universal coding protocol for all sensors as far as the global time gaps are concerned, which isolates multiple distinct frequency signals. At the same time, frequency space ($F(x, y) + iG(x, y)$) grouping protocol for visual, auditory, taste, smell and touch sensory data should be very different from each other. This kind of dual policy is essential to avoid mixing of particular sensory data (note that everything is written in the brain in terms of wave trains of rectangular pulses, so there is no visible difference!), and at the same time, frequency based groups in a particular sensory data should be necessarily isolated. This isolation protocol should be identical for all sensors, only then a fixed synchronization protocol would combine groups from different sensory data into a single new fractal.

This is a very interesting transformation. When different kind of energies falls onto the sensors, we get two different kinds of pulses one is in the positive direction and another in the negative direction. A complex composition of these two pulses generates all possible phase, intensity and growth/decay rates to encode the complex information content in the neurons. This is just like two variables x and y described in $F(x, y) + iG(x, y)$, when function of x and y construct the basic pixel in the frequency space of F and G we get a fractal. The functional behavior of F and G is defined at the sensor itself, composition of positive and negative pulses defines sensory data which are eventually transformed in terms of frequency of a rectangular wave train. Once we get F and G for the input frequency fractal $F(x, y) + iG(x, y)$, the next phase of transformation is to reconstruct F and G into F1 and G1 using the natural resonance frequencies of the nano cylinder (a microtubule analogue) that constructs the artificial axon part of the synthetic neuron that we are constructing. This transformation occurs naturally in the hardware. The necessity of the transformation is that we would not require additional power source for computation in the entire brain. Materials natural vibration would perform the job. Only power consumption in the brain would be for evolving the neuron circuits. Therefore, the sensor systems will be fundamentally different in its design principle.

**(iii) Power supply:** Power supply for our brain-like computer comes from the chemical nano-battery, but it should be noted strictly that our organic molecular systems are not alive, for computation, we do not even require to switch on and off the conducting states just like conventional computer, we simply need power to maintain a potential difference between the protein molecule and every single oscillator. This is a simple condition, however, it is a critical challenge. In the biological brain, we have neuron firing and the membrane potential helps all oscillators in the layers below and above the neuron scale, at every layer, attaching a small battery to individual oscillators is an impossible task. We follow the same trick developed by nature for our artificial brain too. Inside the spherical cavity of the neuroglia and in the neuron we implant the nano-batteries, alternately, ionic conduction protocols could also be implemented just like the real biological brain. The wireless power absorption systems are kept inside neuron and the glial cells and since we mix these synchronizers in a 2:1 ratio, homogeneously throughout the brain-matrix, we expect that these nano-devices will absorb the power sent from outside the brain cavity and work as a source of energy for neurons and glia's synchronization activities. All neurons get power from neighboring glias wirelessly only when they undergo firing, otherwise nano-batteries remain silent. From sensor connected power supplies the signal is transmitted as a 2D frequency pattern in the form of a fractal, which propagates via non-radiative coupling between neurons in the wiring to the core of the brain and pumps all associative "If-then" arguments. Our material design follows a very particular kind of femto-watt power processing technology that we have discovered in microtubule. The necessary power for wireless communication and computation is negligible. However, firing or antenna burst of neurons to reconstruct the circuit would require powering a few microwatts, since firing or bursts happen only at a limited number of neurons. The entire computer is designed to operate a few hundreds of watts.

## 10. The Software Analogue of our Human Brain Model: CMOS Route within Turing Paradigm (AjoChhand-Soft)

While we admit that creating the complex column of arguments (real arguments residing in the base and imaginary arguments along the column) where the bottom layer exists physically in the system and the top higher layers could not be written in any program, however, we can at least implement the cognitive principles described above to demonstrate the proof of concept of the brain at a very primitive scale. Thus, we create a generic machine to summarize the above finding by mutating the basic concepts of cellular automaton and partially correlating our experimentally derived artificial brain model with the existing computer science theories. We have included basic free-energy principle to comprehensively integrate the data produced in the brain [92]. Though we also believe that "simultaneity" cannot have a software analogue, yet we made a simulator that follows the philosophy of "frequency fractal".

The software version of AjoChhand brain uses two parallel approaches. First, we construct an evolved version of the cellular automaton concept; we call it "CubeNet" using which we have constructed an entire artificial brain. On the other hand, we have created behavioral neuroscience based software, wherein we have introduced several unique pattern recognition protocols to generate the entire artificial brain. Note that both the approaches are fundamentally different in the implementation philosophy; however, there is a common point. Consider both that neural pulses of equal width and amplitude constructs the basis of a frequency pattern in the complex domain $F(x, y) + iG(x, y)$ and that it is a pure pattern based computing that executes all kinds of processing and decision making in the brain. While in the CubeNet we consider that each cell of the cellular automaton is made of pulses, the behavioral software uses mathematical set theory to implement the principle of resonant coupling. As we have discussed above, coupling of the basic geometrical patterns as shown in Figure 21, in the frequency space $F(x, y) + iG(x, y)$ works as the foundation for the artificial brain construction. Therefore, we name it NeuBeSet protocol. The basic reason for exponential speed up in quantum computing is the existence of an associative matrix, which is automatically taken care of in the fractal representation, we make sure in the program the associative matrix survives [93].

**Figure 21.** A diagram how an image is decomposed in terms of fractal seeds.

**(i) CubeNet based complete brain model:**
We have made five fundamental changes in the basic definition of the cellular automaton based on our research on experimentally generated Cellular Automaton circuits [93–97]. We have described above the 64-bit cube representing the basic unit of information, our cellular automaton model is the software analogue of this 64-bit cube. The protocols are described in the Figure 22a–c.

**Variable Cell Size:** First change that we have made is changing the cell size and the neighborhood. When a series of input images are given, then, from the average intensity variation of those sets of images it is determined which cells do not change their intensity. Note that each cell in the cellular universe represents a pulse; equal intensity means all peaks have the same height or cell value (cell state). Then we fuse all equal intensity cells and make it into a single cell. As a result, we drastically reduce the amount of information we need to process in an image. This step enables an automatic detection of "open" and "close" features in a pattern.

**Multiple Cell State:** Depending on the intensity of the peaks, the cell value or the assigned color to a cell changes, the fused cells with identical intensities are isolated as new groups and moved to a separate Cube. Therefore, one Cube generates another Cube, we have devised several protocols to fuse multiple Cubes also as par the fractal protocol described above. In this way, we replicate the fractal's fusion/fission/modification by writing software algorithm. Software arguments are analogous to synchronization.

**Figure 22.** (**a**) Fractal seeds are searched in a grid following four rules, simultaneous evolution of multiple states (top-left), multiple neighborhoods (top-right), multiple states of a single switch (bottom left) and optimizing number of repititive rules (bottom right); (**b**) The size of grid is checked by seeing dynamics; (**c**) Multi-layered image decomposition.

**Parallel Grid Geometry:** In the bottom layer of the Cube, 64 cells exist; they should represent both hexagonal close packing and rectangular packing simultaneously. At every layer, both hexagonal and rectangular close packing are considered together, we need to apply this duality because our software analysis of pattern recognition suggests that if we analyze a pattern simultaneously in both kinds of environments then it is easy to resolve the complexity. This particular feature of the simultaneous existence of a dual universe is must for another reason. We are creating both kinds of software brains with the reason that it will learn patterns by itself and after learning it will save the rules. Our simulation shows that dual grid captures versatile dynamics more efficiently. Since we cannot create simultaneity using software, therefore, we have conceptually created two parallel universes of 64 bit Cube, which means always we use two cubes separately for the same image. One for the hexagonal grid wherein each cell looks like a circle and the other is for a rectangular grid where each cell looks like a square. Two Cubes when operate together generate the complete concept of patterns hidden in a single image.

**Multi-layer Grid:** In the conventional cellular automaton, we have only one grid and that is our universe. Now, the question is why do we consider a Cube instead of a single 2D grid? The Cube is made of multiple layers of 2D Grids stacked one above another. The bottom grid takes the real image and then the weighted average of the input pattern is sent one layer above, this reduction continues until the top layer. Since we have parallel universes, only the bottom layers of rectangular and hexagonal cubes are the real information and all layers above, in the two cubes coexist simultaneously. The concepts of a Cube instead of a 2D grid, the duality of the Cube grid, evolving cell size, fractal like fusion and fission of the Cubes to generate an entire network spontaneously forces us to change the name of Cellular Automaton and rename it as a CubeNet protocol.

**Simultaneous co-existence of multiple neighborhood:** The third basic modification is to consider multiple neighborhoods at a time. From the extensive behavioral neuroscience study described below (NeuBeSet), we have found that the nature of neighborhoods enables the software to isolate any patterns into a composition of simplest patterns. We have identified a unique set of protocols for the multiple neighborhoods. These neighborhoods we apply simultaneously to detect the correlation rules between elementary patterns generated in the grids that could explain the relationships of groups inside a set of patterns in the Cube when we move from bottom to top and then top to the bottom in a cyclic manner, repeatedly.

Cubes learn the correlation rule from Visual, Audio or any other kinds of data converted into 2D patterns following principles described in Figure 21. Due to the inherent property of the Cubes it creates a communication network like fractals among all kinds of sensory data and arguments. This network is very different from Ersatz brain building project [98]. We have constructed the software algorithms that replicate the functions of the brain components and following those particular rules the decisions are delivered for a particular problem. The biggest obstacle to this brain is the use of memory space and the search time. We are continuously developing these two fundamental limitations by adopting better protocols. Currently we are planning to attach this self-learning software to the net or keep it in a public place to learn by itself. We have included certain features in the program to track how this brain learns and forms the neural network, how it adds neurons using the memory space of the stored computer whenever necessary, we also track how and for which purpose this brain uses neurons.

**(ii) NeuBeSet based complete brain model:**

Based on the behavioral neuroscience reported in the literatures [99,100], we have developed a direct pattern recognition and learning program, which was initially created for visual images. However, for any sensor input, similar to the CubeNet based soft brain, our simulator would run on any sensory data converted into a 3D pattern form (2D grid and the z axis denotes intensity of the peaks). Below we describe how exactly we carry out pattern recognition and learning rules. For the brain operation, we have written a separate program that generates the "if-then" arguments automatically and following unique software based "self-assembly" protocol we store those arguments and phase transformation protocols in the form of a program. In this case, also, we map how the complexity increases and how the neural network is formed.

The brain component programs used in the CubeNet and in the NeuBeSet are very different. The reason is that for the CubeNet our basic element is a Cube, however, for NeuBeNet, we do not care about any machine perception at all. We concretely believe, just like a normal software program, if we write algorithms on the functions noted in the AjoChhan-BioMod, we should be able to generate a complete software brain. Therefore, there is a fundamental difference between CubeNet and NeuBeSet approaches. While the first uses a software program to create a machine which behaves like a new kind of hardware and tries to build a brain out of it. On the other hand, the latter believes that the new kind of brain operational principles are sufficient to create the entire artificial brain we do not need to include hardware features.

First, we divide any input image to this brain into three groups, (i) the original image (OG); (ii) non-contact groups (NCG); and (iii) all curves which are in contact (having one or more junctions) into separate groups (CG). For pattern (i) blur to the maximum levels and slowly decrease to find most important 3/4 groups. For pattern (ii), if non-contact groups (say, isolated circles) reside inside any larger groups determined by implementing step (i) on OG, then keep them in larger groups. Other non-contact open groups (say randomly distributed "u" shape) should be closed separately and processed in two parallel ways. First, couple the neighboring contact groups to generate new group and repeat step (i) blur and group. Second, for non-contact closed groups, the centers are found and a curved contact object is formed. For all three cases of step (ii), we create higher-level perception layers for all such groups and for each group run PRP program detailed below. For pattern (iii) (contact groups) we search for closed loops, remove the closed loops keeping the junctions intact, then remove straight line like open parts, dots, one by one and thus create a set of groups from each contact group. In these cases, we also check whether we can create more groups when two or more objects share a common line. For pattern (iii), we run PRP program and convert the groups into transformed shapes. Our research suggests that the human brain uses very particular kinds of pattern recognition protocols (PRP) [101].

(1) **Recognition by components (RBC theory) theory to OC10AO theory:** We have modified the RBC theory of pattern recognition (Recognition by Components) [102], it was said that there are only 36 geons or geometric shapes [103], which construct everything that we see around us. In contrast, we consider that we use only 10 objects to define everything around us, 3 of them closed, triangle, circle and square [104]. We consider 5 open forms of these closed curves open triangle and rectangle, three open curves of circle (Figure 21, top row). Then there are straight line and junction of two open lines. Our brain is particularly interested in the junctions and isolates basic objects from junctions very efficiently and basic categories of all patterns are two types "open" and "close" (OC). For all types, we check angular orientations (AO). Thus, we construct OC10AO theory. Any given pattern is then converted into a composition of objects, preserving the orientation, which could easily be stored spending only a few bits of memory such that rotation does not have any impact on the detection process. The most important is the role of "search grid" described in Figure 21, the grids identify the components in any image.

(2) **Bottom-up and top down processing: CSBSL theory:** We modify every single pattern by connecting some open sections (C), Smooth it (S), Blur the images into various levels and transform into a contour image (B), make its skeleton (S). This CSBS protocol runs in a loop (L) to generate an analytic matrix, wherein each element is an image. Statistical analysis of this matrix delivers most probable choices and we start with that to implement OC10AO theory for the transformation of every objected in terms of composition of 10 objects.

(3) **Automated noise removal: JIDT protocol:** We introduce a unique Junction invariance de-grouping technique (JIDT), which enables our brain to recover an important pattern from noise that cannot be removed without prior knowledge that it is not important. For example, we are given an image "a" and asked to detect it with a straight line crossing the letter through the middle. How do we ignore the straight line and capture the information? There could be an astronomically large number of possibilities by which one can introduce simple sets of noises, it is impossible to categorize all kinds of noise and then program a preventing measure, therefore we have devised a protocol to classify the noise. There are two kinds of noise, contact and non-contact structures and these are automatically separated when we apply the isolation in the three steps.

(4) **Extrapolation and group isolation (GCSLCC):** Recognition protocols need learning the global category of the images that we are planning to detect. Therefore, we start detection simply by converting the original image into straight lines (SL) and detecting if a large number of lines in the image has gaps but same slopes, then we connect them and create a background line or structures, isolate the groups and start the isolation process in the beginning. Secondly, we take the same original image and convert the entire image in terms of curves and draw complete circles out of those curves (CC), if there is a matching then we construct that as a group by simply extrapolating the missing parts and proceed with the three steps of pattern recognition noted in the beginning. Both straight lines and curves help us to identify global correlations (GC) between multiple objects hidden in the images.

(5) **Projection from memory to improve cognition (PGBEP):** In some cases, several objects are placed so much in close contact, or multiple different objects are connected with each other that it is very difficult to isolate them using the protocols described above. Therefore, we are left with no choice to create a projection grammar book (PGB) [105], which is a collection of extremely simple patterns of the objects are kept. We write a program which tries to make certain artificial modifications by implementing those previously learnt pattern evolutions principles (EP, a series of higher-level rules are connected, works as a search string) and thus enhance the detection process. This particular approach is extremely helpful in recovering an entire pattern from a part of the structure.

Once pattern recognition protocols (PRP) are implemented, the process begins to activate the arguments via triggering the memory parts where associations are stored. Note that even for the associations we generate patterns and pattern similarity rules. Under any circumstances for CubeNet based brain and the NeuBeSet brains we do not use logical algorithms for decision making, it is always pattern recognition and pattern matching for any job we want to perform, at any level of perception any function we want to incorporate in the Soft brain keeping in view the conflict between supervised and self-learning [106,107]. Our multi-layered learning architecture is inspired by simultaneously co-existing fractal's compressed and expanded forms [108]. Finally, we implement a benchmark problem of "instant learning and predicting" of a chaotic non-linear time series, which would require astronomically large space to construct circuits and astronomically large time to write an algorithm. We could retrieve any part of the time series and several unique similarities of the chaotic process with already stored dynamics both sequentially and parallely [109].

## 11. A New Class of Fourth Circuit Element "Hinductor" Based Route: Beyond CMOS Human Brain Analogue (AjoChhand-InOrg)

Above we have described the human brain model and then analysed the possible realization routes via organic and software or computer programming protocols. The organic route is a slow, scaling up to a comprehensive scale would require an astronomically large amount of time, but the organic path is an accurate development towards a comprehensive realization of our human brain analogue, realizing the magic of computing and demonstrating those great principles to the industry would take a long time. In that respect the software route shows the incredible magic in the cognitive abilities of the brain though even with immense computing power, the computability cannot be scaled up, due to the nature of the complexity of out Human brain model. Therefore, we opt for the third and the last option, wherein we prepare a new kind of device that does not have any relevance to neuron, however, could be used to implement our Human brain model. We have invented a new device to implement the Human brain model, its advantage is that we can print the circuit and therefore scaling up does not require an enormous time as in the organic route and also it does not require enormous software programming or computing power.

This unique version of our Human brain model will be a millimeter scale Inorganic Brain equivalent of the organic nanotechnology based version described above (AjoChhand-Org). We tried to implement the unique features of AjoChhand artificial brain building using conventional circuit elements, resistor, capacitor, inductor OP-AMP *etc.*; however, it was not possible. The reason is that we do not require a simple memory-switch and non-linear features. In our AjoChhand-Bio Mod protocol, we need a processor that generates fractal and stores Fractal seeds. In addition, we have to keep in mind that we cannot use significant power in constructing neurons or other devices, if we do, then, integration of 100 billion devices would require megawatts of power supply, several power plants and that massive wiring are impossible to comprehend with a practical hardware geometry. Even if wiring is achieved, the power management would be extraordinary and the kind of evolving architecture we want to build, conventional power management schemes would be fatal to the formation of a chain of resonance band. Considering all critical limits, we find that we need to confine power expenditure of entire Human brain hardware within a few kilo-watts and, therefore, the power consumption in a single neuron equivalent device cannot be more than a few nano-watts or microwatts. Such a frequency-fractal processor device was never proposed before. We have to invent a fractal-processing device, for that purpose we start with the fourth circuit element, which is claimed to have been invented (as memristor) but we believe Leon O Chuas proposal in 1962 for the memristor as fourth circuit element is a wrong proposal [110]. Memristor is not a fourth circuit element, any device that cannot generate magnetic flux is not a fourth circuit element, also, Chua himself wrote in 1962 that all memristor properties could be derived using L, C and R, hence memristor is not a fundamental circuit element.

Therefore, we have devised a new class of fourth circuit element, which truly uses the basic features of a multiple resonance band oscillations just like the triplet resonance band we have described above while explaining our human brain model and using this giant sized electronic devices (giant compared to the size of our organic neuron) we will create a wire based electronic hardware of the entire brain model described above for the fractal model. Memristor is a known fourth circuit element, which has three versions, memristor, meminductor and memcapacitor [111]. We have created a completely new version of the fourth circuit element, which looks like an inductor but entire path is made of capacitors as shown in the Figure 23a. The helicity of the geometry is a robust concept, helical does not always mean to be a cylinder like spring. Some basic structures that we have explroed are shown in the Figure 23b. Since all properties of memristor, memcapacitor and meminductor could be derived using conventional circuit elements, we do not consider this as a fundamental element at all, in contrast no combination of elementary circuit elements could generate the electronic properties of our fourth circuit element, therefore, it is unique. We have filed a patent on this particular invention [112].

The basic electronic properties apart from "stored charge proportional to the magnetic flux" produced in the device are that the devices are as follows. (i) It operates between an upper and a lower frequency limits (Figure 23c); (ii) it switches faster than any equivalent capacitor, inductor or resistor (Figure 23d); (iii) it modulates signal out of phase at particular conditions; (iv) it operates in a very low power domain (nano-watts) *etc.* This is a pure fourth circuit element, which we name as Hinductor (short term is H). We consider the same neuron model noted above, therefore we need to carry out several modifications in the basic engineering designs of the Hinductor device. We need frequency fractal modulation and for that purpose, we need to have positive and negative resonance peaks which will function as x and y. We have already solved this particular criterion, the elementary capacitor which self-assembles in a helical assembly should be made as a composition of at least two resonance circuits, then if those circuits are oriented inside the capacitor hardware such that one allows left to right passage of signal at a particular frequency of an electromagnetic signal and the other circuit does just the opposite at an another distinct frequency then our desired condition is met. By extensive interaction, x and y would overlap and generate functions like F and G, once this is done, we can convert any image into a fractal seed (iterative function generator system). For this purpose we introduced leaky interconnect between the helical loops. To mimic the axon type modulation of signal demonstrated above, we have constructed frequency modulation by combining multiple helical loops in a suitable fractal geometry. It is obvious now, that we move radically to an unique electronic engineering while creating an equivalence of resonance band chain, we compromise with the technological hardware engineering obvious in the real Human brain and in our organic version, we concentrate more on the replication of the information processing protocols of our Human brain model, not as strictly as we do in our AjoChhand-soft version, but we create the "if-then" argument column most seriously in this version of the brain, since this is easily feasible here.

**Figure 23.** (**a**) Hinductor is organic, or inorganic or mixture of organic and inorganic materials helical device, the same way charge increases with voltage, the magnetic flux also increases; (**b**) The helical symmetry does not always mean tube-like structures, here several alternate helical symmetries in 2D and 3D planes are shown; (**c**) 180° out of phase signal production by Hinductor device (top-left), threshold operational frequency (top-right), Hinductor has upper and lower frequency limits of operation, lossless perfect square current voltage features (bottom left), current voltage under ac (bottom-right); (**d**) Temporal response of a Hinductor device, H = Hinductor, it is compared with a conventional L and C devices.

In order to create the hardware version of the if-then argument column, above described fractal geometry inspired assembly of the Hinductor is a sufficient choice, however, if we do not allow neuron circuit evolution, it is not possible to induce self-learning, in other words the column of arguments will be created however, it will be a static column. For this purpose, we introduce another engineering which enables us to evolve the circuit connections. Two ends of this modified axon-replica circuit is connected to a fractal antenna and a receiver which enables it to rotate the axon antenna with the help of a controller circuit that can orient the antenna to the most suitable direction. For that purpose, the entire circuit system is kept inside a sphere wherein it can rotate and change direction by spending only a few microwatts of power. Similar to organic versions of neuroglia we have designed a multi-directional antenna and a wireless power transmission device, this antenna systems are also kept inside a sphere. Figure 1b shows an example of such prototype architecture. Synchronization of pulsed peaks has been used for neural computation, here we perform the same for the frequency fractal computation [113].

Similar to the organic version of the brain components we can generate fractal modulation circuits, fractal sensor circuits, and fractal fusion and fission circuits, which are connected with the fractal antenna, the resultant devices work as an independent fractal frequency processor. In addition to the brain components, we also need synchronizers. Though the basic Hinductor device works as an oscillator, which synchronizes with other oscillators vibrating with similar resonance frequencies, we need synchronization sensors attached to the circuits, which would control the 3D orientation of the fractal antenna connected to the artificial circuit, which is basically the neuron of AjoChhand-InOrg brain. It is not functionally challenging to replicate the sensor-associated circuits of all the necessary components as described in the AjoChhand-BioMod protocol, just like the CubeNet based AjoChhand-Soft brain. The main objective of the sensor processing circuit is to convert external sensory signals acquired using a pair of camera, microphone, artificial touch and taste sensor devices, into a unique form of fractals. Therefore, signal to fractal conversion for sensory input processing and fractal to pulse based signal creation using suitable fractal converter circuits for developing artificial Cerebellum like circuit are two vital operations and we can design the circuits for these two purposes. In this way, we can construct all equivalent circuits of the brain components and connect them following the principles derived in the AjoChhand-BioMod protocol.

One of the most remarkable features of our AjoChhand Human brain model is that multiple fractal operations start at the beginning when data enters into the brain from the environment and entire processing of the brain ends with fractal. Therefore, how a series of circuit hardwares could maintain evolution, transformation and creation of new fractals is the logistics we have built within the framework of AjoChhand-BioMod, but one interesting issue is that for InOrg version of the AjoChhand brain we cannot use fractal like hardware just like we did in the Organic version. For the organic version of AjoChhand brain, we can design complete synthesis routes, which looks like a fractal and then generates fractal-based evolution of frequency. Now, in this Inorganic version we try to observe scaling up beyond level 3 of our Human brain model, where this weakness leads us to a failure and rectify those systematically. The power consumption for this brain is around a few kilowatts, the largest amount of energy is spent in rotating the antennas inside the cortex region of the brain.

## 12. Conclusions

We set a new computing paradigm expanding the concept of Gödel's incompleteness argument into an engine, it could be realized using combination of classical, semi-classical and quantum mechanics together by creating multilayered worlds with distinct clocks. Our artificial brain-building project differs from all others in the world because (i) we do not use logic gate based computing within the framework of Turing, our decision-making protocol is not a logical reduction of decision rather projection of frequency fractal operations in a real space, it is an engineering perspective of Gödel's incompleteness theorem; (ii) We do not need to write any software, the argument and basic phase transition for decision-making, "if-then" arguments and the transformation of one set of arguments into another self-assemble and expand spontaneously, the system holds an astronomically large number of "if" arguments and its associative "then" situations; (iii) We use "spontaneous reply back", via wireless communication using a unique resonance band coupling mode, not conventional antenna-receiver model, since fractal based non-radiative power management is used, the power expense is negligible; (iv) We have carried out our own single DNA, single protein molecule and single brain microtubule neuro-phisiological study to develop our own Human brain model presented here as AjoChhand-BioMod and created an organic version of it namely AjoChhand-Org. We have created a functional analogue model for the entire brain using software (AjoChhand-Soft) and a new kind of electronic device model (fourth circuit model) and then we have proceeded with its practical realization by designing an unique multilayered 3D wireless circuitry, it is named as AjoChhand-InOrg. Though we are using an organic supramolecular system namely brain-jelly for the brain-like computing, our brain-building task has largely simplified because we are not using chemicals, which are must for the survival of the body. From our computer we do not need to introduce the massive complexity of hormones, enzymes, and vital tasks of the mid brain to safeguard the living body, we are replicating their machine analogue functions only those which require computational decision-making. Our exploration would deliver a human like brain for the robots [114] or creative and intelligent handy companion for industrial use and at the same time would open up a new physical world of biology squarely parallel to the chemical only genetic and molecular biology existing today.

**Acknowledgment**

Authors acknowledge Batu Ghosh, Cameron Keys, Andres Kucera, Michal Cifra for extensive review of the manuscript and concepts. The authors acknowledge the contribution of the engineers Eiichiro Watanabe and Daiju Tsuya of Nanotechnology Innovation Station, NIMS Sengen-site Nano-foundry sponsored by Ministry of Science, Education, Culture and Sports (MEXT), Government of Japan. The current research work is funded by Asian office of Aerospace R&D, Government of USA FA2386-11-1-0001AOARD104173 and FA2386 -10-1-4059 AOARD-10-4059

**Conflicts of Interest**

The authors declare no conflict of interest.

**References and Notes**

1. Turing, A.M. On computable numbers, with an Application to the Entscheidungs problem. *Proc. Lond. Math. Soc. Ser. 2* **1937**, *42*, 230–265.

2. Martin, D.; Sigal, R.; Weyuker, E.J. *Computability, Complexity, and Languages and Logic: Fundamentals of Theoretical Computer Science*, 2nd ed.; Academic Press, Harcourt, Brace & Company: San Diego, CA, USA, 1994.

3. Minsky, M. *Computation: Finite and Infinite Machines*; Prentice-Hall, Inc.: Upper Saddle River, NJ, USA, 1967; Chapter 8, Section 8.2 "Unsolvability of the Halting Problem".

4. Hubert, D. *What Computers Can't Do*; MIT Press: New York, NY, USA, 1972.

5. Jung, S.Y. A topographical method for the development of neural networks for artificial brain evolution. *Artif. Life* **2005**, *11*,Sung Young, J. A Topographical Development Method of Neural Networks for Artificial Brain Evolution, Artificial Life; MIT Press: New York, NY, USA, 2005; Volume 11, pp. 293–316. This is an article from a journal called Artificial Life, not a book. Please follow our format for journal: see nonhighlighted part.. Author is from Korea and family name is Jung, not Sung Yong, which is first name.

6. Kurzweil, R. *The Singularity is Near*; Viking Press: New York, NY, USA, 2005.

7. A List of Some artificial brain building projects around the globe. Available online: http://www.artificialbrains.com/ (accessed on 15 January 2014). We would like to have some description about the link

8. Hoppensteadt, F.C.; Izhikevich, E.M. Oscillatory neurocomputers with dynamic connectivity. *Phys. Rev. Lett.* **1999**, *82*, 2983–2986.

9. Holographic Neural Net Project. Available online: http://www.andcorporation.com/ (accessed on 15 January 2014).

10. Adleman, L.M. Molecular computation of solutions to combinatorial problems. *Science* **1994**, *266*, 1021–1024.

11. Kuhnert, L.; Agladze, K.I.; Krinsky, V.I. Image processing using light-sensitive chemical waves. *Nature* **1989**, *337*, 244–247.

12. Ellenbogen, J. A Brief Overview of Nanoelectronic Devices, 1998. Available online: http://www2.mitre.org/tech/nanotech/ourwork/nano_papers.html#nanoelectronics (accessed on 15 January 2014).

13. Waldner, J.-B. *Nanocomputers and Swarm Intelligence*; ISTE: London, UK, 2007; pp. 173–176.

14. Domino Computer from Wikipedia. Available online: http://en.wikipedia.org/wiki/Domino_computer (accessed on 15 January 2014).

15. Drechsler, R.; Wille, R. Reversible Circuits: Recent Accomplishments and Future Challenges for an Emerging Technology. In Proceedings of the International Symposium on VLSI Design, Automation and Test, Hsinchu, Taiwan, 23–25 April 2012.

16. Ridge, E.; Kudenko, D.; Kazakov, D.; Curry, E. Moving Nature-Inspired Algorithms to Parallel, Asynchronous and Decentralised Environments. In *Self-Organization and Autonomic Informatics (I)*; Czap, H., Unland, R. Brank, C., Tianfield, H., Eds.; IOS press: Amsterdam, The Netherlands, 2005; pp. 35–49.

17. Quantum computing is getting matured as one does not require now to know solution beforehand (Zhou, X.-Q.; Kalasuwan, P.; Ralph, T.C.; O'Brien, J.L. *Calculating unknown eigen values with a quantum algorithm. Nat. Photonics* **2013**, *7*, 223–228), still logical steps in the form of arguments need to be constructed. Standard ISSN abbreviation, no comma after journal name

18. Neumann, J. The General and Logical Theory of Automata. In *Cerebral Mechanisms in Behavior—The Hixon Symposium*; Jeffress, L.A., Ed.; Wiley: New York, NY, USA, 1951; pp. 1–31.

19. Wolfram, S. Statistical mechanics of cellular automata. *Rev. Mod. Phys.* **1983**, *55*, 601–644.

20. Jaeger, H.; Haas, H. Harnessing nonlinearity: Predicting chaotic systems and saving energy in wireless communication. *Science* **2004**, *304*, 78–80.

21. Home page of Reservoir Computing. Available online: http://www.reservoir-computing.org/ (accessed on 15 January 2013).

22. Hinaut, X.; Dominey, P.F. Real-time parallel processing of grammatical structure in the fronto-striatal system: A recurrent network simulation study using reservoir computing. *PLoS One* **2013**, *8*, e52946.

23. Copeland, J.; Proudfoot, D. Alan Turing's forgotten ideas in computer science. *Sci. Am.* **1999**, *280*, 99–103.

24. Brun, T.A. Computers with Closed Timelike Curves Can Solve Hard Problems, 2008. Available online: http://arxiv.org/pdf/gr-qc/0209061v1.pdf (accessed on 15 January 2014). Our format

25. Gödel, K. An example of a new type of cosmological solution of Einstein's field equations of gravitation. *Rev. Mod. Phys.* **1949**, *21*, 447.

26. De B. Pereira, H.B.; Zebende, G.F.; Moret, M.A. Learning computer programming: Implementing a fractal in a Turing Machine. *Comput. Educ.* **2010**, *55*, 767–776.

27. Hilbert, M.; Lopez, P. The world's technological capacity to store, communicate, and compute information. *Science* **2011**, *332*, 60–65.

28. Siegelmann, H.T. Computation beyond the Turing limit. *Science* **1995**, *268*, 545–548.

29. Wegner, P. Why interaction is more powerful than algorithms. *Commun. ACM* **1997**, *40*, 80–91.

30. The Translation from the German Paper of Gödel. Available online: http://www.research.ibm.com/people/h/hirzel/papers/canon00-goedel.pdf (accessed on 15 January 2014).; The second order logic is relevant of this paper, we provide the Wiki link, Available online: http://en.wikipedia.org/wiki/Second-order_logic (accessed on 15 January 2014); And several classes of incompleteness theorem. Available online: http://en.wikipedia.org/wiki/G%C3%B6del's_incompleteness_theorems (accessed on 15 January 2014).

31. Bandyopadhyay, A.; Miki, K. Fabrication of a memory chip by a complete self-assembly process using state-of-the-art Multilevel Cell (MLC) Technology. *Adv. Funct. Mater.* **2008**, *18*, 1173–1177.

32. Lloyd, S. Quantum search without entanglement. *Phys. Rev. A.* **1999**, *61*, 10301.

33. Jozsa, R.; Linden, N. On the role of entanglement in quantum computational speed up. *Proc. R. Soc. Lond. A* **2003**, *459*, 2011–2032.

34. Bandyopadhyay, A.; Miki, K.; Wakayama, Y. Writing and erasing information in multilevel logic systems of a single molecule using scanning tunneling microscope (STM). *Appl. Phys. Lett.* **2006**, *89*, 243506. From IEEE website, we did not find title including this abbreviation for last 3 words.

35. Bandyopadhyay, A.; Wakayama, Y. Origin of negative differential resistance in molecular Junctions of Rose Bengal. *Appl. Phys. Lett.* **2007**, *90*, 023512.

36. Bandyopadhyay, A.; Acharya, A. A 16 bit parallel processing in a molecular assembly. *Proc. Natl. Acad. Sci. USA* **2008**, *105*, 3668–3672.

37. Kurs, A.; Karalis, A.; Moffatt, R.; Joannopoulos, J.D.; Fisher, P.; Soljačić, M. Wireless power transfer via strongly coupled magnetic resonances. *Science* **2007**, *317*, 83–86.

38. Nakamura, S.; Ajisaka, S.; Koma, R.; Hashimoto, H.; Lee, B.H. Electro-Magnetic Resonance Coupling Sensing for Secure, Comfortable and Energy-Saving Space. In Proceedings of the 2011 8th Asian Control Conference (ASCC), Kaohsiung, Taiwan, 15–18 May 2011; pp. 725–730.

39. Mirollo, R.E.; Strogatz, S.H. Synchronization of pulse coupled biological oscillators. *SIAM J. Appl. Math.* **1990**, *50*, 1645–1662.

40. Feige, U.; Goldwasser, S.; Lovász, L.; Safra, S.; Szegedy, M. Approximating Clique is almost NP-Complete. In Proceedings of the 32nd IEEE Symposium on Foundations of Computer Science, San Juan, Puerto Rico, 1–4 October 1991; pp. 2–12; doi:10.1109/SFCS.1991.185341.

41. Bandyopadhyay, A.; Sahu, S.; Fujita, D.; Wakayama, Y. A new approach to extract multiple distinct conformers and co-existing distinct electronic properties of a single molecule by point-contact method. *Phys. Chem. Chem. Phys.* **2010**, *12*, 2198–2208.

42. Ghosh, S.; Dutta, M.; Sahu, S.; Fujita, D.; Bandyopadhyay, A. Nano molecular-platform: A protocol to write energy transmission program inside a molecule for bio-inspired supramolecular engineering. *Adv. Func. Mater.* **2013**, doi:10.1002/adfm.201302111.

43. Hohlfeld, R.G.; Cohen, N. Self-similarity and the geometric requirements for frequency independence in antenna. *Fractals* **1999**, *7*, doi:10.1142/S0218348X99000098. No need to capitalize

44. Izhikevich, E.M. Polychronization: Computation with spikes. *Neural Comput.* **2006**, *18*, 245–282.

45. Vitiello, G. Fractals, Dissipation and Coherent States. In *Quantum Interaction*; Lecture Notes in Computer Science; Springer: Berlin and Heidelberg, Germany, 2012; pp. 68–79.

46. Lie Alzebra from Wikipedia. Available online: http://en.wikipedia.org/wiki/Lie_algebra (accessed on 15 January 2014).

47. Carbone, L.; Freyn, W.; Lee, K.-H. Dimensions of imaginary root spaces of hyperbolic Kac-Moody Algebras. Available online: http://www.math.uconn.edu/~khlee/Papers/ImagRootMult.pdf (accessed on 15 January 2014).

48. Anderson, C.C. Defining physics at imaginary time: Reflection positivity for certain Riemannian manifolds. Available online: http://www.math.harvard.edu/theses/senior/anderson/anderson.pdf (accessd on 15 January 2014).

49. Izhikevich, E.M. Weakly pulse-coupled oscillators, FM interactions, synchronization, and oscillatory associative memory. *IEEE Trans. Neural Netw.* **1999**, *10*, 508–526.

50. Barnsley, M.; Hutchinson, J.; Stenflo, Ö. A fractal valued random iteration algorithm and fractal hierarchy. *Fractals* **2005**, *13*, 111–146.

51. Oraizi, H.; Hedayati, S. Combined Fractal Geometries for the Design of Wideband Microstrip Antennas with Circular Polarization. In Proceedings of Progress In Electromagnetics Research Symposium (PIERS), Suzhou, China, 12–16 September 2011.

52. Barnsley, M.; Hutchinson, J.; Stenflo, Ö. V-variable fractals: Fractals with partial self similarity. *Adv. Math.* **2008**, *218*, 2051–2088.

53. Xue, G.; Dong, Q.; Chen, C.; Lu, Z.; Mumford, J.A.; Poldrack, R.A. Greater neural pattern similarity across repetitions is associated with better memory. *Science* **2010**, *330*, 97–101.

54. World of Fractal. Available online: http://www.math.nus.edu.sg/aslaksen/gem-projects/ maa/World_of_Fractal.pdf (accessed on 15 January 2014).

55. Durstewitz, D.; Seamans, J.K.; Sejnowski, T.J. Neurocomputational models of working memory. *Nat. Neurosci.* **2000**, *3*, 1184–1191.

56. Pandey, S.K.; Yadav, M. Fractint formula for overlaying fractals. *J. Inf. Syst. Commun.* **2012**, *3*, 347–352.

57. Dubey, S. Undecidable problems in Fractal geometry. *Complex Syst.* **1993**, *7*, 423–444.

58. Deutsch, D. Quantum theory, the Church-Turing principle and the universal quantum computer. *Proc. R. Soc. Lond. A* **1985**, *400*, 97–117.

59. MIT Physics Demo—Tuning Forks: Resonance & Beat Frequency. Available online: http://video.mit.edu/watch/tuning-forks-resonance-a-beat-frequency-11447/ (accessed on 15 January 2014).

60. Beats. Available online: http://www-math.mit.edu/daimp/Beats.html (accessed on 15 January 2014).

61. Eliasmith, C.; Stewart, T.C.; Choo, X.; Bekolay, T.; DeWolf, T.; Tang, Y.; Rasmussen, D. A large scale model of the functioning brain. *Science* **2012**, *338*, 1202–1257.

62. Izhikevich, E.M. Weakly connected quasiperiodic oscillators, FM interactions, and multiplexing in the brain. *SIAM J. Appl. Math.* **1999**, *59*, 2193–2223.

63. Eguiluz, V.M.; Chialvo, D.R.; Cecchi, G.A.; Baliki, M.; Apkarian, A.V. Scale free brain functional networks. *Phys. Rev. Lett.* **2005**, *94*, 018102.

64. Cohen, N. Fractals new era in millitary antenna design. *Defense Electronics*, 2005; Available online: http://rfdesign.com/mag/508RFDSF1.pdf (accessed on 15 January 2014).

65. Werner, D.H.; Werner, P.L. Frequency-independent features of self-similar fractal antennas. *Radio Sci.* **1996**, *31*, 1331–1343. ISSN abbreviation

66. Caludea, C.S.; Paun, G. Bio-steps beyond Turing. *BioSystems* **2004**, *77*, 175–194.

67. Achlioptas, D.; Souza, R.D.; Spencer, J. Explosive percolation in random networks. *Science* **2009**, *323*, 1453–1455.

68. Alexander, J.C.; Yorke, J.A.; You, Z.; Kan, I. Riddled basins. *Int. J. Bifurc. Chaos* **1992**, *2*, 795–813.

69. Ott, E.; Alexander, J.C.; Kan, I.; Sommerer, J.C.; Yorke, J.A. The transition to chaotic attractors with riddled basins. *Physica D* **1994**, *76*, 384–410.

70. Izhikevich, E.M.; Desai, N.S.; Walcott, E.C.; Hoppensteadt, F.C. Bursts as a unit of neural information: Selective communication via resonance. *Trends Neurosci.* **2003**, *26*, 161–167.

71. Izhikevich, E.M. Resonance and selective communication via bursts in neurons having subthreshold oscillations. *BioSystems* **2002**, *67*, 95–102.

72. Izhikevich, E.M. Resonate-and-fire neurons. *Neural Netw.* **2001**, *14*, 883–894.

73. Sahu, S.; Ghosh, S.; Hirata, K.; Fujita, D.; Bandyopadhyay, A. Multi-level memory-switching properties of a single brain microtubule. *Appl. Phys. Lett.* **2013**, *102*, 123701.

74. Sahu, S.; Ghosh, S.; Ghosh, B.; Aswani, K.; Hirata, K.; Fujita, D.; Bandyopadhyay, A. Atomic water channel controlling remarkable properties of a single brain microtubule: Correlating single protein to its supramolecular assembly. *Biosens. Bioelectron.* **2013**, *47*, 141–148.

75. Churchland, M.M. Neural population dynamics during reaching. *Nature* **2012**, *487*, 51–56.

76. Bandyopadhyay, A.; Fujita, D.; Pati, R. Architecture of a massive parallel processing nano brain operating 100 billion molecular neurons simultaneously. *Int. J. Nanotech. Mol. Comp.* **2009**, *1*, 50–80.

77. Bandyopadhyay, A.; Sahu, S.; Fujita, D.; Smallest artificial molecular neural-net for collective and emergent information processing. *Appl. Phys. Lett.* **2009**, *95*, 113702.

78. Bandyopadhyay, A.; Pati, R.; Sahu, S.; Peper, F.; Fujita, D. Massively parallel computing on an organic molecular layer. *Nat. Phys.* **2010**, *6*, 369–375.

79. Adamatzky, A. Molecular computing: Aromatic arithmetic. *Nat. Phys.* **2010**, *6*, 325–326.

80. London, M.; Schreibman, A.; Häusser, M.; Larkrum, M.E.; Segev, I. The information efficacy of a synapse. *Nat. Neurosci.* **2002**, *5*, 332–340.

81. Spillmann, L.; Werner, J.S. Long-range interactions in visual perception. *Trends Neurosci.* **1996**, *19*, 428–434.

82. Walling, P.T.; Hicks, K.N. Non-linear changes in brain dynamics during emergence from sevoflurane anesthesia. *Anesthesiology* **2006**, *105*, 927–935.

83. Tsuda, I.; Yamaguti, Y.; Kuroda, S.; Fukushima, Y.; Tsukada, M. A mathematical model for the hippocampus: Towards the understanding of episodic memory and imagination. *Prog. Theor. Phys. Suppl.* **2008**, *173*, 99–108.

84. Prévost, C.; McNamee, D.; Jessup, R.K.; Bossaerts, P.; O'Doherty, J.P. Evidence for model-based computations in the human amygdala during pavlovian conditioning. *PLoS Comput. Biol.* **2013**, *9*, e1002918.

85. Tian, X.; Xiao, Z.G. Functional Model of Brainstem-Cortex-Thalamus Circuit. In Proceedings of the 2005 Neural International Conference on Networks and Brain, ICNN&B '05, Beijing, China, 13–15 October 2005; Volume 3, doi:10.1109/ICNNB.2005.1614873.

86. Oku, M.; Aihara, K. A mathematical model of planning in the prefrontal cortex. *Artif. Life Robot.* **2008**, *12*, 227–231.

87. Schmahmann, J.D.; Caplan, D. Cognition, emotion and the cerebellum. *Brain* **2006**, *129*, 290–292.

88. Beiser, D.G.; Hua, S.E.; Houk, J.C. Network models of the basal ganglia. *Curr. Opin. Neurobiol.* **1997**, *7*, 185–190.

89. Ritch, S. Mathematical Model of Basal Ganglia and Oscillatory Neuron Clusters in Relation to Parkingsonian Tremors. Available online: http://www.math.duke.edu/mathbio/documents/reu-posters/rich_poster.pdf (accessed on 15 January 2014).

90. Nugent, A. Fractal Memory and Computational Methods and Systems Based on Nanotechnology. U.S. Patent 7502769 B2, 10 March 2009.

91. Ryu, J.H.; Tang, L.; Lee, E.; Kim, H.J.; Lee, M. Supramolecular helical columns from the self-assembly of chiral rods. *Chemistry* **2008**, *14*, 871–881.

92. Friston, K. The free energy principle: A unified brain theory? *Nat. Rev. Neurosci.* **2010**, *11*, doi:10.1038/nrn2787.

93. Golubitsky, M.; Lauterbach, R. Bifurcations from synchrony in a homogeneous networks. *SIAM J. Appl. Dyn. Syst.* **2009**, *8*, 40–75.

94. Sahu, S.; Oono, H.; Ghosh, S.; Bandyopadhyay, A.; Fujita, D.; Peper, F.; Isokawa, T.; Pati, R. Molecular Implementations of Cellular Automata. In *Cellular Automata*; Lecture Notes in Computer Science, Volume 6350; Springer: Berlin and Heidelberg, Germany, 2010; pp. 650–659.

95. Sahu, S.; Bandyopadhyay, A.; Fujita, D. Remarkable Potential of Pattern Based Computing on an Organic Molecular Layer Using the Concept of Cellular Automata. In Proceedings of International Symposium on Intelligent Signal Processing and Communication Systems, 2009, ISPACS 2009, Kanazawa, Japan, 7–9 January 2009; pp. 425–428. Our format, do not change

96. Bandyopadhyay, A.; Bhartiya, R.; Sahu, S.; Fujita, D. Investigating Universal Computability of Conventional Cellular Automata Problems on an Organic Molecular Matrix. In *Natural Computing*; Springer: Tokyo, Japan, 2010; pp. 1–12.

97. Sahu, S.; Oono, H.; Ghosh, S.; Bandyopadhyay, A.; Fujita, D.; Peper, F.; Isokawa, T.; Pati, R. On Cellular Automata rules of molecular arrays. *Nat. Comput.* **2012**, *11*, 311–321.

98. Anderson, J.A. A Brain-like Computer for Cognitive Software Applications: The Ersatz Brain Project. In Proceedings of the 4th IEEE International Conference on Cognitive Informatics, ICCI 2005, Irvine, CA, USA, 8–10 August 2005; doi:10.1109/COGINF.2005.1532612.

99. Friston, K. Hierarchical models in the brain. *PLoS Comput. Biol.* **2008**, *4*, e1000211.

100. Hoppensteadt, F.C.; Izhikevich, E.M. Thalamo-cortical interactions modeled by weakly connected oscillators: Could brain use FM radio principles? *BioSystems* **1998**, *48*, 85–94.

101. Kudo, M.; Sklansky, J. Comparison of algorithms that select features for pattern classifiers. *Pattern Recognit.* **2000**, *33*, 25–41.

102. Sternberg, R.J. *Cognitive Psychology*, 4th ed.; Thomson Wadsworth: Belmont, CA, USA, 2006.

103. Eyseneck, M.W.; Keane, M.T. *Cognitive Psychology: A Students Handbook*, 6th ed.; Psychology Press: Hove, UK, 2010.

104. Hung, N.V. Polygon Fractal. Ph.D. Thesis, Ernst-Moritz-Arndt-Universität, Greifswald, Germany, January 2007.

105. Milewski, R.; Govindaraju, V. Binarization and cleanup of handwritten text from carbon copy medical form images. *Pattern Recognit.* **2008**, *41*, 1308–1315.

106. Wolpert, D.H. *The Supervised Learning No Free Lunch Theorems*; Technical Report MS-269-1; NASA Ames Research Center: Mountain View, CA, USA, 2001; Available online: http://www.no-free-lunch.org/Wolp01a.pdf (accessed on 15 January 2014).

107. Andrew, N.; Dean, J. Building high-level features using large scale unsupervised learning, 2012. Available online: http://arxiv.org/pdf/1112.6209.pdf (accessed on 15 January 2014).

108. Utgoff, P.E.; Stracuzzi, D.J. Many-layered learning. *Neural Comput.* **2002**, *14*, 2497–2529.

109. Smolensky, P. Information Processing in Dynamical Systems: Foundations of Harmony Theory. In *Parallel Distributed Processing: Explorations in the Microstructure of Cognition*; Rumelhart, D.E., McClelland, J.L., CORPORATE PDP Research Group, Eds.; MIT press: Cambridge, MA, USA, 1986; Volume 1, pp. 194–281.

110. Chua L.O. Memristor—The missing circuit element. *IEEE Trans. Circuits Theory* **1971**, *18*, 507–519.

111. Di Ventra, M.; Pershin, Y.V.; Chua, L. Circuit elements with memory: Memristors, memcapacitors and meminductors. *IEEE Proc.* **2009**, *97*, doi:10.1109/JPROC.2009.2021077.

112. Sahu, S.; Fujita, D.; Bandyopadhyay, A. An inductor made of arrayed capacitors. JP-096217 (world patent filed), 2010.

113. Cosp, J.; Madrenas, J.; Alarcon, E. Synchronization of non-linear electronic oscillators for neural computation. *IEEE Trans. Neural Netw.* **2004**, *15*, 1315–1327.

114. Bandyopadhyay, A.; Miki, K. A vertical parallel processor. JP-5187804, 2013.